\documentclass[12pt,a4paper,english,twoside,nofootinbib
]
{book}
\usepackage{babel}
\usepackage{latexsym,amsmath,amssymb}
\usepackage{graphicx,overpic}
\usepackage{mathpazo,courier}\usepackage[scaled=.92]{helvet}
\normalfont
\usepackage[T1]{fontenc}\usepackage{textcomp}
\usepackage{graphicx,color,xspace}
\usepackage{hyperref}\hypersetup{pdfstartview=FitH,pdfpagemode=None}
\usepackage{footmisc}
\usepackage{rotating}
\usepackage{epstopdf}
\usepackage{fancyhdr}
\usepackage[sort&compress,square]{natbib}

\hyperbaseurl{http://arxiv.org/abs}

\newcommand{\figcaptionl}[2]{
    \begin{center}\begin{minipage}{.92\textwidth}
      \caption{\small   #2}\label{#1}
    \end{minipage}\end{center}}

\def\g{\text{\sf\itshape g}}

\def\d{{\mathrm{d}}}
\def\x{{\mathbf x}}
\def\v{{\mathbf v}}
\def\r{{\mathbf r}}
\def\k{{\mathbf k}}

\def\im{{\rm i}}

\def\det{{\mathrm{det}}}

\def\bnabla{\mbox{\boldmath$\nabla$}}
\def\Box{\kern0.5pt{\lower0.1pt\vbox{\hrule height.5pt width 6.8pt
  \hbox{\vrule width.5pt height6pt \kern6pt \vrule width.3pt}
  \hrule height.3pt width 6.8pt} }\kern1.5pt}

\makeatletter
\def\cleardoublepage{\clearpage\if@twoside \ifodd\c@page\else
    \hbox{}
    \thispagestyle{plain}
    \newpage
    \if@twocolumn\hbox{}\newpage\fi\fi\fi}
\makeatother \clearpage{\pagestyle{plain}\cleardoublepage}

\newcommand{\eplanck}{ E_* }

\providecommand{\abs}[1]{\lvert#1\rvert}

\numberwithin{equation}{section}

\graphicspath{{figures/}}

\begin{document}
\thispagestyle{empty}
\begin{titlepage}
\begin{center}
 {\large Departamento de F\'{i}sica Te\'{o}rica II} \\[15pt]
 {\large \textbf{Facultad de Ciencias F\'{i}sicas}} \\[15pt]
 {\large \textbf{Universidad Complutense de Madrid}} \\[15pt]
\vspace*{5mm}
\resizebox{30mm}{!}{\includegraphics{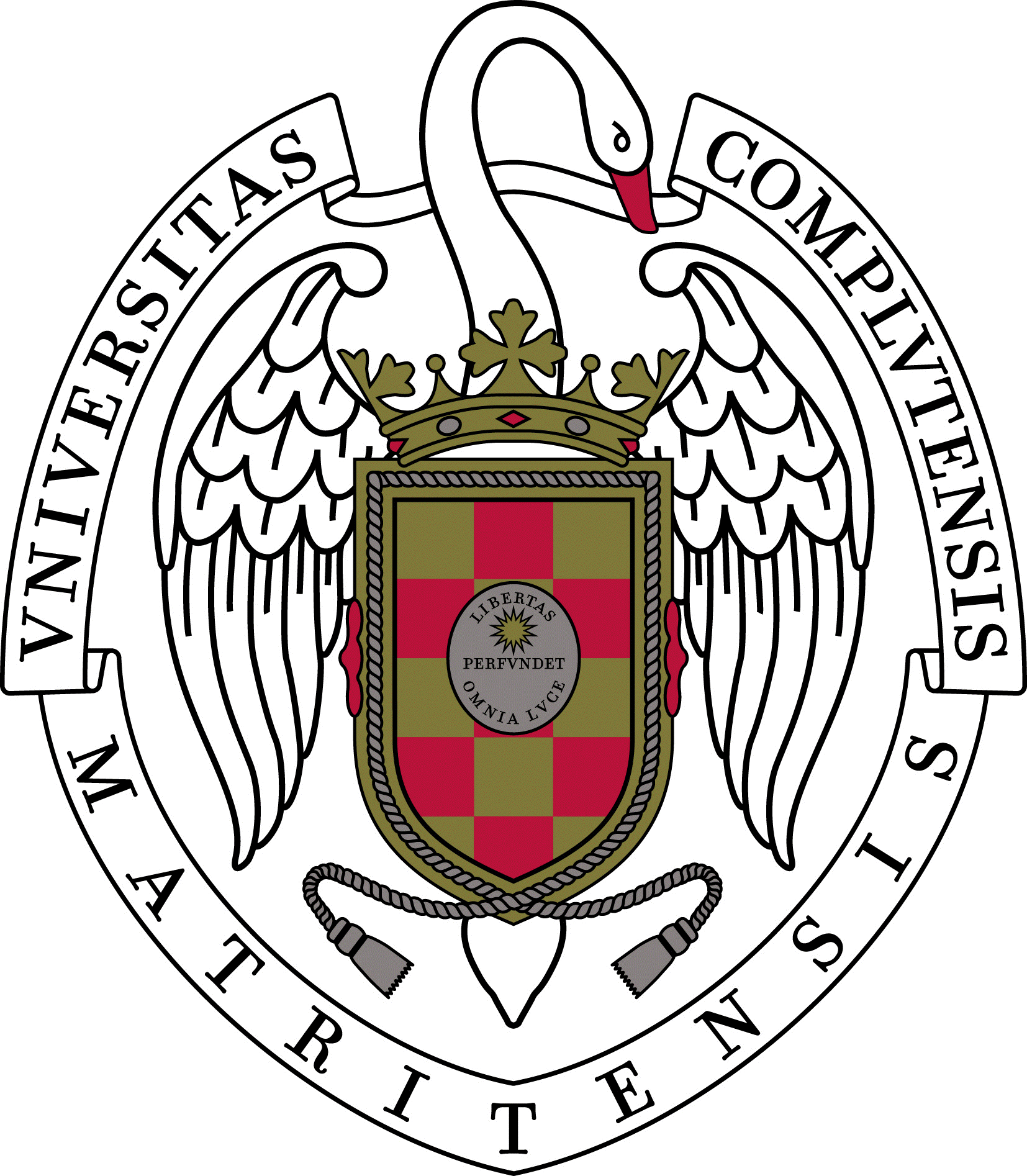}}
\end{center}
\vspace*{10mm}\par
\begin{center}
{\LARGE \textbf{EMERGENT GRAVITY :}} \\ [15pt] {\LARGE \textbf{THE BEC PARADIGM}}

\vspace*{20mm}\par
{\Large \textbf{Gil Jannes$^{1,2}$}}
\end{center}

\vspace*{5ex}
\begin{center}
\bf{PhD Thesis}
\end{center}

\vspace*{5ex}

\parbox{3in}{
Research advisors:\\
Dr.\ Carlos Barcel\'{o} Ser\'{o}n$^1$} \par Dr.\ Luis J. Garay Elizondo$^{3}$

\vspace*{-14mm}\par
 \hspace*{1mm} \hfill
\parbox{2.2in}{
Madrid, 2009}

\enlargethispage{10mm}
\vfill\par \noindent\small
 $^1$ Inst.\ de Astrof\'{i}sica de Andaluc\'{i}a, CSIC, Camino Bajo de Hu\'{e}tor 50, 18008 Granada, Spain.\\
 $^2$ Inst.\ de Estructura de la Materia, CSIC, C/ Serrano 121,
 28006 Madrid, Spain.\\
 $^3$ Depto.\ de F\'{i}sica Te\'{o}rica II, Universidad Complutense de
 Madrid, 28040 Madrid, Spain.\normalsize
\end{titlepage}

\newpage
\thispagestyle{empty}\hspace*{1mm}

\newpage
\thispagestyle{plain}

\newpage
\chapter*{Acknowledgements}

I first and foremost thank Carlos Barcel\'{o} and Luis Garay, my research advisors. It has been a tremendous pleasure to work with two persons who are not only outstanding scientists, but moreover extraordinary personalities. And I immediately wish to extend this description to Guillermo Mena, even though we have not (yet) directly collaborated in terms of science. At all times, I felt like a colleague and friend more than (just) a PhD student, and that is the greatest compliment you could have paid me. I hope that this thesis will mark a starting point rather than an endpoint.

Many thanks to Stefano Liberati, Grisha Volovik and Matt Visser, for their hospitality and for many illuminating scientific discussions. Particular thanks to Matt for the extensive gastronomic and touristic tour of Wellington and surroundings.

There is a long list of people with whom I had plenty of discussions that were totally irrelevant in any scientific way, but nonetheless very entertaining. Rather than attempting to give an exhaustive list and perhaps offending some by forgetting to mention them, I will only name one person explicitly: the one who has paid me the largest amount of beers. That honour goes without a doubt to Jos\'e Antonio S\'anchez Gil. {\it Salud!}

Thanks to the few of my friends who have shown genuine interest in my work. Thanks also to the large majority who, totally understandably, have shown only a marginal interest, or none at all. After all, even if one finds great pleasure in work, being with friends who make you forget work is sometimes an even greater pleasure.

Thanks to my father for being the most avid supporter of my scientific career, and for trying to convince the rest of the family, with variable degrees of success.

Thanks to Wendy for enduring the absurdities and insecurity of a partner in academic life. And of course for being the most marvellous, smart, sweet, funny and sexy woman in the world, thus helping to make my life the wonderful adventure it is.

I would end with a word to Igor, but you wouldn't care, and you are absolutely right. Everything that I have written in this thesis falls into utter meaninglessness compared to the simple pleasure of seeing you smile when I wake you up in the morning. So I will limit myself to apologise for having spent less time playing with you than I should while I was writing up this thesis.

\vfill
\hfill
\parbox{4in}
{{\it Whenever a theory appears to you as the only possible one, take this as a sign that you have neither understood the theory nor the problem which it was intended to solve.}\\
\flushright Karl R. Popper\hspace*{15mm}}

\newpage
\tableofcontents

\newpage
\addcontentsline{toc}{chapter}{Abstract}
\chapter*{Abstract}



We study selected aspects of quantum gravity phenomenology inspired by the gravitational analogy in Bose--Einstein condensates (BECs). We first review the basic ideas and formalism of analogue gravity in BECs, with particular emphasis on the possibility of simulating black holes. We stress that the dispersion relation in a BEC has a relativistic form in the hydrodynamic limit, and acquires non-relativistic, `superluminal' modifications at high frequencies. This makes it a particularly interesting model for many scenarios of quantum gravity phenomenology which consider a possible violation of local Lorentz invariance at high energies. In particular, it allows the study of kinematical corrections that such quantum gravity scenarios could impose on general relativity. 

We present a simple model for a (1+1)-dimensional acoustic black hole configuration in a BEC, and study its {\it dynamical stability}. We find that black hole horizons in a BEC are {\it stable} under quite general boundary conditions, inspired by similar studies of the stability of black holes in general relativity. This allows us to study the stable dynamical modes or quasinormal modes of such configurations, as well as to confirm the long-expected possibility of (at least in principle) simulating Hawking radiation in a BEC. 

We find that there indeed exist short-lived {\it quasinormal modes} in our model, in spite of its (1+1)-dimensionality, and moreover that their spectrum covers a {\it continuous} region of the complex frequency plane. This is in sharp contrast to general relativistic black holes and to acoustic black holes in the hydrodynamic limit, where no quasinormal modes exist in 1+1 dimensions, whereas in higher dimensions the spectrum is discrete. The quasinormal modes that we find are characteristic of a relaxation at the microscopic level, and we trace their presence back to the superluminality of the dispersion relation, and the associated permeability of the horizon. We argue that a similar effect could show up for astrophysical black holes if quantum gravity turns out also to lead to superluminality at high frequencies. 

We then study the impact of superluminal dispersion relations on the {\it Hawking radiation} for a collapsing geometry in which a black hole is created. A crucial aspect of modified dispersion relations is that the horizon becomes a frequency-dependent concept. In particular, with superluminal dispersion relations, at every moment of the collapse process, there is a critical frequency above which no horizon is experienced. If moreover there is an overall saturation level to the collapse (as is certainly the case in BECs due to the interatomic separation, and as is expected in the gravitational case due to quantum gravity resolving the general relativistic singularity), then there is also a global critical frequency, such that higher frequencies never see a horizon and hence do not participate in the thermal radiation process. This leads to several important differences with the standard Hawking radiation. Generally speaking, a finite critical frequency implies that the {\it intensity} of the late-time radiation is weaker than in the standard (Lorentz invariant) case. The {\it Planckian form} of the thermal spectrum is largely preserved (although with a decreased intensity) unless the Lorentz symmetry violation scale (the scale at which non-relativistic deviations in the dispersion relation become important) lies below the critical frequency. In the latter case, high frequencies (above the Lorentz symmetry violation scale) can acquire extremely large surface gravities, leading to {\it important ultraviolet contributions} in the radiation spectrum. Finally, the radiation originating from the collapse {\it dies out} on a relatively short time-scale.

In the final part of this thesis, we discuss some questions related to the possibility of constructing a serious toy model for Planck-scale gravity based on the condensed matter analogy, namely a model of {\it emergent gravity}. The basic idea is that gravity (including its dynamical aspects, i.e., the Einstein equations) might emerge as an effective low-energy macroscopic description of the collective behaviour of the microscopic constituents of a system. A crucial consequence if this turns out to be a correct approach is that there would not be any direct quantisation procedure leading from the macroscopic to the microscopic degrees of freedom. We mention some general motivations for such an approach, in particular the problem of dark energy, and discuss in detail the question of {\it diffeomorphism invariance} in emergent gravity based on condensed matter models. We argue that this question should be split up into a kinematical and a dynamical part. We explain how diffeomorphism invariance at the kinematical level can actually be understood as a low-energy effective symmetry in such models, just like Lorentz invariance. With respect to the dynamical aspects of diffeomorphism invariance, we argue that this is essentially identical to the problem of recovering the Einstein equations in a condensed-matter-like model, and make some remarks with regard to the possibility that Sakharov's induced gravity proposal could solve this issue.

\newpage
\chapter{Introduction}
The fundamental observation underlying the study of models for gravity in Bose--Einstein condensates (BECs) is the following~\cite{Visser:1997ux}. The
equation of motion of acoustic perturbations in a perfect (irrotational, inviscid and barotropic) fluid (such as a BEC---under certain approximations which we will detail extensively in due time) is described by a d'Alembertian equation in curved spacetime: 
\begin{eqnarray}\label{alembert}%
\square \phi\equiv\frac{1}{\sqrt{-\g}} \partial_\mu \sqrt{-\g} \g^{\mu\nu} \partial_\nu \phi =0~.
\end{eqnarray}%
These acoustic perturbations or phonons therefore travel along the null geodesics of the effective metric $\g_{\mu\nu}$, with $\g$ its determinant. This formula is well known from relativistic field theory as the equation of motion for a massless scalar field $\phi$ propagating in a curved spacetime. Moreover, up to a conformal factor, the effective metric has the following form:
\begin{eqnarray}\label{intro-metric}
\g_{\mu \nu} \propto
\begin{pmatrix}
v^2-c^2 && -\v^\text{T}\\
-\v && \mathbb{1}
\end{pmatrix},
\end{eqnarray}
where $\v$ is the velocity vector of the background fluid flow and $c$ the speed of sound in the fluid.\footnote{Throughout this thesis, we will only explicitly use a different notation for the speed of sound and the speed of light when this seems useful to avoid confusion. We will however always write $c$ explicitly rather than setting $c=1$.} Since $\g_{tt}\propto (v^2-c^2)$, the sign of $\g_{tt}$ depends on the fluid regime, indicating that black hole configurations are in principle possible, at least in radial flows. 

This basic observation is the essence of {\it analogue gravity} in condensed matter models, which offers the promising prospect of allowing to better understand and possibly simulate some phenomena related to high-energy physics, and to black holes in particular, as we briefly discuss in the following section. Its interpretation can also be extended in two directions, which we will call {\it quantum gravity phenomenology} and {\it emergent gravity}, in logical order, and which we will introduce thereafter.

\section{Analogue gravity in condensed matter}
The basic observation described above implies that one can study certain aspects of general relativity and quantum field theory, and in particular of black hole physics, by analogy with fluid systems such as a BEC~\cite{Garay:1999sk,Garay:2000jj}. To put it in other words: perfect fluids mimic the {\it kinematical} aspects of general relativity, i.e., the propagation of fields in curved relativistic spacetimes, including the possibility of black hole configurations. Note that there are currently no known analogue systems which reproduce the Einstein equations, which is why the analogy does not immediately extend to the {\it dynamical} level. 

To take maximal advantage of the analogy, the  microscopic physics of the fluid system should be well understood, theoretically and experimentally, preferably even in regimes where the relativistic description breaks down. Then, full calculations
based on firmly verified and controlled physics are possible, even beyond the relativistic regime. Additionally, laboratory
experiments become feasible that could shed light on issues of high-energy physics. 
There exists a wide variety of physical systems, both condensed matter and other (see~\cite{Barcelo:2005fc} for a review), in which curved effective relativistic spacetimes emerge. BECs stand out mainly for their conceptual clarity, both as a source of the gravitational analogy (excitations moving in a mean-field background provided by the collective behaviour of the condensed phase of the condensate) and as a generally well understood and experimentally well controlled system, thereby offering the exciting prospect of directly simulating the analogues of some phenomena predicted in high-energy physics. For instance, BECs are considered a good
candidate for a possible future experimental detection of (phononic) Hawking
radiation~\cite{Barcelo:2001ca,Carusotto:2008ep,Cornell:2009} or of particle creation in expanding spacetimes~\cite{Jain:2007gg,Weinfurtner:2008if}.

\section[From analogue gravity to QG phenomenology]{From analogue gravity to quantum gravity\\ phenomenology}
However, the analogy is not limited to the propagation of relativistic fields in curved background spacetimes. A particularly interesting application is in terms of quantum gravity phenomenology. Indeed, the geometric or relativistic regime described above is not universally valid in a BEC, but only in a certain approximation. Essentially, it is valid for the low-energy modes or phonons, i.e., in the {\it hydrodynamic} approximation. For higher energies, and also in the presence of a horizon (as we will emphasise repeatedly throughout this thesis), the relativistic or hydrodynamic approximation breaks down, giving way to the underlying microscopic description. This behaviour is of course similar to what is expected in the realm of quantum gravity, with the crucial difference that the underlying microscopic description is well known in the case of BECs. Therefore, {\it deviations} from the hydrodynamic or relativistic regime in BECs provide possible insights into the kinematical corrections that a hypothetical quantum theory of gravity might impose on general relativity,\footnote{We again stress the kinematical aspect: the condensed matter analogy does not reproduce the Einstein equations, and hence does not allow to study quantum gravitational corrections to the {\it dynamics} of general relativity beyond the linear regime.} for example on the local Lorentz invariance that characterises general relativity. They thus provide a way of studying quantum gravity phenomenology by analogy. Such extrapolations from condensed matter models to quantum gravity phenomenology are especially relevant if the deviations from classical general relativity imposed by quantum gravity turn out to be roughly similar to those that occur in BECs with respect to the relativistic regime. In BECs, the effective low-energy Lorentz symmetry is broken as follows. The full dispersion relation between the frequency $\omega$ and the wave number $k$ is of the `superluminal' (strictly speaking: supersonic) type:
\begin{equation}
\omega^2=c^2k^2+\alpha^2c^2k^4~,
\end{equation}
with $\alpha$ a parameter (of dimension $[\alpha]=L$) sufficiently small such that a relativistic dispersion relation $\omega^2=c^2k^2$ is recovered at low energies. Note again, as in the metric~\eqref{intro-metric}, that $c$ represents here the (low-frequency) speed of sound.
This dispersion relation indicates that the high-frequency modes move at a speed higher than the speed of sound. Essentially, the gravitational analogy in a BEC is then of particular relevance for quantum gravity phenomenology if quantum gravity (independently of its fundamental structure) modifies the observable low-energy physics in a way that can be written as a modified dispersion relation with respect to some preferred reference frame.\footnote{Note that we use the term `{\it modified} dispersion relation' in the sense traditional in quantum gravity phenomenology, i.e.: modified {\it with respect to a relativistic dispersion relation}.} This is obviously an experimental issue which is far from being settled. Ultimately, it means that the condensed matter model is not a good phenomenological model if quantum gravity turns out to be strictly Lorentz invariant, although even then it could still teach us valuable lessons about the implications of this strict invariance. It is perhaps also of limited use for Doubly Special Relativity and related types of Lorentz modification which can also lead to modified dispersion relations, but in general of a more complicated form and, more importantly, without breaking the equivalence of reference frames~\cite{AmelinoCamelia:2000mn}. However, it is certainly a natural model for effective field theory approaches in general, which form the bulk of current efforts at quantum gravity phenomenology~\cite{Mattingly:2005re,AmelinoCamelia:2008qg}. 

Quantum gravity phenomenology from condensed matter models is therefore an illustration of what one could call a `bottom-up' approach to quantum gravity: a study of the first deviations that could occur with respect to the known low-energy (Lorentz invariant) laws of physics of our universe, based on the kind of deviations that occur in laboratory systems which reproduce low-energy effective Lorentz invariant spacetimes similar to the one of our universe. This approach is therefore complementary to the traditional `top-down' approaches to quantum gravity such as string theory or loop quantum gravity, which postulate a basic structure for our universe or its geometry (or a basic set of mathematical guidelines for arriving at such a structure) and then attempt to connect these basic principles with our known physics. Recovering the (semi)classical limit for our universe (i.e., the ``reconstruction problem'', as it is called in~\cite{Carlip:2001wq}) is precisely one of the main open problems in all such top-down approaches. Actually, at present, it is not clear whether the semiclassical limit is uniquely defined for any of those top-down approaches, and so none of them make any falsifiable prediction with respect to modified dispersion relations, or even with respect to Lorentz invariance in general. Therefore, to bridge the gap, it should be clear that there is a genuine interest for a bottom-up approach such as the one discussed here.

\section{From analogue gravity to emergent gravity}
Finally, the condensed matter analogy with gravity can be taken another step further. If relativistic spacetimes naturally emerge in a multitude of condensed matter systems, then perhaps gravity as such (now, including its dynamical aspects) is actually a generic property of many physical systems, and not necessarily limited to the basic structure of our universe. In other words, gravity might be an effective description emerging at the macroscopic level as a consequence of the collective behaviour of the microscopic constituents of a system. It could even emerge (in some limit) in a wide variety of microscopic systems. If this turns out to be a correct description of gravity, then one important consequence would be that there is no direct quantisation procedure leading from the macroscopic to the microscopic degrees of freedom. 
Part of the motivation for such an approach is to be found in the previous paragraph: contrarily to the case in top-down approaches for quantum gravity, the semiclassical limit is well defined in the condensed matter model. Although not central to this thesis, we will also briefly comment (in chapter~\ref{S:emergent-gravity}) on a second important motivation, namely the dark energy problem. 

However, such an approach also raises a number of important questions. First, as observed repeatedly above, the gravitational analogy in condensed matter systems is currently limited to the kinematical aspects of general relativity, i.e., to the propagation of (classical or quantum) fields in curved relativistic spacetimes. This seems to be an important limitation: any serious toy model for a quantum theory of gravity should lead to the Einstein equations in the adequate regime. Therefore, a crucial question in this respect is the following. Does the condensed matter analogy simply break down at the observation that the Einstein equations do not reproduce in a laboratory, or can we take this further and understand precisely why our universe is ruled by the Einstein equations while laboratory condensed matter systems are not? A second question is related to diffeomorphism invariance and background independence. The effective geometries that emerge in condensed matter models are obviously background dependent: they depend on the presence of the condensed matter atoms, which are localised in the geometry of the laboratory. When extending this to a toy model for the gravitation of our universe, one is naturally led to a structure reminiscent of the ether theories that were popular before the advent of special relativity~\cite{Barcelo:2007iu,einstein1920}, whereby the `spacetime atoms' are localised in an absolute predefined geometry. In particular, this seems to suggest that emergent gravity models inspired by the condensed matter analogy are not diffeomorphism invariant in general, contrarily to other approaches such as loop quantum gravity, in which diffeomorphism invariance is regarded as a crucial requirement for quantum gravity~\cite{Rovelli:2004tv}.

We will argue that the problem of diffeomorphism invariance can be separated in a kinematical part and a dynamical part. Since we have repeatedly stated that condensed matter models reproduce the kinematical aspects of general relativity, it perhaps comes as no surprise that the kinematical aspect of the diffeomorphism invariance problem has a solution, which actually involves more conceptual subtlety than technical machinery. The dynamical aspect of the diffeomorphism invariance question, however, is very much an open problem. We will limit ourselves to point out that it is intimately related to the problem of reproducing the Einstein equations, and discuss a possible way to tackle this problem, based on Sakharov's induced gravity~\cite{Sakharov:1967pk}, together with some of its problems.

\section{A very brief history of analogue gravity}
In this section, we will mention some historical landmarks in the development of analogue models of gravity with particular relevance to the present thesis. A comprehensive review with an extensive history section can be found in~\cite{Barcelo:2005fc}.

Although it had already been known for several years that relativistic metrics such as~\eqref{intro-metric} could be written explicitly for the propagation of sound waves in inviscid fluids~\cite{white1973}, the original observation that this allows (acoustic) black hole configurations, and thereby the possibility of studying high-energy (quantum gravitational) effects on black hole physics is due to Unruh in a seminal paper published in 1981~\cite{Unruh:1980cg}. Perhaps curiously, this paper received little attention, until Jacobson took up Unruh's suggestion to study the transplanckian problem in Hawking radiation ten years later~\cite{Jacobson:1991gr}. This led to a burst of interest and the appearance of several papers on the robustness of Hawking radiation (see chapter~\ref{S:HR}), all based on more or less abstract models for the transplanckian physics. In a crucial paper from 1998~\cite{Visser:1997ux}, Visser established the gravitational analogy on firm foot by deriving~\eqref{alembert} and~\eqref{intro-metric} as a rigorous theorem of mathematical physics and studying in detail the analogue equivalents of quantities associated with black holes such as the horizon, ergosphere and surface gravity.

In parallel, several proposals were made for concrete physical systems in which black hole analogies could be simulated, including (see~\cite{Barcelo:2005fc} and references therein) solid state models, slow light in dielectrics, gravity waves in shallow water, and two key proposals in the area of condensed matter: a series of works by Volovik on gravitational analogies in $^3$He-A, culminating in the appearance in 2003 of the book~\cite{Volovik:2003fe}, and the proposal by Garay {\it et al.} in 2000 to simulate black holes in Bose--Einstein condensates~\cite{Garay:1999sk}. Shortly thereafter, the first explicit discussion of how BEC black holes could be used to simulate Hawking radiation in a laboratory is due to Barcel\'{o}, Liberati and Visser~\cite{Barcelo:2001ca}. A workshop held in Rio de Janeiro in 2000, exclusively dedicated to analogue models of gravity, marked the maturity of the field and led to the publication of the reference book~\cite{Novello:2002qg}.

Finally, to end this brief overview, the 2005 review~\cite{Barcelo:2005fc} firmly settled the field of analogue gravity as an active and well-established research programme within the gravity community, while it seems that the idea of possible laboratory simulations of Hawking radiation has matured sufficiently to be taken very seriously in the condensed matter community, as illustrated for example by~\cite{Carusotto:2008ep,Cornell:2009}.

Note that most of the work appearing on gravitational analogies can indeed be divided into the three categories that we discussed above: analogue gravity as such (with a main focus on laboratory simulations), quantum gravity phenomenology, and emergent gravity:
\begin{itemize}
 \item  An increasingly important number of papers, in the line of~\cite{Barcelo:2001ca} cited above, discuss the possibility of laboratory simulations of phenomena such as Hawking radiation or particle creation in expanding universes. 
\item A second set of works, following the central message of Unruh's original paper~\cite{Unruh:1980cg}, use the gravitational analogy in fluid and other systems mainly as an inspiration for quantum gravity phenomenology, the lion's share being dedicated to the possible influence of transplanckian physics on Hawking radiation. 
\item Finally, Volovik~\cite{Volovik:2003fe}, apart from studying the possibility of laboratory experiments on the gravitational analogy in liquid Helium, is also the main advocate of taking the condensed matter analogy seriously as a candidate toy model for emergent gravity.
\end{itemize}

\section{Synopsis}
Most of the work discussed in this thesis is relevant mainly for the second level of interpretation, namely that of quantum gravity phenomenology. It also has obvious consequences for analogue gravity and laboratory simulations as such, in the following sense. We will see that departures from the hydrodynamic regime in a BEC (at high energies and in the presence of horizons) predict deviations from the standard relativistic phenomena related to black hole physics, such as Hawking radiation. Therefore, if these deviations are sufficiently strong, then apart from providing a possible window to quantum gravity phenomenology, they could also show up spontaneously and even inevitably in laboratory experiments reproducing the BEC analogues of these relativistic phenomena. We will end this thesis by making some comments on the third and strongest level of interpretation, which takes the condensed matter model seriously as a possible toy model for a full theory of quantum gravity.

The structure of this thesis is then the following. In chapter~\ref{S:preliminaries}, we discuss some general preliminaries with respect to effective spacetimes in condensed matter systems and in BECs in particular. Since this thesis is mainly directed at the (quantum) gravitational community, we start with a brief refresher of some basic BEC theory, including the Gross--Pitaevskii equation which describes the evolution of the condensed phase of the BEC in the mean-field approximation. We then describe the obtention of the effective acoustic metric, and stress that this allows the consideration of black hole configurations. Finally, we derive the Bogoliubov dispersion relation and look at the hydrodynamic limit. We emphasise that in this hydrodynamic limit, the dispersion relation acquires a relativistic form, and examine the validity of this hydrodynamic or relativistic limit. 

In chapter~\ref{S:BHconfigurations}, we present a simple model for effectively (1+1)-dimensional black hole configurations in a BEC. The model consists of two homogeneous regions with a step-like discontinuity in the density profile. Although such a model is highly idealised, we argue that it captures the essence of the physics that we are interested in, at least for the purpose of studying the dynamical behaviour of black hole horizons in BECs. We derive the matching conditions that connect the magnitudes describing the condensate, such as its density and phase and their derivatives, across the discontinuity. Finally, we develop a numerical method to find the dynamical modes (instabilities and quasinormal modes) of such configurations. 

This setup is then applied to the dynamical modes of such (1+1)-dimensional black hole configurations in a BEC, namely an analysis of the dynamical instabilities in chapter~\ref{S:instabilities}, and a quasinormal mode analysis in chapter~~\ref{S:QNMs}, respectively. 

In chapter~\ref{S:instabilities}, we first indicate the slightly confusing state of the literature with respect to the dynamical stability of black hole horizons in BECs, and highlight the importance of this issue with respect to the possibility of simulating Hawking radiation in a BEC. We then thoroughly discuss the boundary conditions that are needed to apply the algorithm developed in the previous chapter, and analyse the possible appearance of dynamical instabilities. We discuss the results both for black hole and white hole horizons, and show that black hole horizons are generally stable, although we stress that modifying the boundary conditions (e.g., by introducing a wall or sink into the condensate) could affect this result. For a white hole horizon, the stability or instability is shown to depend even more crucially on the type of boundary conditions applied. In any case, we conclude that stable black hole configurations are possible, at least in principle. This means that studying the stable dynamical modes or quasinormal modes, as well as the Hawking radiation, of BEC black hole configurations indeed makes sense. 

The structure of chapter~\ref{S:QNMs} is the following. After a general introduction on quasinormal modes of black holes (essentially: relaxation modes of the black hole after perturbation), we examine the boundary conditions that are applied in general relativity to calculate the quasinormal modes of a black hole, and discuss how these should be modified in the case of BECs due to the superluminality of the dispersion relation. We emphasise that in general relativity, no quasinormal modes exist in (1+1)-dimensional configurations, whereas a discrete spectrum appears in 3+1 dimensions. The same is true for acoustic black holes in the hydrodynamic limit. We then apply the procedure outlined in chapter~\ref{S:BHconfigurations} to verify the occurrence of quasinormal modes in the simple (1+1)-dimensional model presented earlier when using the full dispersion relation, including non-relativistic corrections. Not only do we find that such quasinormal modes indeed exist, but quite remarkably, their spectrum turns out to cover a continuous region of the complex frequency plane. A rough dimensional estimate shows that these quasinormal modes are short-lived, indicative of a relaxation at the microscopic scale. This leads us to speculate that the discrete spectrum resulting from the standard quasinormal mode analysis in (3+1)-dimensional general relativity will also develop continuous bands if the dispersion relations at high energies undergo superluminal modifications. Such additional modes, though perhaps too short-lived to be detectable directly, could in principle leave a trace of transplanckian (quantum gravitational) physics in the spectrum of gravitational waves emitted by black holes. 

In a brief intermezzo, chapter~\ref{S:intermezzo}, we discuss in slightly more detail the link between modified dispersion relations of the type that occur in BECs, and the Lorentz symmetry violation scenarios that are expected in quantum gravity phenomenology. This link justifies why we speculate that some of the results obtained in the two previous chapters can be directly extrapolated to such scenarios of quantum gravity phenomenology. It also allows us, instead of developing the gravitational analogy starting from the microscopic theory of a BEC as we did in the previous chapters, to directly study the impact of modified dispersion relations. Such an approach is then valid both for quantum gravity phenomenology in a scenario with superluminal dispersion relations, and for analogue gravity in a BEC. 
We adopt this strategy in the next chapter.

Chapter~\ref{S:HR} studies the impact of superluminal dispersion relations on the late-time radiation or Hawking radiation from the formation of a black hole through a dynamical collapse. We start with a detailed motivation of why this study is relevant in spite of the extensive literature already existing with regard to Hawking radiation and the transplanckian problem. In particular, we highlight that most existing studies have focused on stationary black hole configurations and subluminal dispersion. We also stress, as already mentioned at the beginning of this section, that if strong deviations from the standard Hawking picture are predicted, then these could spontaneously show up in laboratory simulations of the phononic equivalent of Hawking radiation in a BEC. We then discuss the classical geometry of our collapse model, and propose two concrete profiles, which differ in the behaviour of their surface gravity. The first profile considers a black hole with a constant surface gravity, while in the second one---characteristic of a Schwarzschild black hole---the surface gravity increases with frequency. This frequency-dependence of the surface gravity can be traced back to the fact that the horizon itself becomes a frequency-dependent concept because of the modified dispersion relation, an issue which we discuss in detail. The main body of the chapter consists in a concise derivation of the standard Hawking radiation, i.e., with relativistic dispersion relations, and a demonstration of how this derivation can be adapted to account for superluminal dispersion relations. We obtain an analytic expression for the late-time radiation from the collapse of a black hole with superluminal high-frequency dispersion. Three important differences with the standard case can be deduced from this expression. First, we highlight the dependence on a critical frequency which appears during the collapse process due to the superluminal dispersion, and above which no horizon is experienced. Second, we stress the importance of the surface gravity, which is shown to become frequency-dependent as well, again due to the superluminality of the dispersion relation. Finally, we point out that the radiation originating from the collapse process dies out as time advances. We then numerically integrate the analytic expression for the late-time radiation, and graphically illustrate the three issues just mentioned. In particular, it is clearly seen that the existence of a critical frequency implies that the radiation is fainter than in the standard case, even in a profile with constant surface gravity, unless this critical frequency goes to infinity (i.e., unless a singularity is formed at the end of the collapse). We also illustrate that the dependence on the surface gravity could lead to important ultraviolet contributions, depending on the relation between the critical frequency and the Lorentz symmetry violation scale. Finally, we show the time evolution of the radiation originating from the collapse and indicate that, for a natural choice of magnitudes, this extinguishes extremely rapidly. We conclude with a comparison of our results with other recent results available in the literature, and speculate that some of the standard assumptions that are taken when arguing for the robustness of Hawking radiation, which are directly extrapolated from the standard case, might be model-dependent and actually broken in a physically realistic collapse scenario with superluminal dispersion. 

Chapter~\ref{S:emergent-gravity} discusses some aspects of how the analogue gravity framework in condensed matter models can be extended to a serious model for emergent gravity. We start with some general considerations on what quantum gravity could or should be, and mention that one of the main motivations for an emergent gravity approach is the dark energy problem. The bulk of this chapter is devoted to the question of diffeomorphism invariance in emergent gravity models based on the condensed matter analogy. We argue that this question should be split up into a kinematical aspect and a dynamical aspect. We discuss why the kinematical aspect of diffeomorphism invariance poses no real problem, since diffeomorphism invariance emerges at the same level as Lorentz invariance, i.e., it is realised as an effective low-energy symmetry. This discussion involves some subtle concepts concerning the relation between internal observers and microscopic physics. We end with some remarks with respect to the dynamical aspect of diffeomorphism invariance. We argue that this is essentially identical to the problem of recovering the Einstein equations in analogue gravity models, and discuss the possibility of using Sakharov's induced gravity to such an end, an issue which is under current investigation.

We round off in chapter~\ref{S:summary} with an extensive general summary and outlook.


\newpage
\chapter{Preliminaries}
\label{S:preliminaries}
\section{General considerations}
It is widely expected that underneath the general relativistic description
of gravitational phenomena there is a deeper layer
in which quantum physics plays an important role. However, at this
stage we do not have sufficient intertwined theoretical and observational
knowledge to formulate an appropriate description of what underlies
gravity. Moreover, starting from
structurally complete quantum theories of gravity, it could still be
very difficult to extract the specific way in which the first
`quantum' modifications to classical general relativity might show
up. This happens for example within the loop quantum gravity
approach~\cite{Ashtekar:2004eh,Rovelli:2008zza}, while in string theory a huge `landscape' of quantum vacua seems to arise~\cite{Susskind:2003kw}, although it is not clear yet which ones---if any---could correspond to our universe~\cite{Kumar:2006tn}.

Analogue models of general relativity in condensed matter systems provide specific and clear examples
in which effective spacetime structures ultimately emerge from
(non-relativistic) quantum many-body systems. As mentioned in the introduction, the key
point is the realisation that some collective properties of these
condensed matter systems satisfy equations of motion formally
equivalent to those of a relativistic field in a curved spacetime. The
simplest example is the equation describing a massless scalar field
$\phi$,
\begin{eqnarray}%
\square \phi\equiv\frac{1}{\sqrt{-\g}} \partial_\mu \sqrt{-\g} \g^{\mu\nu} \partial_\nu \phi =0~,
\end{eqnarray}%
satisfied, for example, by acoustic perturbations in a moving perfect
fluid. These acoustic perturbations therefore travel along the
null geodesics of the acoustic metric $\g_{\mu\nu}$. The finite speed of propagation of the perturbations in the effective fields (in our case, the speed
of sound) plays the role of the speed of light in fundamental
physics.

This example illustrates how for certain
(semiclassical) configurations and low levels of resolution, one can
appropriately describe the physical behaviour of the system by means
of a (classical or quantum) field theory in a curved Lorentzian
background geometry. However, when one probes the system with higher
and higher resolution, the geometrical structure progressively
dissolves into a purely microscopic regime~\cite{Barcelo:2000tg}.
Therefore, although analogue models cannot be considered at this
stage complete models of quantum gravity (since they do not reproduce the
Einstein equations), they provide
specific and tractable models that reproduce many aspects of the
overall scenario expected in the realm of real gravity. 

The main objective of studies such as the ones that we will present in this thesis is then to obtain specific
indications about the type of deviations from the general relativistic behaviour to be
expected when quantum gravitational effects become important. All,
under the assumption that the underlying structure to general relativity is somewhat
similar to that in condensed matter systems. For reasons that we will elaborate in this chapter, we are interested in the case of Bose--Einstein condensates (BECs).

There are two ways of proceeding. The first way is to start from a theoretical description of BECs, elaborate the gravitational analogy and see which type of deviations arise with respect to the relativistic model. By extrapolation, one can then expect the same kind of deviations to occur in gravity if quantum corrections to gravity have a nature similar to the non-relativistic corrections that show up in BECs beyond the hydrodynamic or relativistic limit.

The second way is to directly calculate some possible quantum corrections in gravity, by modifying the standard relativistic description of a (predicted) gravitational phenomenon in a way which immediately incorporates the kind of modifications suggested by the condensed matter analogy. This essentially consists in replacing the relativistic dispersion relations by superluminally modified dispersion relations in a consistent way, as we will discuss in detail further on.

We will apply the first approach to an analysis of the dynamical stability (chapter~\ref{S:instabilities}) and of the quasinormal mode spectrum (chapter~\ref{S:QNMs}) of black hole configurations in Bose--Einstein condensates, based on a (1+1)-dimensional model which we introduce in chapter~\ref{S:BHconfigurations}. Subsequently, the second approach will be applied to Hawking radiation for collapsing configurations (chapter~\ref{S:HR}). First of all, though, in this chapter, we will review some basic elements of BEC theory and of how the gravitational analogy in a BEC is obtained.

\section{Bose--Einstein condensates of dilute atomic gases}
\label{SS:theoryofbecs} 
In this section, a brief general refresher of the theory of Bose--Einstein condensation will be given~\cite{Dalfovo:1999zz,Castin:2001} together with some relevant numbers related to their experimental achievement~\cite{Ketterle:1999,Streed:2006}.

A Bose--Einstein condensate is an aggregation state of matter which arises in a many-body system of bosons, confined through an external potential, when a macroscopic amount of them are led to occupy the ground state determined by this external potential. Such bosonic many-body systems can be described in terms of quantum field operators $\widehat \Psi(\r)$ and $\widehat \Psi^\dag(\r)$ that annihilate and create particles,
together with the appropriate equal-time commutation rules
\begin{eqnarray}%
[\widehat\Psi(\r),\widehat\Psi(\r')]=0~, \quad [\widehat\Psi^\dag(\r),\widehat\Psi^\dag(\r')]=0~, \quad [\widehat\Psi(\r),\widehat\Psi^\dag(\r')]=\delta(\r-\r')~, 
\end{eqnarray}%
and a Hamiltonian typically of the form
\begin{equation}\label{hamiltonian}
\widehat{H} = \int \d\r\,
\widehat\Psi^\dag(\r)h_{0}(\r)\widehat\Psi(\r) +
\frac{1}{2}\int \d\r\,\d\r'\,
\widehat\Psi^\dag(\r)\widehat\Psi^\dag(\r') V(\r'-\r)
\widehat\Psi(\r)\widehat\Psi(\r')~.
\end{equation}
%
Here,
\begin{equation}\label{hzero}
h_0(\r)=-\frac{\hbar^2}{2m}\bnabla^2+V_\text{ext}(\r)
\end{equation}
is the non-interacting single-particle Hamiltonian, $m$ the atomic
mass, $V_\text{ext}$ the external trapping potential, and it has been assumed that the many-body interactions are well modelled by a two-body interaction potential $V(\r'-\r)$.

When the system is cooled below a certain critical temperature $T_\text{crit}$, it 
condenses,\footnote{The condensation of a BEC
takes place in momentum space: all atoms tend to the lowest energy
state characterised by $\k=0$. 
For experimental implementations with atomic gases, the condensation in momentum space is accompanied by a condensation in
position space, see
Fig.~\ref{fig:ketterle4}, due to the use of harmonic traps.} and $\widehat \Psi$ can be separated into a
macroscopic wave function $\psi$ plus quantum fluctuations.
%
%
In atomic systems, the interaction between the atoms implies that some of them are excited even at zero temperature, in contrast to the ideal or pure Bose condensate depicted in Fig.~\ref{fig:ketterle1}. The proportion of excited atoms or {\it quantum depletion} increases from typically $0.1\%$ at zero temperature to still less than $1\%$ at the non-zero temperatures achieved in laboratory experiments. By contrast, in
helium, the main other quantum liquid used for the study of gravitational analogies, the quantum depletion is of the order of
90$\%$, making the theoretical treatment much more complicated.

Note that the theory of Bose--Einstein condensation is in principle applicable to bosonic many-body systems in general. However, most systems would experience a classical phase transition
to liquid and solid state long before the quantum degeneracy occurs, thereby seriously complicating the simple theoretical treatment of the condensation described so far. 
In order to avoid this, the most successful approach has been to use
weakly interacting alkaline atomic gases with extremely low densities, i.e. \emph{dilute atomic
gases}, combined with rapid cooling techniques to prevent solidification. 

For such gases, the evolution equation for the condensed part $\psi$ can be obtained as follows. Binary collisions characterised by the $s$-wave scattering length~$a$ are the main relevant interaction in a cold dilute gas. One can then replace the interaction potential $V(\r'-\r)$ in the Hamiltonian~\eqref{hamiltonian} by the effective prescription
\begin{eqnarray}%
V(\r'-\r)=g\delta(\r'-\r)~,
\end{eqnarray}%
with $g=4\pi\hbar^2 a/m$ the effective coupling constant. The
Hamiltonian becomes
\begin{eqnarray}%
\widehat{H} = \int d^3\r\, [\widehat\Psi^\dag h_0 \widehat\Psi +
\frac{g}{2} \widehat\Psi^\dag \widehat\Psi^\dag \widehat\Psi \widehat\Psi]~,
\end{eqnarray}%
and the evolution equation for $\widehat\Psi$ is
\begin{equation}\label{evol}
i\hbar \partial_t\widehat\Psi = [\widehat\Psi,\widehat H] = (h_0 + g
\widehat\Psi^\dag \widehat\Psi)\widehat\Psi~.
\end{equation}
Since the quantum depletion is small, it makes sense to use a mean-field approach 
and expand the theory as a perturbation series around the macroscopic wave function $\psi$. 
The zeroth-order term can be obtained simply by substituting $\psi$ for $\widehat\Psi$ in eq.~\eqref{evol}, which gives the non-linear Schr\"{o}dinger equation or {\it Gross--Pitaevskii equation}
\begin{equation}\label{gross-pitaevskii}
i\hbar\partial_t\psi=(h_0+g\abs{\psi}^2)\psi~.
\end{equation}
This classical mean-field equation is known to provide an excellent approximation for the evolution of the condensed phase $\psi$ in the case of weakly interacting systems ($s$-wave scattering length $a$ much smaller than the interatomic distance, or in other words: $na^3\ll 1$, with $n$ the particle density) containing a large number of atoms.

\begin{figure}
\begin{center}
\includegraphics[width=.4\textwidth]{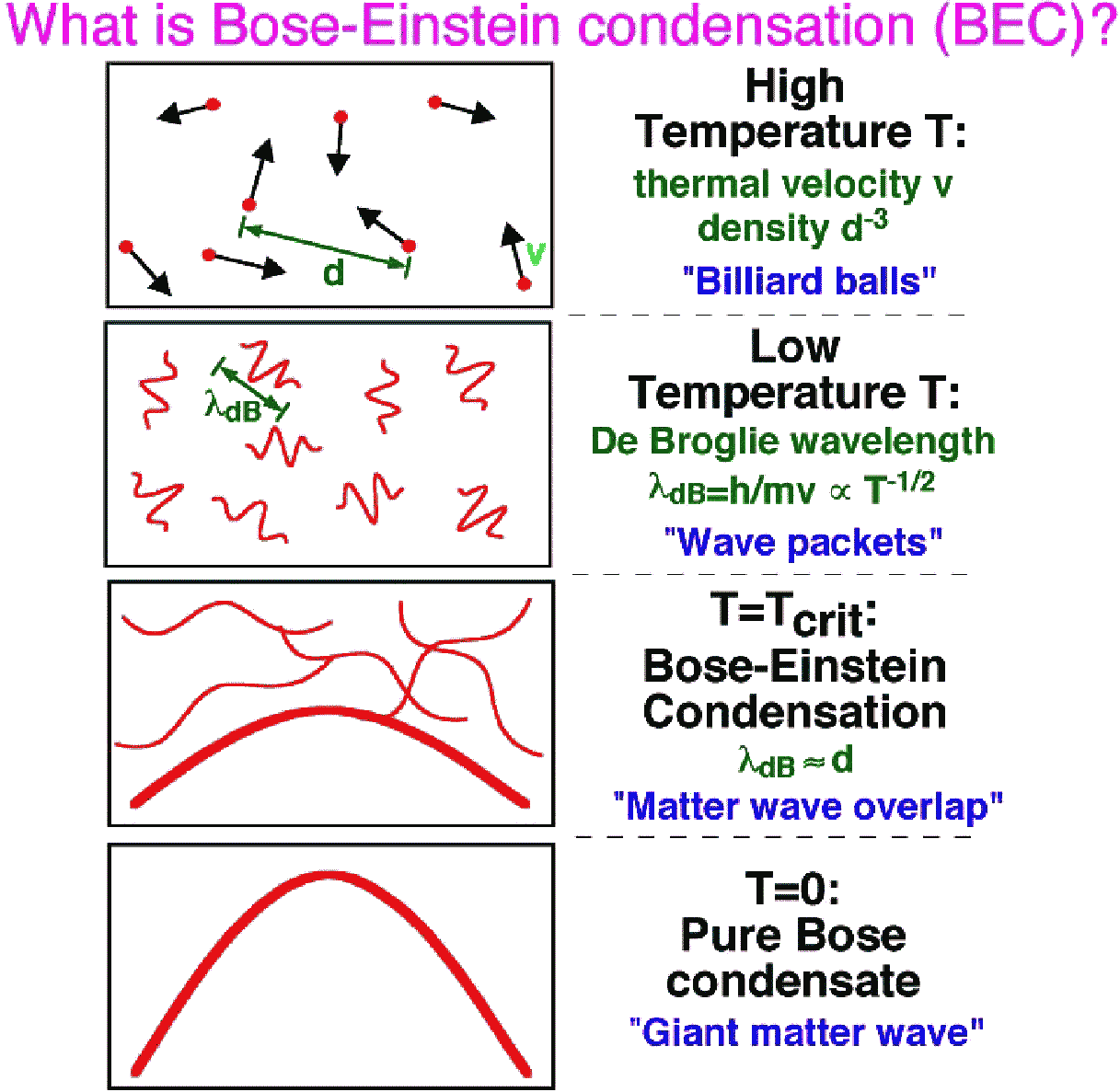}
\end{center}
\figcaptionl{fig:ketterle1}{Criterion for Bose--Einstein
condensation, taken from \cite{Ketterle:1999}. $d$ is the interatomic
distance and $\lambda_{dB}$ the thermal or de Broglie wavelength,
which is a measure for the extent of the wavepackets associated with
the atoms ($\lambda_{dB} \propto T^{-1/2}$).}
\end{figure}

Bose--Einstein condensation was first experimentally observed in 1995
at JILA in rubidium ($^{87}\text{Rb}$), and at MIT in sodium
($^{23}\text{Na}$), see Fig.~\ref{fig:ketterle4}. Since then,
about a dozen other atomic gases including $^7\text{Li}$, $^1$H and
$^{85}\text{Rb}$ have followed, $^{87}\text{Rb}$ and
$^{23}\text{Na}$ still being the favourites from an experimental
point of view~\cite{Streed:2006}. Apart from atomic gases in weakly interacting regimes, Bose--Einstein condensation has also been achieved in strongly interacting regimes~\cite{Cornish:2000zz} by tuning the scattering length through the use of a magnetic Feshbach resonance~\cite{Feshbach:1958,Inouye:1998}.\footnote{Note that condensation of bosons in a strongly interacting regime is also responsible for the superfluidity of $^4$He, a phenomenon which has been observed as early as in 1938~\cite{Kapitza:1938}. However, $^4$He is not well described by the standard mean-field theory of Bose--Einstein condensation since less than 10\% of the atoms are in the condensed state, due to the strong interaction between the atoms.} This technique was also used to suddenly switch from a weak repulsive to a strong attractive interaction, producing a violent collapse followed by a partial explosion process called ``Bose nova'' because of its similitude with astrophysical supernova explosions~\cite{Donley:2002}. As another striking example, in magnons, a collective excitation of the atomic spin structure of a crystal in the presence of an external magnetic field, Bose--Einstein condensation has recently been achieved at room temperature~\cite{Demokritov:2006}.
\begin{figure}[t]
\begin{center}
\includegraphics[width=.7\textwidth]{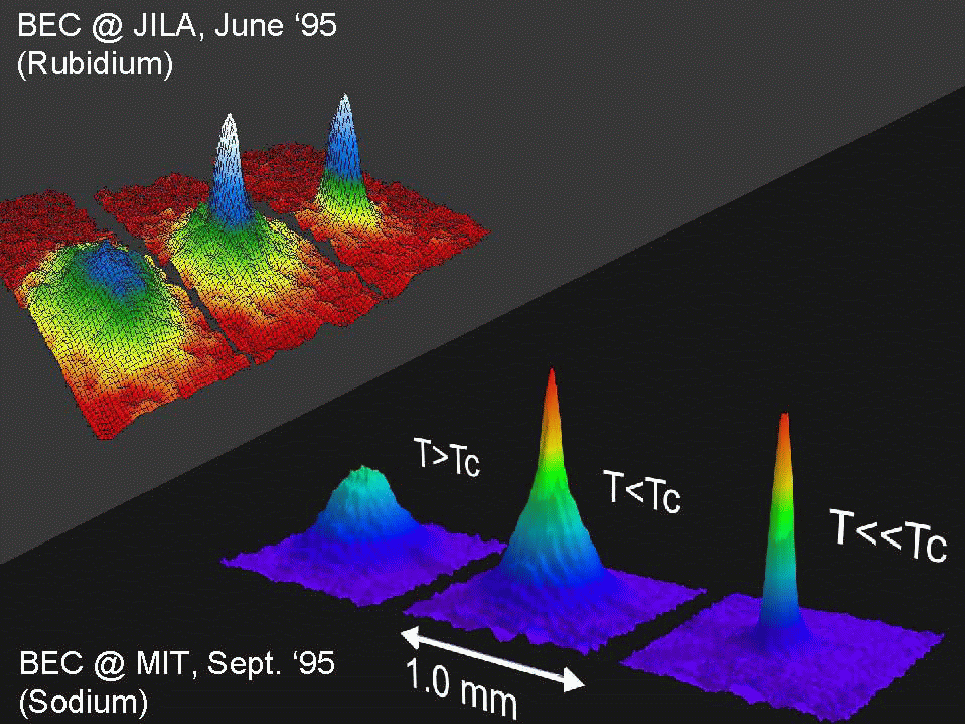}
\end{center}
\figcaptionl{fig:ketterle4}{Experimental observation of
Bose--Einstein Condensation in
$^{87}$Rb at JILA (above) and in $^{23}$Na at MIT (below). 
In both pictures, the left frame represents the gas at a temperature
just above the critical condensation temperature, the centre frame
the same gas just after the condensation and the
right frame a nearly pure condensate. The upper picture is a
computer image of the momentum distribution, the lower one of the
spatial distribution.}
\end{figure}

A few
illustrative numbers associated with BECs in atomic gases are the following~\cite{Anglin:2002,Streed:2006}. The critical temperatures
involved are typically from a few hundred nK to a few $\mu$K, at
densities between $10^{14}$ and $10^{15}$ atoms$/\text{cm}^3$. The
total amount of atoms in a BEC ranges from only a few hundred to
roughly 100 million in Na and Rb and about 1 billion
in atomic hydrogen (note that larger amounts are desirable mainly
because they make the experiment more robust with respect to
impurities and misalignments), with typical condensate sizes varying from 10 $\mu$m to a few hundred $\mu$m. Finally, two parameters that we will regularly need in this thesis are the speed of sound $c$ and the healing length $\xi$, which, roughly speaking, is the distance needed for the condensate to smoothen out a sharp inhomogeneity in the atomic density. Typical values for the velocity of sound $c$ in BECs range between 0.5mm/s and 10mm/s, while the healing length $\xi$ usually lies
between 100nm and 2$\mu$m.

Apart from being a very active field of research in themselves, Bose--Einstein condensates have also become an ``ultralow-temperature laboratory for atom optics, collisional
physics and many-body physics''~\cite{Anglin:2002}. And one might add:
``for high-energy physics'', as illustrated not only by the many works
on analogue gravity in BECs, but also for example on braneworlds in
warped spacetimes~\cite{Barcelo:2002xw} and a suggestion to
probe certain aspects of supersymmetry in a Bose--Einstein condensate~\cite{Snoek:2005yt}.

Here, we will focus on the gravitational analogy in BECs in regimes that can be described by mean-field theory.


\section{Gross--Pitaevskii equation and quantum potential}
\label{SS:GP and QP}
In the previous section, we arrived at the Gross--Pitaevskii equation
\begin{eqnarray}
 \im \hbar \; \frac{\partial }{\partial t} \psi(t,\r)= \left(
 - \frac{\hbar^2}{2m} \nabla^2
 + V_{\rm ext}(\r)
 + g \; |\psi(t,\r)|^2 \right) \psi(t,\r)~,
\label{GP}
\end{eqnarray}
which describes the evolution of the condensate's macroscopic wave
function or order parameter $\psi$, and where we have now explicitly indicated the spatial and temporal dependencies.\footnote{Note that the external potential $V_{\rm ext}$ can also be time-dependent, and can even be used to tune the coupling constant $g$ through a Feshbach resonance, as mentioned in the previous section. For simplicity, we will assume both $V_{\rm ext}$ and $g$
time-independent.} We remind the reader that $m$ is the boson mass, $V_\text{ext}$ the external potential and $g$ a coupling constant which is related to the
corresponding $s$-wave scattering length $a$ through $g ={4\pi \hbar^2 a
/m}$. 

The Gross--Pitaevskii equation can be expressed in terms of hydrodynamic quantities
such as the local speed of sound $c$ and the velocity of the fluid
flow $\mathbf v$.  This proceeds by first introducing the hydrodynamic or Madelung
representation for the order parameter in the Gross--Pitaevskii equation:
\begin{eqnarray}
\psi = \sqrt{n}e^{\im \theta/\hbar} e^{- \im \mu t/\hbar }~.
\label{madelung}
\end{eqnarray}
Here $n$ is the condensate density, $\mu$ the chemical
potential and $\theta$ a phase factor. Substituting in eq.~\eqref{GP} we arrive
at
%
\begin{eqnarray}
\partial _t n &=& - \frac{1}{m} \bnabla \cdot (n \bnabla \theta)~, \label{GP_n_theta_a}\\
\partial _t \theta &=& -\frac{1}{2m}(\bnabla \theta)^2
- g \; n
-V_\text{ext}-\mu -V_\text{quantum}~,
\label{GP_n_theta_b}\end{eqnarray}
\label{GP_n_theta}
%
where the so-called `quantum potential' is defined as
\begin{eqnarray}\label{quantum-potential}
V_\text{quantum}=
-\frac{\hbar^2}{2m}\frac{\nabla^2 \sqrt{n}}{\sqrt{n}}~.
\label{QP}
\end{eqnarray}
In many situations the quantum potential in eq.~\eqref{GP_n_theta_b}
can be neglected, a point to which we will return in section~\ref{SS:validity}. The resulting equations
\eqref{GP_n_theta_a} and~\eqref{GP_n_theta_b} are then equivalent to
the hydrodynamic equations, i.e., a continuity equation and an Euler equation, for a perfect classical
fluid. In this case, it is well known that the propagation of acoustic
waves in the system can be described by means of an effective metric,
thus providing the analogy with the propagation of fields in curved
spacetimes~\cite{Unruh:1980cg,Visser:1997ux}, as we will illustrate in the next section. Given a background
configuration ($n_0$ and $\theta_0$), this metric can be written as
\begin{eqnarray}
\g_{\mu \nu}=\frac{m}{g} c
\begin{pmatrix}
v^2-c^2 && -\mathbf v^\text{T}\\
-\mathbf v && \mathbb{1}
\end{pmatrix},
\label{metric}
\end{eqnarray}
where $\mathbb{1}$ is the $3\times3$ unit matrix, $c^2 \equiv g n_0/m$ and $\mathbf v \equiv \bnabla \theta_0 /m$. These magnitudes, $c$ and $\mathbf v$, represent the local speed of
sound and the local velocity of the fluid flow respectively.

\section{Effective acoustic metric}
\label{SS:metric}
Acoustic waves or phonons are linear perturbations of a background solution to the Gross--Pitaevskii equation. In order to linearise the eqs.~\eqref{GP_n_theta}, we write
%
\begin{eqnarray}
n (\r, t) &=& n_0(\r) + g^{-1} \widetilde n_1(\r, t)~,
\\
\theta(\r, t) &=& \theta_0(\r) + \theta_1(\r, t)~,
\end{eqnarray}
%
where $\widetilde n_1$ and $\theta_1$ are small perturbations of the
density and phase of the BEC (the coupling constant $g$ is included in the linearisation of $n$ for notational convenience). The eqs.~\eqref{GP_n_theta} then
separate into two time-independent equations for the background:
\begin{subequations}\label{GP_background}
\begin{eqnarray}
0&=& - \bnabla \cdot (c^2 \mathbf v)~, \label{GP_BG1}\\
0 &=& -\frac{1}{2}m\mathbf v^2 -m c^2 - V_\text{ext}-\mu
+\frac{\hbar^2}{2m}\frac{\nabla^2 c}{c}~,
\end{eqnarray}
\end{subequations}
plus two time-dependent equations for the perturbations:
\begin{subequations}
\label{GP_lin}
\begin{eqnarray}
\partial _t \widetilde n_1 &=& - \bnabla \cdot
\left( \widetilde n_1 \mathbf v + c^2 \bnabla \theta_1 \right)~,
\label{GP1_lin}
\\
\partial _t \theta_1
&=& -\mathbf v \cdot \bnabla \theta_1 - \widetilde n_1 +\frac{1}{4} \xi^2
\bnabla \cdot \left[c^2 \bnabla \left( \frac{\widetilde n_1}{c^2}\right)\right]~,
\label{GP2_lin}
\end{eqnarray}
\end{subequations}
where $\xi \equiv \hbar / (mc)$ is the {\it healing length} of the condensate, which, as we mentioned in~\ref{SS:theoryofbecs}, can be understood as the distance over which an otherwise homogeneous condensate recovers or `heals' from a sharp inhomogeneity. The effective acoustic metric~\eqref{metric} can then be derived as follows.

In the hydrodynamic or acoustic limit, the last term on the right-hand side of eq.~\eqref{GP2_lin}, which is a direct consequence of the presence of the quantum potential~\eqref{quantum-potential} in eq.~\eqref{GP_n_theta_b}, can be neglected. Then, extracting $\widetilde n_1$ from the remaining terms in eq.~\eqref{GP2_lin} and plugging into eq.~\eqref{GP1_lin}, the result can be written in four-dimensional notation\footnote{We use the convention $x^\mu=(t,\mathbf r)$.} as
\begin{equation}
\partial_\mu(f^{\mu\nu}\partial_\nu\theta_1)=0~,
\end{equation}
where
\begin{equation}\label{premetric}
f^{\mu\nu}(t,\r)\equiv \frac{n}{c^2}\begin{pmatrix}\ -1&&
-v^j\\-v^i&&c^2\delta^{ij}-v^iv^j\end{pmatrix}.
\end{equation}
Define $\sqrt{-\g}\g^{\mu\nu}=f^{\mu\nu}$, where $\g\equiv \det(\g_{\mu\nu})$,
and hence also $\g=\det(f^{\mu\nu})$. By inverting $\g^{\mu\nu}$, we obtain
the acoustic metric
\begin{equation}\label{effmetr}
\g_{\mu\nu}(t,\r)
=\frac{n}{c}\begin{pmatrix}\ v^2 -c^2&& -\mathbf
v^\text{T}\\-\mathbf v&&\mathbb{1}
\end{pmatrix}.
\end{equation}
Note that we have derived this acoustic metric from the Gross--Pitaevskii equation. However, the same result can be obtained directly by combining and manipulating the Euler and continuity equations for an irrotational flow of any non-viscid and barotropic fluid~\cite{Visser:1997ux}. Some indications of why we are nevertheless specifically interested in BECs have already been given in the introduction. We will postpone a more detailed motivation until section~\ref{SS:why-BECs}.

From eq.~\eqref{effmetr}, the wave equation for the propagation of the fluctuation field
$\theta_1$ of the velocity potential (the sound wave), is
\begin{equation}\label{waveq}
\square\theta_1\equiv
\g^{\mu\nu}\nabla_\mu\nabla_\nu\theta_1=
\frac{1}{\sqrt{-\g}}\partial_\mu(\sqrt{-\g} \g^{\mu\nu}\partial_\nu\\
\theta_1)=0~.
\end{equation}
In other words, acoustic perturbations propagate according to a d'Alembertian equation of movement, which is formally identical to the wave equation of a
massless scalar field propagating in a curved spacetime. 

Before moving back to the main discussion, a few remarks might be in place. 
First, note that the effective or acoustic metric has a Lorentzian signature $(-,+,+,+)$. So the collective behaviour of the condensate really gives rise to the emergence of an effective curved background spacetime for the propagation of the phonons.

Second, it is important to stress that we are dealing with a bi-metric system. Of course, each part of the system (background and
perturbations) can only couple to one metric. The first one is the
usual Minkowski metric for flat spacetime
\begin{equation}
\eta_{\mu\nu}=(\text{diag}[-c^2_\text{light},1,1,1])_{\mu\nu}~.
\end{equation}%
This metric governs the background fluid, in which relativistic effects are totally negligible, since for the fluid flow velocities involved: $v\ll c_\text{light}$. The background fluid couples \emph{only} to this
metric. The phonons or acoustic perturbations, on the other hand, do not see
this flat-space metric but are instead governed \emph{only} by the
effective acoustic metric $\g_{\mu\nu}$. It is thus seen that the
classical Newtonian dynamics of the background fluid leads to an
effective low-energy relativistic curved Lorentzian spacetime
metric for the sound waves.

Third, the sign of $g_{tt}$ in~\eqref{effmetr} depends on the values of $c$ and $v$, i.e., on the fluid regime. A change of sign in $g_{tt}$ characterises an ergoregion. For a purely radial flow, ergoregion and horizon coincide. Therefore, if we can manipulate our system in such a way that for some $r_h$, a subsonic outside region and a supersonic inside region are created,
i.e.:
\begin{align*}
c^2(r)&> v^2(r) \quad \text{for}~~\; r>r_h \quad (\text{subsonic outside region})~,\\
c^2(r)&< v^2(r) \quad \text{for}~~\; r<r_h \quad (\text{supersonic inside
region})~,
\end{align*}
then there will be a horizon at $r_h$:
\begin{equation}c(r_h)=\abs{v(r_h)}~.
\end{equation}
Indeed, for $c^2-v^2>0$, $g_{tt}$ is negative, indicating the
timelike character of the time coordinate, whereas for
$c^2-v^2<0$, $g_{tt}$ becomes positive and so this coordinate
becomes spacelike. As time marches on, any particle in
the supersonic inside region will irrevocably move towards the
centre of the system ($r=0$), where the singularity would lie in the case of a gravitational black hole.

Finally, in general we will be mainly interested in fluctuations on top of a stationary background, with the notable exception of chapter~\ref{S:HR}, where we examine the Hawking radiation in a collapsing geometry. In stationary cases, we can write
\begin{equation}
\partial_t n=\partial_t c=\partial_t \mathbf v=0.
\end{equation}
As a further simplification, we will focus on effectively (1+1)-dimensional configurations and write $c(t,x)$ and $\v(t,x)$. This means that we consider perturbations propagating in a condensate in such a way that the transverse degrees of freedom are effectively frozen. In other words, the only allowed motions of both the perturbations and the condensate itself are along the $x$-axis.

We now move back to the main line of discussion.

\section{Dispersion relation}
\label{SS:dispersion}
A crucial element in the quantum gravitational analogy is the phononic dispersion relation in a BEC and how it compares to its relativistic counterpart.


To derive the dispersion relation, consider a
region in which the condensate is homogeneous (with $c$ and $v$
constant), and seek for solutions of eqs.~\eqref{GP_lin} in the form
of plane waves:
\begin{subequations}\label{plane-waves}\begin{eqnarray}
\widetilde n_1(t,x)&=&
A e^{i(kx - \omega t)}~, \\
\theta_1(t,x)&=& B e^{i(kx - \omega t)}~,
\end{eqnarray}\end{subequations}
where $A$ and $B$ are constant amplitude factors. We stress that the
frequency $\omega$ and the wave number $k$ are not restricted to real values, but are allowed to be complex. The imaginary parts of $\omega$ and $k$ will encode the dynamical behaviour of the system. 
Substituting into eqs.~\eqref{GP_lin} we find
\begin{eqnarray}
\begin{pmatrix}
i(\omega- vk)  && c^2 k^2 \\ \\
1 +\frac{1}{4} \xi^2 k^2 && - i(\omega-vk)
\end{pmatrix}
\begin{pmatrix}
A\\ \\
B\end{pmatrix}=0~.
\end{eqnarray}
For a non-trivial solution to exist, the determinant of the above
matrix must vanish. This immediately leads to the Bogoliubov dispersion relation.
\begin{eqnarray}\label{dispersion_xi}
(\omega - vk)^2= c^2k^2 + \frac{1}{4} c^2 \xi^2 k^4~.
\end{eqnarray}
This equation is globally valid in homogeneous condensates. In non-homogeneous configurations, it is still valid as a strictly {\it local dispersion relation}, provided that the quantities involved ($c$, $v$, $\xi$ and $k$) all vary slowly with $x$.

The full dispersion relation~\eqref{dispersion_xi} is \textit{quartic} in $k$. It should be compared to the hydrodynamic dispersion relation obtained by neglecting the quantum potential:
\begin{eqnarray}\label{quadr_dispersion}
(\omega - vk)^2= c^2k^2~.
\end{eqnarray}
Such a hydrodynamic dispersion relation is of the usual \textit{quadratic relativistic} form. This fundamental observation allows us to refer to the {\it hydrodynamic} or the {\it relativistic} limit interchangeably.

However, we are not so much interested in this hydrodynamic or relativistic limit, but precisely in deviations from it. Therefore we will always work with the full dispersion relation, and only use the hydrodynamic limit as a means of comparison. The key point is that, because the full dispersion relation is quartic, when writing the mode $u_\omega$ corresponding to a certain frequency $\omega$:
\begin{equation}\label{bogoliubov_modes}
u_\omega=e^{-i\omega t}\sum_j A_j e^{ik_jx},
\end{equation}
there are now four contributions to this mode, stemming from the four values of $k$ associated with each value of $\omega$. In the hydrodynamic limit there are obviously only two contributions to each $u_\omega$-mode:
\begin{equation}\label{hydro-modes}
k_{1,2} = \omega/(v \pm c),
\end{equation}
In the configurations that we will analyse, the additional modes will turn out to play a crucial role at all frequencies because of the presence of black hole horizons, as we will discuss in the next section.

Another crucial aspect of the non-hydrodynamic corrections at high frequencies is that these make the dispersion relation `superluminal' (i.e., superphononic or supersonic). Indeed, the group velocity for the modes which behave like eq.~\eqref{hydro-modes} in the hydrodynamic limit can in general be written as
\begin{eqnarray}\label{v-eff}
v_g\equiv \frac{\d \omega}{\d k} =v\pm c_k~,
\end{eqnarray}
with an effective propagation speed
\begin{eqnarray}\label{eff-sos}
c_{k}=c\frac{1+\frac{1}{2}\xi^2k^2}{\sqrt{1+\frac{1}{4}\xi^2k^2}}~,
\end{eqnarray}
(see Fig.~\ref{Fig:c_xi}). For sufficiently large wave numbers, $c_k$ becomes significantly larger than $c$, i.e., it becomes effectively superphononic.

%
\begin{figure}
\begin{center}
\includegraphics[width=0.5\columnwidth]{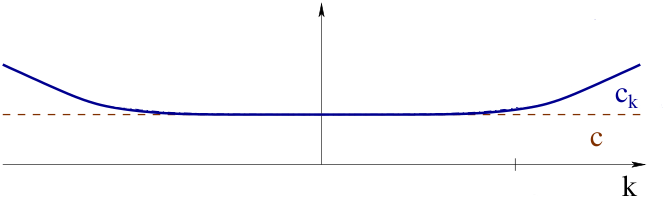}
\put(-57,3){2/$\xi$}
\end{center}
\figcaptionl{Fig:c_xi}{Behaviour of the effective speed of sound $c_k$ as a function of the wave number $k$ in a BEC. Due to high-frequency non-hydrodynamic (non-relativistic) corrections, the effective speed of sound starts to increase significantly around the Lorentz symmetry breaking scale $k_L\equiv 2/\xi$, and becomes effectively superphononic ($c_k>c$).
}
\end{figure}
%
The connection with quantum gravity phenomenology is the following. In a wide range of phenomenological scenarios for quantum gravity, violations of the local Lorentz symmetry are expected at high frequencies. When these Lorentz violations are associated with the existence of privileged reference frames, they typically lead to modified dispersion relations of the kind
\begin{eqnarray}
\omega^2= c^2k^2 \pm c^2\frac{k^4}{k_L^2}~,
\end{eqnarray}
with $k_L$ the local Lorentz symmetry breaking scale (and obviously $c$ represents the speed of light in this last equation). BECs therefore provide a concrete example of such a dispersion relation with `superluminal' Lorentz-breaking corrections, characterised by a Lorentz violation scale $k_L\equiv 2/\xi$. We will come back to the connection with quantum gravity phenomenology in chapter~\ref{S:intermezzo}.

\section{Validity of the hydrodynamic approximation}
\label{SS:validity}
The functions $c(t,x)$ and $\mathbf v(t,x)$ completely
characterise the acoustic metric. In general relativity, any metric has to be
obtained by solving the Einstein equations. Here, however, the
magnitudes $c(t,x)$ and $\mathbf v(t,x)$, and so the acoustic
metric, are those satisfying the continuity and Euler equations
of hydrodynamics, eqs.~\eqref{GP_n_theta} without the quantum
potential. Thus, these equations play a role analogous to the
vacuum Einstein equations in general relativity. Of course, at the global non-linear
level these hydrodynamic equations are completely different from the real
Einstein equations. But we are interested in their behaviour when linearised
around a background solution. Although this does not lead to the full complexity that is expected in general relativity (if only because the fluctuations of the metric here are characterised by a scalar or spin-0 field, versus a spin-2 tensor field in general relativity), it still captures much of the essence of a proper linearised general relativistic behaviour.

There exist, however, situations in which the quantum potential
in eq.~\eqref{GP_n_theta} cannot be neglected, and the hydrodynamic approximation is not valid. To analyse this, we start from the full dispersion relation. One can obtain the hydrodynamic approximation by writing $\omega$ as a series expansion in terms of $\xi k$, and truncating the higher order terms, i.e.:
\begin{eqnarray}
\omega &=& \left ( v \pm c\sqrt{1+ {1 \over 4}\xi^2k^2} \right )k \nonumber \\
& \simeq&
(v \pm c)k \mp {1 \over 8} c\xi^2 k^3 + \mathcal{O}(\xi^4k^5)
\simeq  (v \pm c)k + \mathcal{O}(\xi^2k^3). 
\end{eqnarray}
For the series expansion to make sense, one must have $\xi k \ll 1$, i.e., the frequencies must be small compared to the characteristic scale of the system, or in other words the wavelengths must be much larger than the healing length $\xi \equiv \hbar / (mc)$. The truncation moreover requires 
\begin{eqnarray}%
\frac{\xi^2 k^2}{8|1 \pm v/c|} \ll 1.
\end{eqnarray}%
This second condition is never satisfied around a horizon ($c=|v|$). Therefore, near a horizon, the additional modes that are not present in the hydrodynamical limit are always important, even at low frequencies. Whenever probing the configuration either with high frequencies, or near the horizon (with any frequency), the geometrical picture will dissolve and the underlying microscopic structure will appear to some degree.

Without forgetting this subtlety, we will continue to call eq.~\eqref{metric} the `effective' or `acoustic' metric in the system, even when
analysing the full Gross--Pitaevskii equation. Then, we can consider the equations~\eqref{GP_n_theta} to play a role similar to the vacuum Einstein equations. Their treatment is classical, but they
incorporate corrections containing $\hbar$. Therefore, the
standard treatment of BECs based on the Gross--Pitaevskii equation provides an example of a
way of incorporating quantum corrections to the dynamics of a system
without recurring to the standard procedures of back-reaction.
Again, at the global non-linear level these equations bear
no relationship whatsoever with any sort of `semiclassical' Einstein
equations, nor do they present sufficient complexity to lead to spin-2 fields that could be the analogue of gravitational waves. However, at the linear level, that is, in terms of linear
tendencies of departing from a given configuration, equations~\eqref{GP_n_theta} encode much of the essence of the linearised general relativity
behaviour (a Lorentzian wave equation in a curved background),
semiclassically modified to incorporate a superluminal dispersion
term, as we have already discussed. 

\section{Why Bose--Einstein condensates?}
\label{SS:why-BECs}
We mentioned right after eq.~\eqref{effmetr} that the effective metric could have been derived for any irrotational, non-viscid and barotropic fluid, real or idealised. There are nevertheless several good reasons why we are interested specifically in BECs. 

First of all, real fluids have the obvious advantage over idealised fluids of permitting real experiments. It is not too difficult to simulate a black hole horizon in a real fluid system. For example, engineers routinely produce acoustic horizons in wind tunnels. Actually, at first sight, any moving fluid, including plain water, might serve as the background fluid in which to conduct the gravitational analogy, at least at the level of creating a supersonic flow and associated acoustic horizon. However, water and other `normal' fluids show a high degree of viscosity which
introduces terms that cannot be moulded in the relativistic form described earlier, and hence break the gravitational analogy even at low energy. Furthermore, many quantum effects of interest, such as Hawking radiation, would become indistinguishable
from the background noise and
impurities in normal fluids such as water. Finally, estimates for the Hawking temperature in water 
indicate an order of magnitude of \mbox{1 $\mu$K}, which is clearly
hopeless to detect at temperatures for which water behaves as a fluid.
Because of these problems, the quest for a useful background fluid leads in the
direction of superfluids. Superfluids have a vanishing viscosity,
and some superfluids can be made extremely pure. Two main candidates
stand out: Bose--Einstein condensates in dilute atomic gases and superfluid $^3$He.

BECs have the comparative advantage of being conceptually well understood, and relatively simple to describe theoretically and manipulate experimentally. This is partly due to the extremely low densities usually involved, which means that the interatomic distance is typically a lot
greater than the scattering length,\footnote{Typical values for
$^{23}$Na at a density of approximately $10^{14}$ atoms/cm$^3$ are 3
nm for the scattering length, versus 200 nm for the interatomic
distance~\cite{Ketterle:1999}.} leading to rather weak interactions.
In superfluid $^3$He systems, the density is several orders of
magnitude larger than in the BEC, leading to an interatomic
separation of the same order of magnitude as the scattering length.
As a consequence, only a fraction of all the particles are inside
the condensate, while the others form a second, normally fluid phase. 
Furthermore, the fundamental degrees of freedom in $^3$He are fermionic. The superfluidity arises from
Cooper-pair formation between weakly bound fermions.
These pairs behave as bosons, but some of the properties of the
individual fermions persist. All of this stimulates the existence of
many secondary effects, which make $^3$He a complex and rich model, in many ways much more appropriate for thinking in analogue terms about our universe than the simple model provided by BECs~\cite{Volovik:2003fe}. But in terms of
conceptual simplicity, Bose--Einstein condensates clearly stand out
as the best option. As a simple example, in Bose--Einstein
condensates under normal energetic circumstances, atomic interactions only take place in the $s$-wave. A much higher temperature than the condensation temperature, or a
far higher atom density, would be required to excite higher
multipole waves.\footnote{For the same example of $^{23}$Na at a
density of approximately $10^{14}$ atoms/cm$^3$, the limiting
temperature below which only $s$-wave scattering occurs is $\sim$
1mK, while the condensation temperature is $\sim$ 2$\mu$K~\cite{Ketterle:1999}.} In superfluid $^3$He on the other hand, the size
of the atom, the thermal de Broglie wavelength and the interatomic separation
are all of the same order of magnitude. This
creates a complex and rich situation, but also blurs and mixes
effects which might be interesting to study separately.

Finally, the third reason why we are interested in BECs, is that they provide a real system in which modified dispersion relations arise. As we mentioned in the introduction, and will further detail in chapter~\ref{S:intermezzo}, they therefore offer an interesting model for a prominent line of research within quantum gravity phenomenology. The mere occurrence of Lorentz-breaking modifications of the dispersion relations shows that there exist concrete examples in nature in which Lorentz invariance is realised as a low-energy effective symmetry, thereby strengthening the assumption that this might be the case for the local Lorentz invariance of general relativity as well. Furthermore, such modified dispersion relations also model most types of possible Lorentz symmetry violations at high energies that are expected from a theoretical quantum gravity phenomenological point of view (at least the ones associated with a preferred reference frame). Also, in contrast to subluminal dispersion, which gradually dampens the influence of high frequencies, superluminal dispersion relations such as those that arise in BECs amplifies the effects of high-frequency (`transplanckian') physics. This makes BECs an interesting probe for the possible influence of microscopic corrections to classical gravity, as well as for the robustness of phenomena related to black holes, such as Hawking radiation, see chapter~\ref{S:HR}.

\clearpage
\chapter[Black hole configurations in BECs]{Black hole configurations in Bose--Einstein condensates}
\label{S:BHconfigurations}
\section{A simple (1+1)-dimensional model}
We will now apply the theory developed in the previous chapter to analyse the dynamical behaviour of effectively (1+1)-dimensional BEC configurations with density and velocity profiles containing a single acoustic black hole horizon, see Figs.~\ref{F:profiles} and ~\ref{F:profile_finite_L}. This analysis will be separated into two parts: the dynamical stability will be examined in chapter~\ref{S:instabilities}, and a study of the quasinormal modes is contained in chapter~\ref{S:QNMs}. First, in the present chapter, we will discuss the background configurations and numerical method used in these analyses.

The type of (1+1)-dimensional background
profiles that we will examine consists of two regions each with a uniform
density and velocity, connected through a steplike discontinuity, see Fig.~\ref{F:profiles}. This is obviously an idealised case, obtained from Fig.~\ref{F:profile_finite_L} in the limit when the intermediate transition region $L\rightarrow 0$. We always consider flows moving from right to left, hence a negative value of $v$.

The parameters of the configuration are chosen such that there is an acoustic horizon at the $x=0$ discontinuity, which is the formal equivalent of a general relativistic black hole horizon, as described in section~\ref{SS:metric}. The left-hand asymptotic region ($x \to -\infty)$ mimics a general relativistic black hole singularity in the sense that nothing can escape from it towards the horizon (we will always impose an adequate boundary condition in order to ensure this). Note, however, that in our model, this region is perfectly regular, so the analogy with a general relativistic singularity is limited to the previous aspect.

At each discontinuity, matching conditions apply that connect the
magnitudes describing the condensate at both sides of the
discontinuity. Furthermore, we need a set of boundary
conditions, which determine what happens at the far ends of the
condensate. All these elements determine the
characteristics of the system, and hence its eigenfrequencies.

\begin{figure}[tbp]
\begin{center}
\includegraphics[width=.7\textwidth,clip]{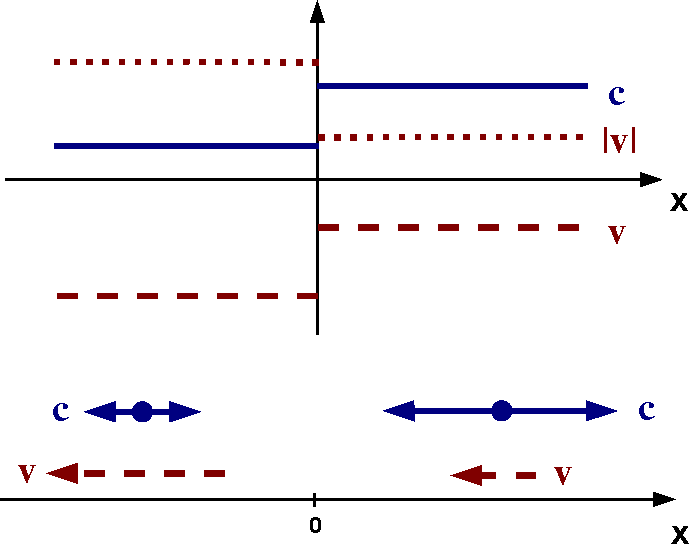}
\end{center}
\figcaptionl{F:profiles}
{Flow and sound velocity profile consisting of two homogeneous regions with a steplike discontinuity at $x=0$ simulating a black hole configuration in a BEC. The solid blue line represents the speed of sound $c$, the dashed red line the fluid velocity $v$. In the upper part of the picture, the negative value of $v$ indicates that the fluid is moving leftwards. For $x>0$, the fluid is subsonic since $c>\lvert v \rvert$. At $x<0$ it has become supersonic. At $x=0$, there is a sonic horizon. This picture can be considered the idealised limiting case of Fig.~\ref{F:profile_finite_L} when $L\rightarrow 0$.}
\end{figure}

\begin{figure}[tbp]
\begin{center}
\includegraphics[width=.7\textwidth,clip]{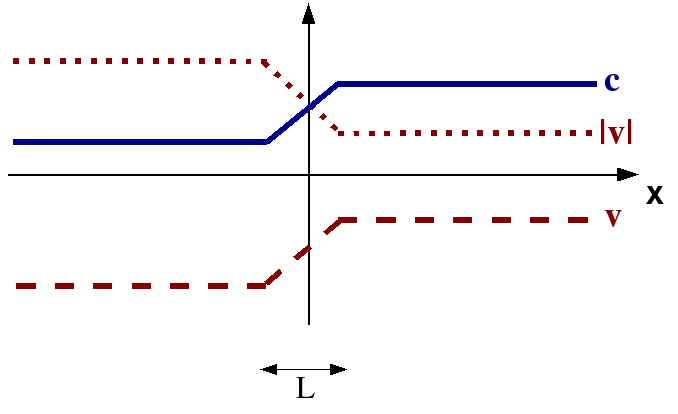}
\figcaptionl{F:profile_finite_L}{Flow and sound velocity profile consisting of two homogeneous regions connected through a transition region of width $L$ around the sonic horizon at $x=0$. This profile simulates a black-hole-configuration in a BEC. As in Fig.~\ref{F:profiles}, the solid blue line represents the speed of sound $c$ and the dashed red line the fluid velocity $v$. Actual calculations presented in this work will be based on the limiting case $L \rightarrow 0$, see Fig.~\ref{F:profiles}.
}
\end{center}
\end{figure}

In the remainder of this short chapter, we
will describe this particular model of (1+1)-dimensional configurations with abrupt discontinuities. 

%
\section{Background configuration and numerical method}
\label{SS:numerical}
The profiles that we will study are piecewise homogeneous with a single discontinuity. In order to find their dynamical modes, we will first of all need a way to describe the solutions that characterise the profiles under study. The idealised form of these profiles
allows a simplified description in each homogeneous region. Then, the
various sections have to be linked through matching conditions at the
discontinuity. After finding all the mode solutions, the dynamical modes (instabilities or quasinormal modes, respectively) will be those that moreover satisfy the adequate boundary conditions. We will describe the matching conditions in the current section, together with a straightforward algorithm for
numerical implementation. The boundary conditions depend on the concrete problem, the calculation of instabilities or of quasinormal modes, and will therefore be described in chapter~\ref{S:instabilities} and~\ref{S:QNMs}, respectively.

In order to calculate the matching conditions that link the description of the condensate in two homogeneous regions around a discontinuity, we integrate the differential equations~\eqref{GP_lin} that describe the evolution of the perturbations around this discontinuity (say $x=0$). This results in the following four independent, generally valid, matching conditions:
\begin{subequations}\label{matching}\begin{eqnarray}
&[\theta_1]&=0~,\qquad [v \widetilde n_1+ c^2 \partial_x \theta_1]=0~,\label{matching_a}\\
&\left[\frac{\widetilde n_1}{c^2}\right]&=0~,
\qquad \left[c^2\partial_x
\left(\frac{\widetilde n_1}{c^2}\right)\right]=0~.
\end{eqnarray}\end{subequations}
The square brackets in these expressions denote, for instance,
\mbox{$[\theta_1]=\left.\theta_1\right|_{x=0^+} -\left.
\theta_1\right|_{x=0^-}$}. We can simplify the second condition
in eq.~\eqref{matching_a} to $[c^2 \partial_x \theta_1]=0$ by noting
that $[v \widetilde n_1]=0$ because of the background continuity
equation~\eqref{GP_BG1}, while for our choice of a homogeneous
background the last condition becomes simply $[\partial_x \widetilde
n_1]=0$.

For a given frequency $\omega$, the general solution of
eqs.~\eqref{GP_lin} can be written as
%
\begin{eqnarray}\label{j-1to8}
\widetilde n_1 = \begin{cases}
\displaystyle\sum _{j=1}^4 A_j e^{i(k_jx - \omega t)}& (x<0),\\
\displaystyle\sum _{j=5}^8 A_j e^{i(k_jx - \omega t)}& (x>0),
\end{cases}
\qquad \qquad \theta_1 =
\begin{cases}
\displaystyle\sum _{j=1}^4 A_j \frac{\omega - v_\text{L}
k_j}{ic^2_\text{L} k ^2_j} e^{i(k_jx- \omega t)}  & (x<0),\\
\displaystyle\sum _{j=5}^8 A_j \frac{\omega - v_\text{R} k_j}{ic^2_\text{R} k ^2_j}e^{i(k_jx- \omega t)}& (x>0),
\end{cases}
\end{eqnarray}
where $\{ k_j\}$ are the roots of the corresponding dispersion
equations (four roots for each homogeneous region), and the subscripts $L$ and $R$
indicate the values of $c$ and $v$ in the left-hand-side and
the right-hand-side region respectively. The constants $A_j$
have to be such that the matching conditions~\eqref{matching} are satisfied. We can write down
these conditions in matrix form as $\Lambda_{ij}A_j=0$, where
\begin{eqnarray*}
\left (\Lambda_{ij} \right )=
\begin{pmatrix}
\frac{\omega-v_\text{L} k_1}{c^2_\text{L}k ^2_1} &
\frac{\omega-v_\text{L} k_2}{c^2_\text{L}k ^2_2} &
\frac{\omega-v_\text{L} k_3}{c^2_\text{L}k ^2_3} &
\frac{\omega-v_\text{L} k_4}{c^2_\text{L}k ^2_4} &
-\frac{\omega - v_\text{R} k_5}{c^2_\text{R}k ^2_5}&
-\frac{\omega - v_\text{R} k_6}{c^2_\text{R}k ^2_6}&
-\frac{\omega - v_\text{R} k_7}{c^2_\text{R}k ^2_7}&
-\frac{\omega - v_\text{R} k_8}{c^2_\text{R}k ^2_8}
\\ \\
\frac{\omega}{k_1}&\frac{\omega}{k_2}&\frac{\omega}{k_3}&\frac{\omega}{k_4}&
- \frac{\omega}{k_5}&- \frac{\omega}{k_6}&- \frac{\omega}{k_7}&- \frac{\omega}{k_8}
\\ \\
\frac{1}{c^2_\text{L}}&\frac{1}{c^2_\text{L}}&\frac{1}{c^2_\text{L}}&\frac{1}{c^2_\text{L}}&
- \frac{1}{c^2_\text{R}}&- \frac{1}{c^2_\text{R}}&- \frac{1}{c^2_\text{R}}&- \frac{1}{c^2_\text{R}}&
\\ \\
k_1 &k_2 &k_3 &k_4 &- k_5&- k_6&- k_7&- k_8&
\\
\end{pmatrix}.
\label{lambda-matrix}
\end{eqnarray*}

Since there are eight free parameters $A_j$ for each discontinuity (four in each region),
and only four conditions, matching is in principle always
possible. However, there are additional conditions: the boundary
conditions which we will discuss in the next chapters. These boundary
conditions can be added as additional rows into the matrix $\Lambda_{ij}$, to obtain a
$(4+N)\times 8$ matrix $\widetilde\Lambda_{ij}$, with $N=N(\omega)$ the number
of constraints on the coefficients $A_j$ (i.e., the number of modes that do not fulfil the boundary conditions and hence should be forbidden).\footnote{To give a simple example, assume that a boundary condition consists in the requirement: $v_g>0$ at the right asymptotic infinity. One can then simply check, for the four right-hand-side modes, whether they fulfil this condition. If not, the mode in question is forbidden and $N(\omega)$ is incremented by one.} Depending on the value of $N(\omega)$, we can have the following
cases.
\begin{itemize}
\item If $N(\omega)<4$, then there will always exist a non-trivial solution ${A_j}$, since there are more degrees of freedom than constraints. Therefore, a region of the complex $\omega$-plane where $N<4$ will represent a continuous region of eigenfrequencies. 
\item If $N(\omega)=4$, then we have an $8 \times 8$ system of equations and there is a non-trivial solution ${A_j}$ only if $\det(\widetilde\Lambda)=0$. Generally speaking, this means that in a region where $N=4$, we might at the very most expect isolated discrete eigenfrequencies, but no continuous zones of solutions.
\item If $N(\omega)>4$ then the system can be split into two (or more) $8\times 8$ subsystems, each with a subdeterminant $\lambda_i$. For a non-trivial solution to exist, each of these subsystems should fulfil the previous condition $\det(\lambda_i)=0$.
\end{itemize}
To summarise, we can define a non-negative function $F(\omega)$ such that
\begin{itemize}
\item $F(\omega)=0~~$ if $~~N(\omega)<4$,
\item $F(\omega)=|\det(\widetilde\Lambda)|~~$ if $~~N(\omega)=4$,
\item $F(\omega)=\sum|\det(\lambda_i)|~~$ if $~~N(\omega)>4$.
\end{itemize}
The eigenfrequencies of the system are those values of $\omega$ for which $F(\omega)=0$. Instabilities correspond to eigenfrequencies with Im$(\omega)>0$ and are therefore located in the upper half of the complex $\omega$ plane. Quasinormal modes on the other hand are relaxation modes, i.e., they have Im$(\omega)<0$ and will therefore be located in the lower half complex $\omega$ plane.

The essential ingredient that is still missing to calculate the instabilities and the quasinormal modes in these BEC black hole configurations is the set of boundary conditions. Since these are different for both problems, we will postpone them to the relevant chapters.

%
\clearpage
\chapter{Analysis of dynamical instabilities}
\label{S:instabilities}
Now that we have set down the general framework, it is time to examine the linear dynamical stability of various (1+1)-dimensional configurations in Bose--Einstein condensates with steplike sonic horizons, as illustrated in Fig.~\ref{F:profiles}. We will examine both black hole and white hole horizons. We will start with a motivation for the problem. Since most of the framework has been discussed in the previous section, we will then immediately discuss the boundary conditions, with particular attention to their meaning in gravitational terms. We highlight that the stability of a given configuration depends not only on its specific geometry, but especially on these boundary conditions. Under boundary conditions directly extrapolated from those in standard general relativity, both black hole and white hole configurations are shown to be stable. However, we show that under other (less stringent) boundary conditions, configurations with a black hole horizon remain stable, whereas white hole configurations develop instabilities associated with the presence of sonic horizons.

\section{Motivation}
\label{SS:inst_intro}
The stability analysis presented in~\cite{Garay:2000jj} for acoustic black hole configurations with fluid sinks
in their interior concluded that these configurations were
intrinsically unstable. However, the WKB analysis of the stability
of horizons in~\cite{leonhardt2003} suggested that black hole
horizons might be stable, while configurations with white hole
horizons seem to inevitably possess unstable modes. Regarding configurations in
which a black hole horizon is connected with a white hole horizon in
a straight line, the analysis presented in~\cite{Corley:1998rk}
concluded, also within a WKB approximation, that these
configurations were intrinsically unstable, producing a so-called
``black hole laser''. However, when the white hole horizon is
connected back to the black hole horizon to produce a ring, it was
found that these configurations can be stable or unstable depending
on their specific parameters~\cite{Garay:1999sk,Garay:2000jj}. This suggests that
periodic boundary conditions can eliminate some of the instabilities
associated with the black hole laser. So there seems to exist a variety of different results, and hence a bit of confusion in the related literature.

The importance of this issue is easy to see when one bears in mind that BECs have long been speculated to be good candidates for an experimental detection of Hawking radiation~\cite{Barcelo:2000tg,Barcelo:2001ca}, see chapter~\ref{S:HR}. In this context, dynamical stability is an important condition. Indeed, Hawking radiation is (by definition~\cite{Hawking:1974sw}) the `late-time' radiation of a black hole, once all transitory phenomena have vanished. However, if dynamical instabilities unavoidably exist, then these will necessarily start dominating from a certain time on. Depending on the relation between the time scales characteristic for the onset of Hawking radiation and the one for the dynamical instabilities, it might still be  possible to achieve a `metastable' experiment, i.e., one with a sufficient lifetime to detect the black hole radiation in spite of the existence of dynamical instabilities. However, this might require additional fine-tuning, and would in any case add an important source of background `noise', thereby seriously complicating the detection of the black hole radiation.
Therefore, if one could show that there exist black-hole configurations in BECs which are in principle devoid of dynamical instabilities, this would remove an important obstacle in the experimental realisation of Hawking radiation.

In this chapter, we will therefore try to shed some light on these issues
and clear up some of the apparent contradictions. To simplify matters, we will consider (1+1)-dimensional profiles that are piecewise uniform with a single steplike
discontinuity, as described earlier (see Fig.~\ref{F:profiles}). In terms of dynamical (in)stability, there seems to
be no crucial qualitative difference between the present case and a
profile with smooth but narrow transitions between regions with an
(asymptotically) uniform density distribution, at least not from a theoretical point of view~\cite{Garay:2000jj}. Therefore, here we will not take into account
the effects of the finite size of the transition region and consider
only idealised steplike cases. The specific way to examine the kind
of instability we are interested in, consists basically in seeking
whether, under appropriate boundary conditions, there are complex
eigenfrequencies of the system which lead to an exponential increase
with time of the associated perturbations, i.e., a dynamical
instability. Throughout the section we will use a language and
notation as close to general relativity as possible. In particular, we will use
boundary conditions similar to those imposed in the standard
stability and quasinormal mode analysis of black holes in
general relativity~\cite{Nollert:1999ji,Kokkotas:1999bd}. 

\section{Boundary conditions}
\label{SS:inst_boundary}

In order to extract the possible intrinsic instabilities of a BEC
configuration, we have to analyse whether there are linear mode
solutions with positive $\text{Im}(\omega)$ that satisfy outgoing
boundary conditions.\footnote{Recall that we follow the convention $u_\omega=\sum_j A_j e^{i(k_jx-\omega t)}$, see eq.~\eqref{bogoliubov_modes} or~\eqref{j-1to8}.\label{f:sign-convention}} By `outgoing' boundary conditions we mean that
the group velocity is directed outwards (toward the boundaries of the
system). The group velocity for a particular $k$-mode is defined as
\begin{equation}\label{group-velocity}
v_g \equiv \text{Re} \left( \frac{d \omega}{dk}\right) =
\text{Re}\left(
\frac{c^2 k + \frac{1}{2} \xi^2 c^2 k^3}{\omega - vk} + v
\right),
\end{equation}
where we have used the full Bogoliubov dispersion relation~\eqref{dispersion_xi}.\footnote{Note that this definition is equivalent to the simpler definition~\eqref{v-eff} only for the modes that behave like eq.~\eqref{hydro-modes} in the hydrodynamic limit.} The
physical idea behind this outgoing boundary condition is that we are only examining the intrinsic instabilities originating inside the system, since these are equivalent to dynamical instabilities in general relativity.

To illustrate this assertion, let us look at the classical linear
stability analysis of a Schwarzschild black hole in general relativity~\cite{Kokkotas:1999bd,Nollert:1999ji}. When
considering outgoing boundary conditions both at the horizon and in
the asymptotic region at infinity, only
negative $\text{Im}(\omega)$ modes (the quasinormal modes) are found,
and thus the black hole configuration is said to be stable. If the presence of
ingoing waves at infinity were allowed, there would also exist
positive $\text{Im}(\omega)$ solutions. In other words, the
Schwarzschild solution in general relativity is stable when considering only internal
rearrangements of the configuration. If instead the black hole was
allowed to absorb more and more energy from outside the system (i.e., ultimately coming from infinity), its
configuration would continuously change and appear to be unstable.

The introduction of modified dispersion relations adds an
important difference with respect to the traditional boundary
conditions used in the linearised stability analysis of a general relativistic black hole. In
a BEC black hole, one boundary is the standard asymptotic region,
just like in general relativity. To determine the other boundary, the following must be taken into consideration. For a relativistic or hydrodynamic (quadratic)
dispersion relation, nothing can escape from the interior of a sonic
black hole: the acoustic behaviour is analogous to linearised general relativity. 
But due to the superluminal corrections, the high-frequency effective co-moving speed of sound $c_k$ has a higher value than the zero-frequency speed of sound $c$, see~\ref{eff-sos} and Fig.~\ref{Fig:c_xi}. These superluminal corrections are incorporated in the full Bogoliubov dispersion relation but not in the relativistic approximation. Due to this superluminal character, information from the interior of the acoustic black hole can escape through the horizon and affect its exterior. Therefore, since we are precisely interested in deviations from the relativistic regime, this permeability of the horizon must be taken into consideration. The second
boundary then is not the black hole horizon itself, but the internal `singularity', which corresponds to $x \rightarrow - \infty$ in our model.
We repeat that, in our model, all magnitudes are regular in this region, so there is no singularity in the strict sense. The outgoing boundary condition there reflects the
fact that no information can escape from the singularity.

There is another complication that deserves some attention. In the
case of a hydrodynamic dispersion relation, the signs of $v_g
\equiv v \pm c$ and $\text{Im}(k) \equiv (v \pm c)\text{Im}(\omega)$
coincide for $\text{Im}(\omega)>0$. For example, in an asymptotic \mbox{$x
\to +\infty$} subsonic region, an outgoing $k$-mode has $v_g>0$, so that
$\text{Im}(k)>0$. Such a mode is therefore damped towards this
infinity, giving a finite contribution to its norm. In the
linear stability analysis of black hole configurations, stable dynamical modes therefore correspond to non-normalisable perturbations, while unstable modes 
correspond to normalisable ones. 
When considering modified dispersion relations, however, this
association no longer holds. In particular, with the BEC
dispersion relation, in an asymptotic \mbox{$x \to +\infty$} region,
among the unstable ($\text{Im}(\omega)>0$) outgoing \mbox{($v_g>0$)}
$k$-modes, there are modes with $\text{Im}(k)>0$ as well as modes
with $\text{Im}(k)<0$. An appropriate
interpretation of these two possibilities is the following. The
unstable outgoing modes that are convergent at infinity (those with
$\text{Im}(k)>0$) are associated with perturbations of the system
that originate in an internal compact region of the system.
Unstable outgoing modes that are divergent at infinity are
associated with initial perturbations acting also at the boundary at
infinity itself.

Take for example a black hole configuration of the form
described in Fig.~\ref{F:profiles}. The right asymptotic region,
which can be interpreted as containing a `source' of BEC gas in
our analogue model, simulates the asymptotic infinity outside the
black hole in general relativity. The convergence condition at this right-hand side then implies
that the perturbations are not allowed to affect this asymptotic
infinity initially. However, for the left asymptotic region, this
condition is less obvious. In our BEC configuration, this left
asymptotic region can be seen as representing a `sink'. It
corresponds somehow to the general relativistic singularity of a gravitational black hole. The
fact that in general relativity this singularity is situated at a finite distance
(strictly speaking, at a finite amount of proper time) from the
horizon, indicates that it might be sensible to allow the
perturbations to affect this left asymptotic region from the start.
We will therefore consider two possibilities for the boundary
condition at $x \rightarrow -\infty$. (a) Either we impose
convergence in both asymptotic regions, thereby eliminating the
possibility that perturbations have an immediate initial effect on
the sink, or \mbox{(b) we} allow the perturbations to affect the sink right
from the start, i.e., we do not impose convergence at the left
asymptotic region. The option of imposing convergence at the
left asymptotic region could be interpreted as excluding the
influence of the singularity on the stability of the system. In
other words, condition (b) would then be equivalent to examining the
stability due to the combined influence of the horizon and the
singularity, while under condition (a) only the stability of the
horizon is taken into account.

As a final note to this discussion, since we are interested in
the analogy with gravity, we have assumed an infinite system at the
right-hand side. In a realistic condensate other boundary conditions could
apply, for example taking into account the reflection at the ends of
the condensate (see e.g. \mbox{\cite{Garay:1999sk,Garay:2000jj})}.

\section{Case by case analysis and results}
\label{SS:case-by-case}

Now that we have discussed the boundary conditions, we have all the ingredients necessary to perform the numerical analysis described in the previous section. 
Since we are looking for instabilities (Im$(\omega)>0$), we  plot the function $F(\omega)$, defined in Sec.~\ref{SS:numerical}, in the upper half of the complex plane,
and look for its zeros. Each of these zeros indicates an unstable
eigenfrequency, and so the presence (or absence) of these zeros will
indicate the instability (or stability) of the system.

All figures are given in dimensionless units. Typical values for the velocity of sound $c$ in BECs range from 0.5 to 10mm/s, while the healing length $\xi$ typically lies between 100nm and 2$\mu$m. Taking, for concreteness, $c$=1mm/s and $\xi$=1$\mu$m, our numerical results can be translated to realistic physical numbers by using milliseconds as natural units. For example, the typical timescale for the development of an instability with \mbox{Im$(\omega) \simeq 0.1$} would be about 10ms.
We have checked within a reasonable range of values that our results do not depend on
the particular quantities chosen for the velocities of the system.

We will now discuss the specific
configurations that have been analysed case by case.

\subsection{Black hole configurations}
\label{SS:bh}

\begin{figure*}[tbp]\centering
\begin{tabular}[t]{l p{2pt} c p{5pt} c}

&& 
\hspace*{-50pt}
{\bf Black hole}

&&

\hspace*{-5pt}
{\bf White hole}

\\

&&
\hspace*{-45pt}
{\small $c_\text{super}=0.7$,~ $c_\text{sub}=1.8$.}

&&

{\small $c_\text{sub}=1.8$,~ $c_\text{super}=0.7$.}

\\

\hspace*{-55pt}\begin{tabular}{c}
\vspace{-160pt}
\\
(a)
\\
\vspace{45pt}
\\
(b)
\\
\end{tabular}
&&
\hspace*{-50pt}
\includegraphics[height=.38\textwidth,clip]{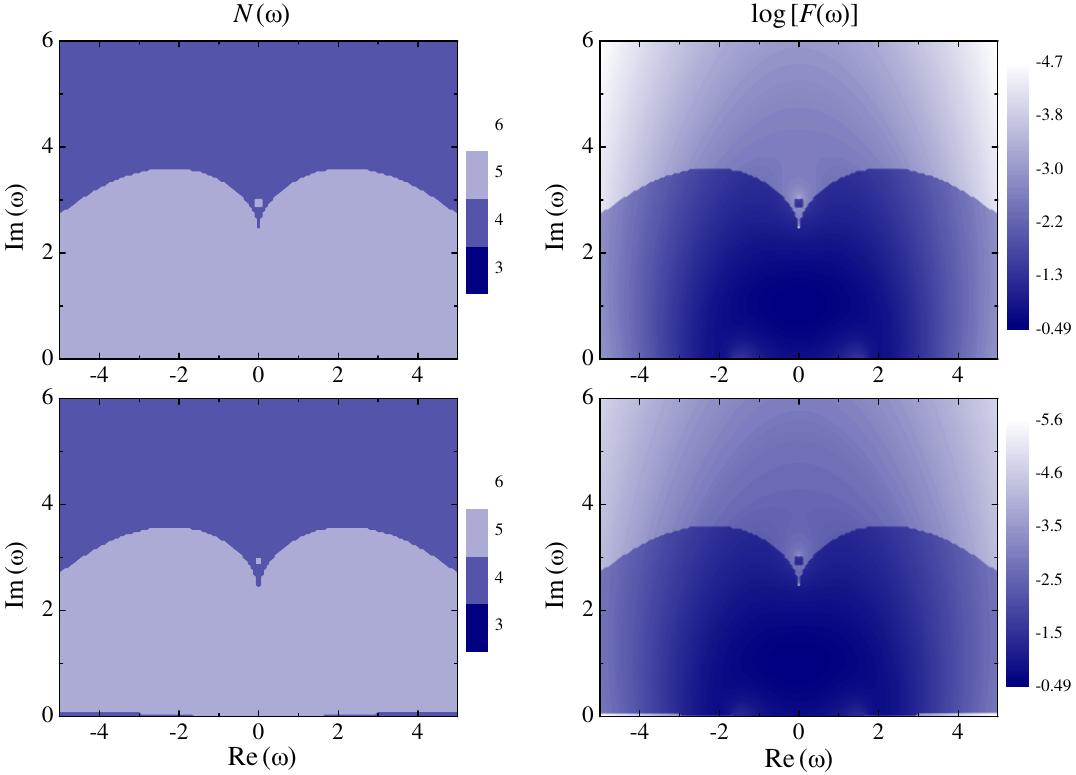}

&&
\hspace*{-7pt}
\includegraphics[height=.38\textwidth,clip]{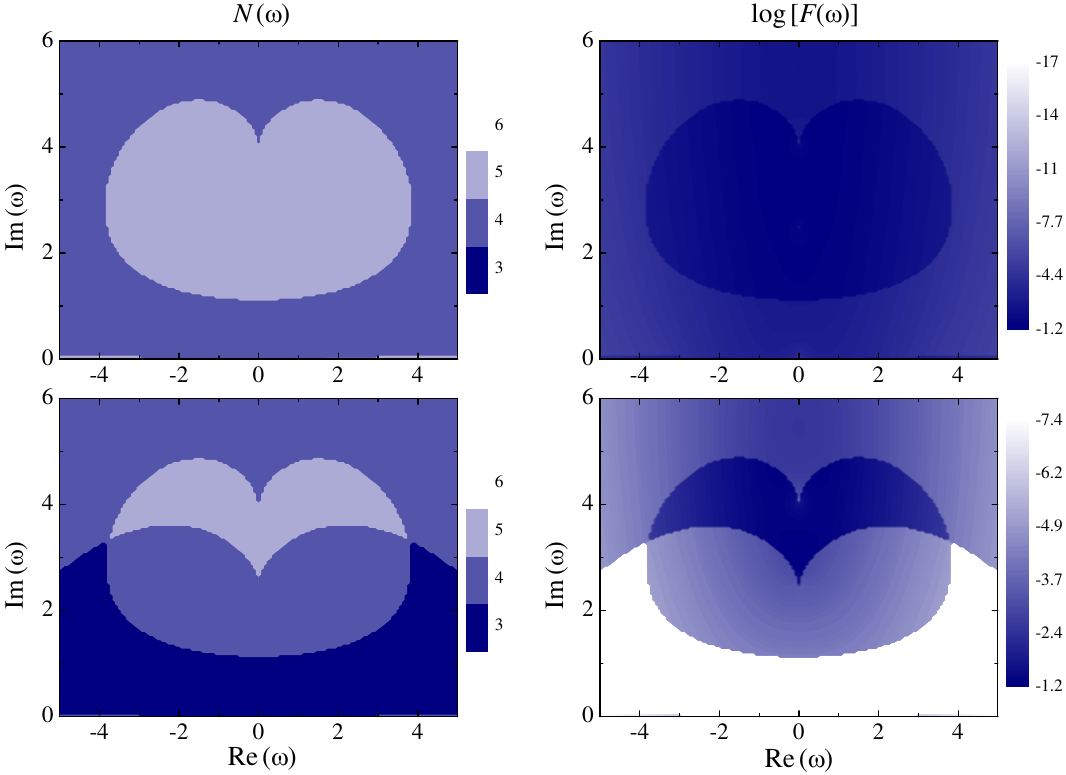}

\\ 
\\ \end{tabular}

\caption{
Plots of $N(\omega) $ and $\log F(\omega) $ for black hole (left) and white hole (right)
configurations (white points or regions, where $F(\omega)\to 0$, represent instabilities). The
speed of sound $c$ is indicated for each region and the velocity $v$ is
then obtained from the constraint $vc^2=1$. Note that $c>1$ corresponds to
a subsonic region, $c<1$ to a supersonic region. In the upper
pictures [case (a)], convergence has been imposed in both asymptotic
regions. In the lower pictures [case (b)], convergence has been
imposed only in the upstream asymptotic region, but not at the sink or `singularity' downstream. It is seen that black
hole configurations are stable in both case (a) and (b), as are
accelerating subsonic flows in general (see Fig.~\ref{F:1stepsSub} below). White hole configurations are stable in
case (a), but develop a huge continuous region of instabilities in
case (b). Only a small strip of instabilities subsists in the
decelerating subsonic flow (see Fig.~\ref{F:1stepsSub} below), indicating that the major part of this
unstable region is a genuine consequence of the existence of the white
hole horizon. Note that continuous regions of instability correspond
to $N(\omega)<4$.}
\label{F:1stepsSuper} 
\end{figure*}

Consider a flow accelerating from a subsonic regime on the right-hand side to a
supersonic regime on the left, see Fig.~\ref{F:profiles}. For observers at the right-hand side, this configuration possesses an acoustic black hole horizon. For such
configurations with a single black hole horizon, when requiring
convergence in both asymptotic regions [case (a)], there are no zeros
(see Fig.~\ref{F:1stepsSuper}), except for two isolated points on the
imaginary axis (see Fig.~\ref{F:Zoom}, which is a zoom of the
relevant area in Fig.~\ref{F:1stepsSuper}). Note that we always check the
existence of a zero by zooming in on the area around its location up 
to the numerical
resolution of our program. These two isolated points seem to be of a rather special
nature. They are located at the boundary between regions with
different number $N$ of forbidden modes in the asymptotic regions. The
zeros that we will find for other configurations are of a totally
different nature: they are either sharp vanishing local minima of
$F(\omega)$ situated well inside an area with a constant value of $N$
($N=4$ to be precise), or continuous regions where the algebraic system represented by the matrix $\widetilde\Lambda_{ij}$ is underdetermined (and hence $N<4$, see Sec.~\ref{SS:numerical}). A technical analysis of these
special points can be found in~\cite{Barcelo:2006yi}. Here, let it
suffice to mention that points of this kind are always present in any
flow, independently of whether it reaches supersonic regimes or
not. Hence it is clear that they do not correspond to real physical
instabilities, since otherwise any type of flow would appear to be
unstable. Accordingly, in the following, we will not
take these points into consideration. When we assert that a figure is
devoid of instabilities, we will mean that the function $F(\omega)$
has no zeros except for the special ones just mentioned.

Fig.~\ref{F:1stepsSuper} also shows that the system remains
stable even when eliminating the condition of convergence at the
left-hand side, i.e., at the sink simulating the singularity [case (b)].

To sum up, configurations possessing a (single) black hole
horizon are stable under the general boundary conditions that we
have described, i.e., outgoing in both asymptotic regions and
convergent in the upstream asymptotic region, independently of
whether convergence is also fulfilled in the downstream asymptotic
region or not.

\begin{figure*}[htbp]\centering

\begin{tabular}[t]{l p{2pt} c p{5pt} c}

&&
\hspace*{-50pt}
{\bf Accelerating subsonic flow}

&&

\hspace*{-5pt}
{\bf Decelerating subsonic flow}

\\

&&
\hspace*{-45pt}
{\small $c_\text{sub1}=1.8$, $c_\text{sub2}=1.9$.}

&&

{\small $c_\text{sub1}=1.9$, $c_\text{sub2}=1.8$.}

\\

\hspace*{-50pt}
\begin{tabular}{c}
\vspace{-160pt}
\\
(a)
\\
\vspace{45pt}
\\
(b)
\\
\end{tabular}

&&
\hspace*{-50pt}
\includegraphics[height=.38\textwidth,clip]{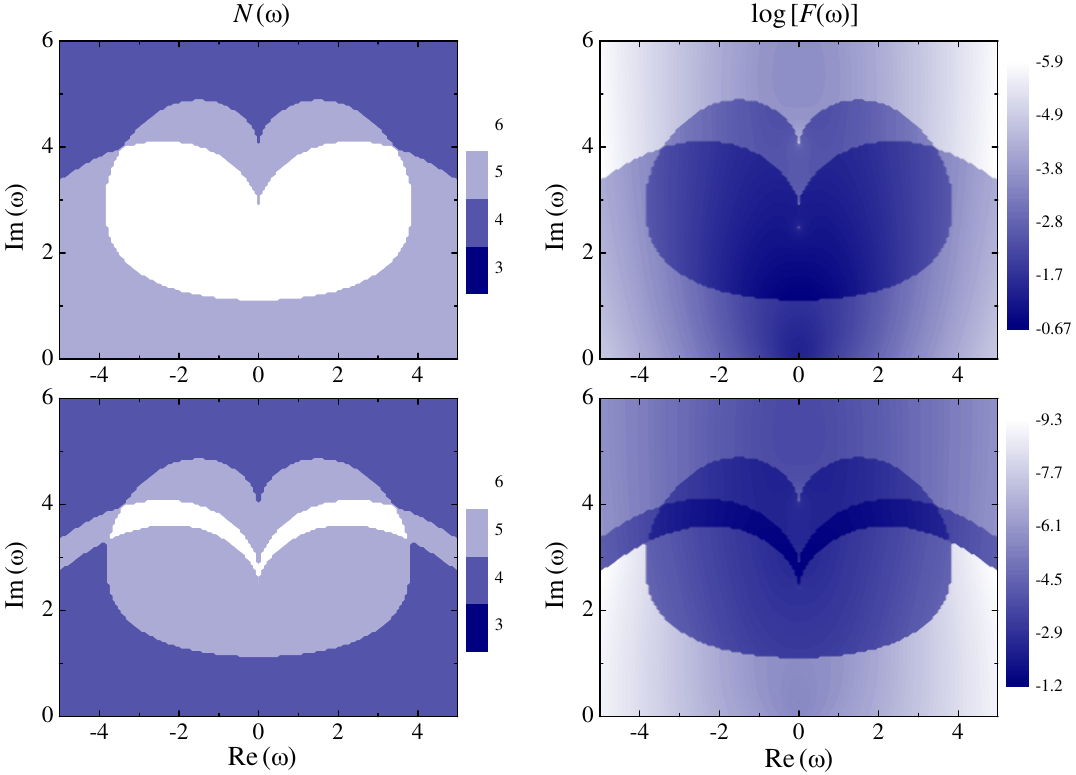}

&&
\hspace*{-7pt}
\includegraphics[height=.38\textwidth,clip]{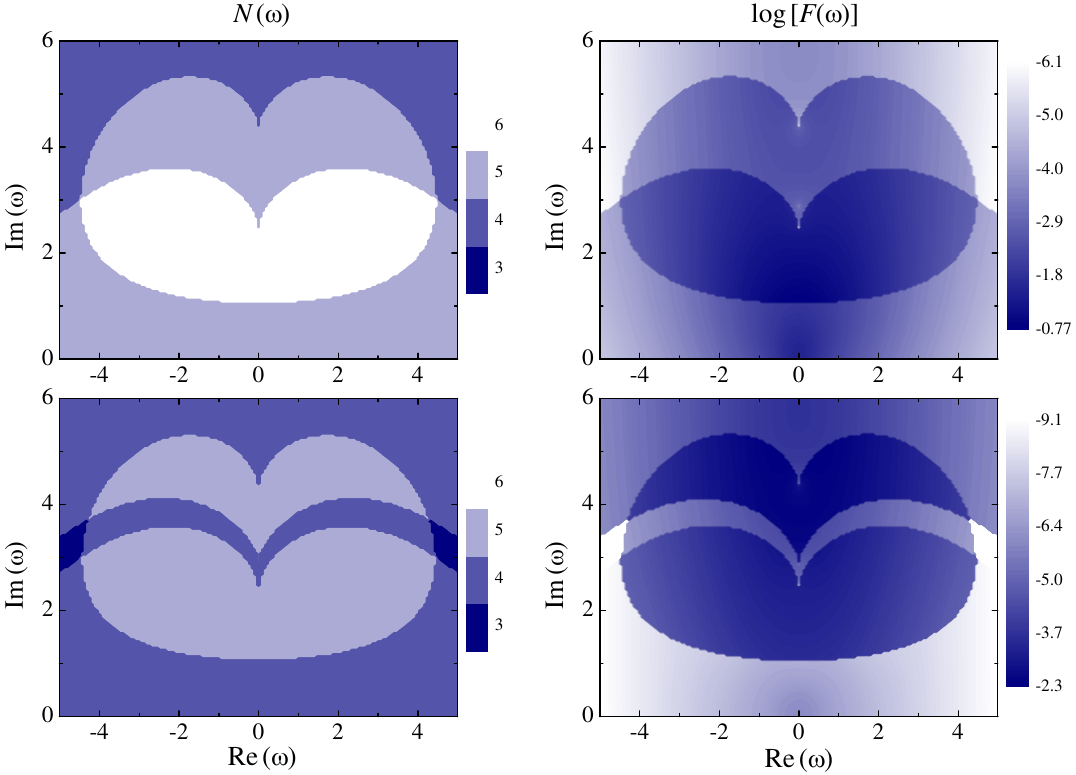}
\\ \end{tabular}

\caption{Plots of $N(\omega) $ and $\log F(\omega) $ for accelerating (left) and decelerating (right)
subsonic flows (i.e., without acoustic horizons). In the upper
pictures [case (a)], convergence has been imposed in both asymptotic
regions. In the lower pictures [case (b)], convergence has been
imposed only in the upstream asymptotic region. The
decelerating subsonic flow exhibits a small strip of instabilities in case (b), which was already found in the white
hole case (note that these continuous regions of instability correspond
to $N(\omega)<4$).}
\label{F:1stepsSub} 
\end{figure*} 

\begin{figure}[tbp]
\begin{center}
\includegraphics[width=.3\textwidth,clip]{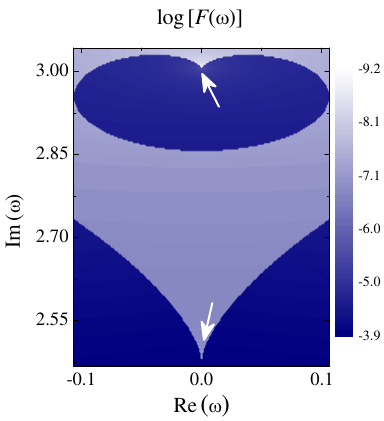}
\figcaptionl{F:Zoom}{Two special zeros of the function $F(\omega)$ appear in the stability 
analysis of a black hole configuration (this plot is a zoom of
the corresponding region in Fig.~\ref{F:1stepsSuper}). They are
located at the boundary between regions with different number $N$ of
prohibited modes. These points appear in the stability analysis of all configurations studied, including purely subsonic ones, and hence do not seem to represent real 
physical instabilities of the system.}
\end{center}
\end{figure}

\subsection{White hole configurations}
\label{SS:wh}

Let us now consider flows decelerating from a supersonic regime at the right-hand side to a subsonic one on the left. From the point of view of observers at the left-hand side, the
geometric configuration possesses a white hole horizon. In general relativity, a white
hole corresponds to the time reversal of a black hole. Therefore,
in general relativistic terms, unstable modes of a white hole configuration would correspond to
stable (quasinormal) modes of the black hole. From this point of view, any
quasinormal mode found in the analysis of a black hole configuration (see next chapter)
could immediately be interpreted as an instability of the corresponding white hole. Note that such a point of view implies that the boundary conditions appropriate for the analysis of
white hole configurations correspond to only allowing {\it ingoing} waves
at the boundaries, due to time reversal. 

But from the point of view
of acoustic models in a laboratory, the analysis
of the intrinsic stability of the flow (under the outgoing
boundary conditions described above) is also interesting. This analysis also has particular relevance with regard to configurations with two horizons (see~\cite{Barcelo:2006yi}).

In Fig.~\ref{F:1stepsSuper} we see that under outgoing boundary conditions the
flow is stable when convergence is required in both asymptotic
regions [case (a)], but exhibits a continuous region of
instabilities at low frequencies when convergence is fulfilled only at
the right-hand side [case (b)]. Indeed, in this continuous region, $N=3$,
in other words the algebraic system
$\widetilde\Lambda_{ij}A_{ij}=0$ is underdetermined and any frequency
is automatically an eigenfrequency.

When looking at the case of a completely subsonic flow undergoing a
deceleration (see Fig.~\ref{F:1stepsSub}), we find something similar. The
system is clearly stable when convergence is imposed at the left-hand side,
i.e. in case (a). Without convergence at the left-hand side, case (b), there is a
small continuous strip of instabilities which corresponds, as in the
white hole case, to a region where $N=3$. However, this region
is located at relatively high frequencies, disconnected from
$\omega \sim 0$. It seems that part of the continuous region of
instabilities found in the white hole configuration has its origin merely in the
deceleration of the flow (giving rise to this high-frequency strip).
But there is still a complete region of instabilities
that is a genuine consequence of the existence of a white hole horizon. In fact, by
decreasing the healing length parameter $\xi$, the strip associated with the mere deceleration moves up to higher and higher frequencies, becoming less and less important as one
approaches the hydrodynamic or acoustic limit. The main continuous region of
instabilities that can be associated with the horizon, i.e., the white region expanding from $\omega \sim 0$, on the other hand, does not change its
character in this process.

We can therefore conclude the following with regard to
decelerating configurations. When convergence is fulfilled
downstream, the configuration is stable, regardless of whether it
contains a white hole horizon or not. When this convergence
condition is dropped, there is a tendency to destabilisation. In the
presence of a white hole horizon, the configuration actually becomes
dramatically unstable, since there is a huge continuous region of
instabilities, and even perturbations with arbitrarily
small frequencies destabilise the configuration. In the absence of such a
horizon, only a small high-frequency part of this unstable region subsists.

\subsection{Black hole configurations with modified boundary conditions}
\label{SS:sink-bh}

We have seen earlier that configurations with a black hole horizon do not possess instabilities under either of the sets of boundary conditions that we considered. However, instabilities can show up when instead of extending the black hole configuration towards $- \infty$, a wall (or sink) is introduced at a finite distance inside the supersonic region, leading to a possible bouncing and amplifying effect of some modes inside the black hole. Such a case would be described by boundary conditions different from the ones we have considered so far. For example, by replacing the  boundary conditions at the left-hand side by $\theta|_{x=-L}=0$,
it can be seen from Fig.~\ref{F:Sink} that a discrete set of
unstable modes appears. 

\begin{figure}[tp]\centering
\includegraphics[width=.45\textwidth,clip]{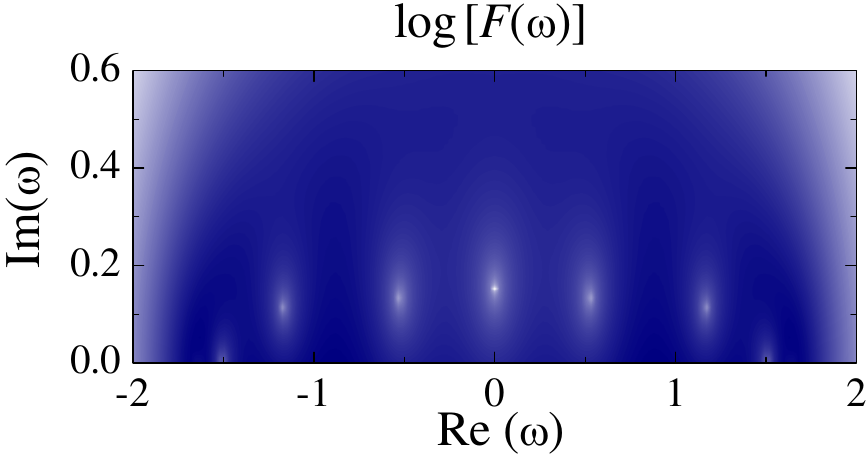}
\figcaptionl{F:Sink}{A discrete set of instabilities appears in a black hole 
configuration when the left asymptotic region simulating the
singularity is replaced by a wall or sink at a finite distance inside the black hole. In this plot we have used
$c_{\rm sub}=1.9$ and $c_{\rm super}=0.7$ (with the corresponding
$v=1/c^2$). In addition we have taken $L=6$ as the size of the finite internal region.}
\end{figure}

\section{Discussion}
\label{SS:stab_discussion}

Let us start by discussing the stability of configurations with an acoustic
black hole horizon in analogue systems that incorporate
superluminal dispersion relations. We have seen that by requiring
purely outgoing and convergent boundary conditions in both asymptotic
regions, these configurations do not show any signs of instability.
The same applies when dropping the convergence condition
downstream (i.e., on the left-hand side simulating the singularity). This seems to contradict the results in~\cite{Garay:2000jj}. There, the existence of a future (spacelike)
singularity inside the black hole, from which no information is
allowed to escape, was implemented by introducing a sink in the
supersonic region at a finite distance from the horizon. Then, it was
found that there were discrete instabilities in the system. However,
these instabilities correspond to the following particular set of
boundary conditions: i) At the asymptotic region, only convergent
boundary conditions were imposed, without any condition about the
direction of propagation (in- or outgoing) of the perturbations;
\mbox{ii) At} the sink, two types of boundary conditions were
required, specifically designed for dealing with symmetric and
anti-symmetric configurations. In our language these boundary conditions
correspond to $\{ \theta'|_{\rm sink} = 0, n'|_{\rm sink}=0 \}$ and $\{
\theta|_{\rm sink} = 0, n|_{\rm sink}=0 \}$, respectively. Comparison of the results obtained with these boundary conditions~\cite{Garay:2000jj} with the results that we obtained here, leads to two  important conclusions. On the one hand, the unstable modes found in~\cite{Garay:2000jj} have ingoing contributions at the asymptotic region. On the other hand, the boundary conditions at the
sink are such that they combine outgoing and ingoing contributions. The sink implementation causes waves reaching the sink to bounce back
towards the horizon. These two facts are responsible for the unstable
behaviour of such black hole configurations. If no energy is
introduced into the system from the asymptotic region (in other words,
if only outgoing perturbations are allowed) and moreover any bouncing
at the sink is eliminated, then these configurations
are stable. This is in agreement with the result found in~\cite{leonhardt2003}.

In the case of configurations with a white hole horizon, we
have seen that with outgoing and convergent boundary conditions in
both asymptotic regions, there are no instabilities in the
system. However, when eliminating the convergence condition in the
downstream asymptotic region (at the sink), one finds a continuous region of
instabilities surrounding $\omega=0$. Thus, we see that these white
hole configurations are stable only when the boundary conditions are
sufficiently restrictive.

When analysing configurations connecting two different subsonic
regions, i.e., in the absence of acoustic horizons, we have also seen that, again, when convergence is required
at the left-hand side, they are stable. But when this convergence condition is
relaxed, globally decelerating configurations tend to become unstable,
whereas globally accelerating ones remain stable. The instabilities of
these decelerating configurations without horizons (i.e., purely
subsonic ones) show up, however, in a small strip at high
rfequencies. In contrast, white hole configurations present
instabilities for a wide range of frequencies, starting from
arbitrarily small values. This shows that the presence of a white
hole horizon drastically augments the instability of the configuration.

\section{Summary and conclusions}
To sum up, we have shown the high sensitivity of the stability not
only to the type of configuration (the presence of a horizon, the accelerating or decelerating character of the fluid), but particularly to the boundary conditions. In general, the boundary conditions that we have considered are directly inspired by similar analyses in general relativity, and so we have always imposed {\it outgoing} boundary conditions, i.e., only the possible existence of instabilities arising {\it within} the system should be taken into account. For similar reasons, we argued that convergence must also be imposed at the upstream (right-hand side) asymptotic region, i.e., at the source of the condensate.

When moreover requiring convergence at the left-hand or downstream asymptotic region (which can be interpreted as simulating a singularity, since nothing can escape from it towards the horizon), both black hole and white hole
configurations are stable (and also the combination of both into a
black hole--white hole configuration, see~\cite{Barcelo:2006yi}). When relaxing this convergence
condition at the left-hand side, configurations with a single black hole horizon
remain stable, whereas white hole 
configurations (and black hole--white hole configurations, see~\cite{Barcelo:2006yi}) develop instabilities not present in flows 
without horizons.

Some final comments with relevance for the remainder of this thesis might be useful. We have seen that black hole configurations are stable under rather general boundary conditions. This implies the following. 

First of all, we have explicitly worked with the Bogoliubov dispersion relation valid for a BEC. However, since we have checked a wide range of parameters, the results that we have obtained can be extended to superluminal dispersion relations of the Bogoliubov type in general. One can then tentatively conclude that adding such superluminal corrections does not radically alter the dynamical stability of a relativistic black hole, whether it is an acoustic black hole or a gravitational one. 

Note, however, that this result is `fragile' in the following sense: configurations with a black hole horizon and a bounce inside the system can become unstable. Also, the stability is guaranteed if only ingoing modes are allowed. Although this is a sensible condition from a theoretical analogue-gravity point of view, it is not so clear that this condition makes sense in a BEC experiment. Nevertheless, even in a realistic BEC experiment, it should be possible to make a condensate sufficiently long in order for reflection at its boundaries to be negligible for the duration of the experiment~\cite{Garay:2000jj,Cornell:2009}. Let us also remark that dynamical instabilities are not the only instabilities that can appear in an experiment. Surface instabilities, for example, could also appear. Such surface instabilities are sometimes a desired effect because they can lead to vortex formation. However, they are also generally easy to avoid, in particular in effectively one-dimensional condensates~\cite{Pogosov:2006}.

Second, since we have shown that there can exist dynamically stable acoustic black hole configurations, it indeed makes sense to analyse the stable dynamical modes of such configurations, i.e., the quasinormal modes, as we will do in the next chapter. 

Finally, it also means that such configurations can indeed be expected to reproduce analogue (i.e., phononic) Hawking radiation. Although this obviously does not solve all experimental issues, at least it removes the need to recur to metastable configurations, fine-tuned in the sense that the dynamical instabilities would not disrupt the configuration nor obfuscate the black hole radiation in the time necessary to achieve the experiment.

\clearpage
\chapter{Quasinormal modes}
\label{S:QNMs}
In this chapter, we perform a quasinormal mode analysis of black hole configurations
in Bose--Einstein condensates (BEC). In this analysis we use the full
Bogoliubov dispersion relation (see Sec.~\ref{SS:dispersion}), not just the hydrodynamic or 
relativistic approximation. We restrict our attention to (1+1)-dimensional
black hole configurations in BEC with steplike discontinuities of the type illustrated in Fig.~\ref{F:profiles}. For this case we show 
that in the hydrodynamic or relativistic approximation, quasinormal
modes do not exist. The full dispersion relation, however, allows the existence of short-lived quasinormal modes. Remarkably, the spectrum of these modes is not discrete but continuous.


\section{Introduction}
\label{SS:QNM_intro}

In the previous chapter, the stability of several
(1+1)-dimensional configurations in BECs was studied. These
configurations were chosen in analogy with the geometries
associated with gravitational black holes. Here we will elaborate further on black hole
configurations, which were shown to be devoid of instabilities,
and examine their quasinormal or relaxation modes. The analysis we
carry out is analogous to the standard quasinormal mode (QNM)
analysis in gravity
\cite{Nollert:1999ji, Kokkotas:1999bd}. According to general relativity, when black holes are
perturbed, they emit gravitational waves in the following way (see Fig.~\ref{F:grav_radiation}). The
initial regime depends mainly on the concrete form of the
perturbation. Then, an oscillatory decay phase is attained in which the
precise form of the emitted wave depends only on the properties of the
black hole itself. Finally, a polynomial tail characterises the return
to equilibrium. In the intermediate oscillatory decay phase, a discrete set of complex
frequencies is excited: the quasinormal modes of our concern. Since
quasinormal modes are modes of decay, i.e., of energy dissipation,
they are outgoing: they move towards the exterior of the spacetime region connected to an asymptotic observer. In general relativity this means that their group velocities are
directed outwards both in the asymptotic region and at the
horizon. Moreover, as we will illustrate, in general relativity this also
automatically implies that they are non-normalisable or divergent.

\begin{figure}[tbp]\centering
\includegraphics[width=0.5\textwidth,clip]{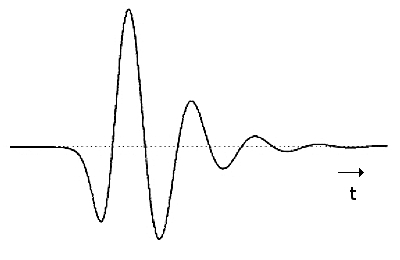}
\figcaptionl{F:grav_radiation}{Typical wave form of gravitational radiation emitted by a perturbed black hole as a function of time $t$. The initial pulse depends on the concrete form of the perturbation. Then, a damped oscillation consisting of a discrete set of frequencies, the quasinormal modes, sets in. (Often, as in this figure, a single frequency clearly dominates). Finally, a power-law tail takes over at late time. 
}
\end{figure}

As we will briefly point out, in standard general relativity in 1+1 dimensions, quasinormal modes do not exist. In higher dimensions, a discrete spectrum appears \cite{Nollert:1999ji, Kokkotas:1999bd}, see Fig.~\ref{F:QNM_discrete_spec}. A general investigation of quasinormal modes in analogue black holes was carried out in \cite{Berti:2004ju}. Results qualitatively identical to the general relativistic case were found. Here the essential element that we add is the influence on these
QNMs of a non-relativistic contribution to the dispersion relation. 

\begin{figure}[t]\centering
\includegraphics[width=0.7\textwidth,clip]{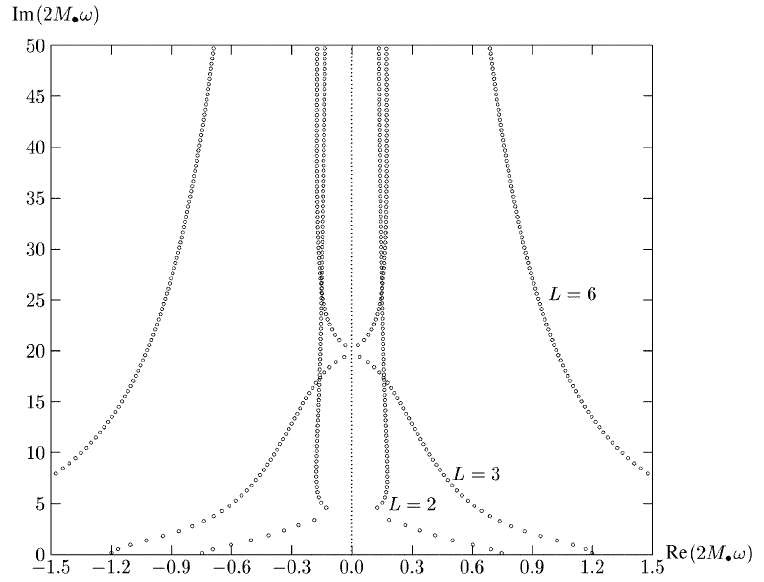}
\figcaptionl{F:QNM_discrete_spec}{Discrete set of quasinormal modes in the complex frequency-plane for a Schwarzschild black hole and different values of the angular momentum index $L$. Figure taken from~\cite{Nollert:1999ji}. (Note that we use the opposite convention for the sign of Im($\omega$).)}
\end{figure}

\section{Boundary conditions}
\label{SS:QNM_boundary}
The general framework to calculate the quasinormal modes in these BEC black hole configurations is the same as for the calculation of the instabilities, and has been described in Chapter~\ref{S:BHconfigurations}. However, we still need to complement this by the adequate boundary conditions. We will first take a look at the boundary conditions that are used to calculate QNMs in general relativity. This can be immediately extrapolated to BECs in the hydrodynamic limit. As we will show, in this hydrodynamic limit in 1+1 dimensions, there exist no QNMs. Finally, we will come to the boundary conditions with the full dispersion relation.
%
\subsection{Quasinormal modes of a gravitational black hole}
\label{SS:QNM_gravity}
Quasinormal modes of a gravitational black hole are the relaxation modes or modes of energy dissipation that characterise the pulsations of the black hole after perturbations initiated internally in the spacetime surrounding the black hole. 

Since the QNMs are relaxation modes, they must decay in time and moreover be outgoing. The first condition means that, for the sign convention followed here (see e.g. eq.~\eqref{bogoliubov_modes}), they can be represented by complex frequencies $\omega$ with Im$(\omega)\!<\! 0$. The outgoing character must be imposed in the asymptotic region. At the horizon, QNMs are also required to be outgoing because in general relativity nothing can escape through the horizon.\footnote{As we mentioned in the introduction to this chapter, in QNM terminology, `outgoing' means directed towards the exterior of the spacetime region connected to an asymptotic observer. In other words, at the horizon, `outgoing' means directed towards the singularity.} So the boundary conditions that are imposed in general relativity to find the QNMs of a black hole are simply that these modes should be outgoing both in the asymptotic region and at the horizon.

Because QNMs of a gravitational black hole are outgoing, they are also divergent. Indeed, due to the quadratic form of the relativistic dispersion relation, $\omega^2\propto k^2$, the group velocity $v_g\equiv \d \omega/\d k$ and the imaginary part of $k$ are related by 
\begin{equation}
\text{Im}(k)\propto\text{Im}(\omega) /v_g~,
\end{equation}
and therefore, since QNMs have Im$(\omega)<0$: 
\begin{equation}
\text{sign}[\text{Im}(k)]=-\text{sign}[v_g]~.
\end{equation}
%

\subsection{Hydrodynamic limit in 1+1 dimensions: absence of QNMs}
\label{SS:hydrodynamic_QNM}

In BECs in the hydrodynamic limit, the dispersion relation is also quadratic: $(\omega-vk)^2=c^2k^2$, see eq.~\eqref{quadr_dispersion}, and therefore decaying outgoing modes are also automatically divergent. Indeed,
\begin{equation}
v_g=v \pm c~,\qquad
\text{Im}(k)=\text{Im}(\omega)/v_g~,
\end{equation}
and therefore $\text{sign}[\text{Im}(k)]=-\text{sign}[v_g]$ as before. Let us discuss whether QNMs can exist in this case.

There are two ways to deal with this question. Either one can define a connection matrix based on the hydrodynamic limit of the matching conditions~\eqref{matching}, and see that the appropriate outgoing boundary conditions can never be fulfilled, see~\cite{Barcelo:2007iu}. Alternatively, one can reason as follows. Because of the presence of the horizon, we know that, in the hydrodynamic limit, none of the modes at the left-hand side can move upstream across the horizon, i.e., they cannot connect to outgoing modes at the right-hand side. In other words, when the $(x<0)$ region is supersonic, it becomes causally disconnected from the subsonic region $(x>0)$. But then, just like in general relativity, the requirement that QNMs must be outgoing should be imposed at the horizon, and not at the left asymptotic infinity. Because of the homogeneity of the profile, the mode solutions for each particular frequency $\omega$ can be written as piecewise sums of plane waves:
\begin{eqnarray}
\widetilde n_1 &=& A_1 e^{i(k_1x - \omega t)} + A_2 e^{i(k_2x - \omega t)}~,\\
\theta_1 &=& B_1 e^{i(k_1x- \omega t)} + B_2 e^{i(k_2x- \omega t)}~.
\end{eqnarray}
Because of this plane-wave character, none of these modes can be outgoing both at the right asymptotic region and at the horizon. QNMs can therefore not exist in the hydrodynamic limit in 1+1 dimensions. 

Although the argument presented here seems to depend on the homogeneity of the profiles, it is actually more general. Indeed, in the hydrodynamic limit, the geometric description developed in section~\ref{SS:metric} is valid. It is well known from general relativity that all (1+1)-dimensional metrics are conformally flat. In addition, the d'Alembertian equation in two
dimensions is conformally invariant~\cite{Birrell:1982ix}. Therefore,
all the solutions of the system are conformally equivalent to plane waves, and so they do not satisfy the requirement of being outgoing both in the asymptotic region and at the horizon. Hence QNMs cannot exist in (1+1)-dimensional relativistic configurations, and this argument is valid both for general relativity and for acoustic black holes in the hydrodynamic limit. The situation is completely different in 3+1 dimensions, in which both general relativistic and hydrodynamic acoustic black holes exhibit a discrete QNM spectrum, as we mentioned earlier.

%
\subsection{Boundary conditions with the full dispersion relation}
\label{SS:boundary_conditions}
A crucial feature of the superluminal character of the dispersion relation is that it makes the horizon permeable, so that outgoing boundary conditions should be imposed at both asymptotic infinities (rather than at the right asymptotic infinity and at the horizon). Moreover, since we have two homogeneous regions, the ingoing or outgoing character of a mode is conserved along each side of the horizon ($x=0$). 
An outgoing mode is then one that has a group velocity $v_g<0$ for $x<0$ (and in particular for $x\!\to\! -\infty$) and $v_g>0$ for $x>0$ (and in particular for $x\!\to\! +\infty$). $v_g$ is obtained from the full Bogoliubov dispersion
relation, see eq.~\ref{group-velocity}:
\begin{equation}
v_g(k) \equiv \text{Re} \left( \frac{d \omega}{dk}\right) =
\text{Re}\left(
\frac{c^2 k + {\frac{1}{2}} \xi^2 c^2 k^3}{\omega - vk} + v
\right).
\end{equation}
Note that the immediate connection between the sign of $v_g$ and that of Im$(k)$ which existed in the hydrodynamic limit clearly does not subsist, so the automatic divergent character of outgoing modes is lost. 

In any case, quasinormal modes should be outgoing, and this is the essential boundary condition. 
Additionally, one might wonder whether the QNMs will be convergent, or divergent after all (in spite of the modification of the dispersion relation). An easy way to check this is by applying convergence as an additional boundary condition and comparing the resulting spectrum with the one obtained without this additional condition.

\subsection{Summary}
To sum up, quasinormal modes should be outgoing. In general relativity and in acoustic black holes in the hydrodynamic approximation, this outgoing condition must be imposed at the external (right-hand side) asymptotic infinity and at the horizon. Moreover, the outgoing character of QNMs automatically implies a
divergent behaviour in the asymptotic regions.
However, in both these cases, there are no QNMs in 1+1 dimensions, whereas in 3+1-dimensional black holes, a discrete QNM spectrum appears. 

In BECs with the full modified dispersion relation,
outgoing modes are no longer automatically divergent. The boundary
condition to determine QNMs in the (1+1)-dimensional BEC black hole
configurations under study is simply that they should be
outgoing in both asymptotic regions.

Application of the method developed in chapter~\ref{S:BHconfigurations} then leads to the results that we will now discuss.
%
\section{Results and discussion}
\label{SS:QNM_discussion}
The quasinormal mode spectrum of a (1+1)-dimensional acoustic black hole in a BEC, obtained with the full dispersion relation, is illustrated in Fig.~\ref{F:QNM_outgoing}. Remember from the previous chapter that such a black hole did not possess unstable eigenfrequencies, so it indeed makes sense to look for stable modes. Surprisingly, as can be seen from Fig.~\ref{F:QNM_outgoing}, not only does such a QNM spectrum indeed show up, but this spectrum is actually continuous. Indeed, in the whole continuous region where $N(\omega)\!=\!3$, the algebraic system composed of the matching and boundary conditions is underdetermined. Therefore, according to the procedure outlined in chapter~\ref{S:BHconfigurations}, every frequency in this region is automatically a quasinormal mode. We have checked that no further (discrete) quasinormal modes appear in the $N\!=\!4$ region.
%
\begin{figure}[t]\centering
\includegraphics[width=0.6\textwidth,clip]{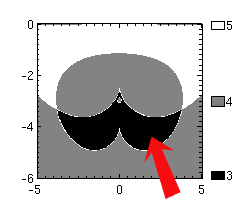}
\put(-160,0){\bf\Large{Re($\omega$)}}
\put(-155,218){\bf\Large{N($\omega$)}}
\put(-270,95){\begin{sideways}{\bf\Large{Im($\omega$)}}\end{sideways}}
\figcaptionl{F:QNM_outgoing}{Plot illustrating the quasinormal mode spectrum for a (1+1)-dimensional black hole configuration in a BEC. We represent the number $N$ of constraints (i.e., the number of forbidden modes) following from the boundary condition that QNMs should be outgoing. The QNM spectrum consists of the continuous region where $N(\omega)\!<\!4$, marked with a red arrow. The numerical values used for this plot, in units such that the healing length $\xi=1$ and the speed of sound $c=1$ are $v=0.7$ in the subsonic region and $v=1.8$ in the supersonic region.}
\end{figure}

%
%
We have also checked whether these quasinormal modes are convergent or not. We have found that the quasinormal modes here contain both convergent and divergent contributions, and actually disappear as soon as either type of contribution is prohibited (i.e., as soon as a strict convergence or divergence condition is imposed at either side of the configuration). Overall, this makes these quasinormal modes divergent. However, this divergent character is not obtained in the same automatic sense, i.e., it is not due to a strict relation between the signs of $v_g$ and Im$(k)$, as in the case of the QNMs that are found in the (3+1)-dimensional relativistic and hydrodynamic cases discussed earlier.

%
%

We have furthermore calculated the QNM spectrum for other black-hole-like configurations that were shown to be devoid of instabilities in~\cite{Barcelo:2006yi}, and we have checked that a similar continuous spectrum of QNMs appeared in all of them. This indicates that the continuous character of the QNM spectrum is an essential consequence of the modified dispersion relation, and not just of the particular characteristics of a precise configuration. In fact, the presence of these QNMs can essentially be traced back to the permeability of the horizon due to the superluminal character of the dispersion relation. In our particular model, this is easy to see. Indeed, imposing an outgoing boundary condition at the horizon would prohibit the existence of QNMs due to the piecewise homogeneity of the profile, regardless of whether one considers the full dispersion relation or only its acoustic limit.

Finally, a straightforward dimensional estimate shows that these quasinormal modes are short-lived. Indeed, as can be seen from Fig.~\ref{F:QNM_outgoing}, the obtained quasinormal modes typically have Im$(\omega) \sim 1$ (in dimensionless units). This means that these quasinormal modes are characteristic of a relaxation at the microscopic level, i.e., at the scale of the healing length (which, as we have seen in section~\ref{SS:dispersion}, is also characteristic of the Lorentz symmetry breaking scale). Adding dimensions, the lifetime of such quasinormal modes is of the order of $\xi/c$, the healing length of the condensate divided by its speed of sound, typically around 1$\mu$m and 1mm/s, respectively, resulting in lifetimes $t_\text{QNM}\sim$1ms. 
%
\section{Summary and conclusions}
We have examined the quasinormal modes of (1+1)-dimensional black hole configurations in BECs, and laid particular emphasis on the importance of the quartic form of the full dispersion relation, compared to the quadratic form of a hydrodynamic or relativistic one. In the latter case, outgoing relaxation modes are automatically divergent. The modification of the dispersion relation means that this relation is no longer automatic. Nevertheless, the quasinormal modes in the systems discussed here are also divergent. 

More importantly, the QNM spectrum that was obtained in these systems with a modified dispersion relation consists of a continuous region in the complex frequency plane. The lifetime of these QNMs was estimated to be of the order of $\xi/c$, the healing length of the condensate divided by its speed of sound. These quasinormal modes are therefore characteristic of a relaxation at the microscopic scale, with typical values of the lifetime in the millisecond range.

The importance of this result lies in the following observation. Both in general relativity and in BECs in the hydrodynamic limit, QNMs simply do not exist in 1+1 dimensions, while in higher dimensions they form a discrete spectrum. So due to the modification of the dispersion relation, the QNM spectrum in (1+1)-dimensional BEC black holes changes from non-existent to a continuous region of frequencies. It should be noted that this result does not depend on fine-tuned choices of the parameters of the configuration, but is a generic consequence of the permeability of the horizon in the presence of superluminally modified dispersion relations. It is therefore a straightforward speculation that the discrete spectrum which is obtained in
the usual QNM analysis in standard 3+1-dimensional general relativity will
also develop continuous regions if it turns out that there exist superluminal modifications of the
dispersion relations at high energies. The lifetimes of these additional, continuous bands of quasinormal modes will probably be extremely short, since they would characterise a relaxation at the microscopic or `(trans)planckian' level, after which a relaxation at the classical level of general relativity takes over. Their existence could therefore in principle be a signal of transplanckian (quantum-gravitational) physics encoded in the 
spectrum of gravitational waves emitted by black holes.

\clearpage
\chapter[Lorentz violation and modified disp.\ relations]{Intermezzo: Lorentz violation and modified dispersion relations}
\label{S:intermezzo}
In the previous chapters, we have studied the stability and quasinormal modes of BEC black holes, and speculated that the results could be extrapolated to quantum gravity. In this brief intermezzo, we will make a more precise argument for this extrapolation to quantum gravity in scenarios with Lorentz symmetry breaking---see~\cite{Mattingly:2005re,Jacobson:2005bg} for general introductions on Lorentz violations at high energy and their possible detection.

From a phenomenological point of view, one of the first modifications that quantum gravity could impose on classical physics is a modification of the dispersion relations for particles. Assuming that this effect is perturbative at low energy, this can be written in the following way (in a given observer's frame). The usual locally Lorentz invariant energy-momentum dispersion relation 
\begin{equation}\label{relDR}
 E^2= m^2c^4 + c^2p^2
\end{equation}
is replaced by a modified dispersion relation
%
%
%
\begin{equation}\label{modDR}
 E^2= m^2c^4 + c^2p^2 +  \sum_i f^{(i)}c^i\abs{p}^i E_L^{2-i}~,
\end{equation}
where $E_L$ indicates the energy scale of local Lorentz symmetry breaking and $f^{(i)}$ the strength of $i^\text{th}$ order Lorentz symmetry breaking at the energy $E_L$. Here, we are in the first place interested in examining whether there exist concrete physical phenomena that break Lorentz invariance around a certain well-defined fundamental energy scale. In order to extract concrete boundaries on the possible existence of such phenomena, eq.~\eqref{modDR} is then the adequate starting point. Actually, from a `naive' quantum gravity perspective, this $E_L$---if it exists---is expected to be around the Planck scale, and so the previous relation is usually written as
\begin{equation}\label{modDR-planck}
 E^2= m^2c^4 + c^2p^2 +  \sum_i \tilde{f}^{(i)}c^i\abs{p}^i E_{Pl}^{2-i}~,
\end{equation}
where $E_L$ has been replaced directly by the Planck energy $E_{Pl}$.

The fact that we immediately wrote, for example, $f^{(2)}p^2$ in the previous expressions rather than the more general $f_{ij}^{(2)}p^ip^j$ means that we are only looking for Lorentz symmetry breaking in the boost sector. The motivation is the following. Breaking the rotational subgroup of the Lorentz group would either entail observable effects at everyday macroscopic scales or imply that the boost subgroup is also broken. The opposite is not true: the boost subgroup can be broken without the rotation subgroup being broken too, and without leading to effects that should have been detected already at energies experimentally available. Therefore, phenomenologically speaking, it is sufficient to focus on Lorentz symmetry breaking in the boost sector. Also note that $f^{(i)}$ can be particle- and helicity-dependent.

The scheme just summarised is most adequate for effective field theory approaches (EFTs)~\cite{Mattingly:2005re,Jacobson:2005bg}. There are two reasons~\cite{AmelinoCamelia:2008qg} why we are interested here in studying Lorentz symmetry breaking from an EFT perspective rather than for instance Doubly Special Relativity (DSR).\footnote{Doubly Special Relativity~\cite{AmelinoCamelia:2000mn,Magueijo:2001cr} consists in the proposal that the Lorentz group might be {\it deformed} at high energies rather than {\it broken}, by the introduction of a second observer-independent quantity (apart from the speed of light), namely a minimum length. DSR leads to modified dispersion relations, though generally more complicated than the form we discuss here, without the equivalence of reference frames being broken.} First, the EFT approach is the simplest one from a phenomenological point of view. It is also an approach for which the framework is sufficiently well understood and precisely defined to allow the  extraction of sensible constraints on concrete sets of physically meaningful parameters. The second reason originates in analogue gravity: the form of Lorentz symmetry breaking discussed here is the one which corresponds to how the effective low-energy Lorentz symmetry in condensed matter and other systems is broken at high energies. In other words, there are plenty of examples in nature where an approximate Lorentz symmetry is obtained in the low-energy limit of an equation of the form~\eqref{modDR} (and in particular,~\eqref{modDR-f4} below). It therefore seems worth examining whether the same scheme could be applicable to gravity as well.

The standard approach within the EFT framework is the following. One constructs a Lagrangian using the standard model of particle physics in combination with Lorentz violating operators of physical interest. These additional operators are classified by their mass dimension. For example, all renormalisable terms that can be added to the standard model Lagrangian (forming the so-called minimal Standard Model Extension or mSME) have been studied more than a decade ago~\cite{Colladay:1998fq}. They are all of mass dimension 3 or 4 and are further classified by their behaviour under CPT transformations. The full set of dimension 5 operators has also been recently classified~\cite{Bolokhov:2007yc}. Observational constraints on the various $f^{(i)}$ for different particle species can be obtained, for example, from ultra-high-energy cosmic rays, in particular by examining effects such as reaction threshold energies, and dispersion or birefringence effects in long distance propagation. These constraints then allow to calculate the corresponding limits on the various Lorentz violating operators. The lowest-dimensional operators which are not essentially ruled out at the Planck level by current observations, are CPT even operators of mass dimension five and six (see~\cite{Mattingly:2007zz,Maccione:2009ju} and references therein). It can be shown that these correspond to a lowest-order allowed term of order $p^4$ in the above framework.

An intuitive way to understand this is the following. $(i)=1,3$ terms are essentially ruled out because they would lead to strong parity violations which are not observed. An $(i)=2$ term would lead to Lorentz violation at all energies, which is also not observed. The first reasonable Lorentz symmetry breaking term is therefore $(i)=4$. As a matter of fact, note that $(i)=1$ terms would be amplified by the prefactor $E_L$ in the simple scheme of eq.~\eqref{modDR}, which should be ruled out even if it did not lead to parity violation. In a more refined scenario, $(i)=1$ would be suppressed, for example by writing $\mu^2 f^{(1)}c^3\abs{p}/M_{Pl}$, with $\mu$ the particle mass and $M_{Pl}$ the Planck mass. Even so, non-zero $(i)=1$ terms are still so tightly constrained from observations as to be essentially ruled out~\cite{Mattingly:2005re}. A similar remark applies to $(i)=2$ terms. These are not strictly ruled out by observations when written in the form $\mu^2f^{(2)}c^2p^2/M_{Pl}^2$, but still so tightly constrained that, for our present purposes, we can essentially forget about them. There exist several additional complications~\cite{Mattingly:2005re,Mattingly:2007zz}. For example, dimension 3 and 4 Lorentz-violating operators are not strictly ruled out, because of the comment just given. They must be extremely small, but it is not excluded that there exists some theoretical scheme which `naturally' accomodates this smallness. Another complication is that non-renormalisable operators of dimension 5 and higher induce lower-order terms (such as $p^2$ contributions) via radiative corrections. CPT odd operators of dimension 5 are very tightly constrained, but again not completely ruled out, at least if one assumes the existence of some additional suppression mechanism (for example, a residual symmetry such as supersymmetry). Note that CPT even dimension 5 and 6 operators would still require some protection mechanism between the TeV scale and the Planck scale to suppress the lower-order radiative corrections.

For our purposes, we can simply conclude that all modifications of the dispersion relation with $(i)\!<\!4$ are essentially ruled out at the Planck level, or at least require some extraordinary fine-tuning, while even the $(i)=4$ terms can already be quite tightly constrained at the Planck level for the most common Lorentz-violating scenarios~\cite{Galaverni:2007tq,Maccione:2008iw,Maccione:2009ju}. In any case, we can concentrate on $(i)=4$ as the lowest-order allowed modification. Since we will focus on massless particles (photons and phonons), we can write a modified dispersion relation of the type
\begin{equation}\label{modDR-f4}
E^2= c^2p^2 + f^{(4)}c^4p^4/E_L^2~,
\end{equation}
or, in terms of frequencies $\omega$ and wave numbers $k$:
\begin{equation}\label{modDR-omega-k}
\omega^2= c^2k^2 + f^{(4)}c^2k^4/k_L^2~.
\end{equation}

Several remarks might be useful.

First, this type of Lorentz symmetry breaking implies a violation of the equivalence of reference frames. Indeed, as mentioned just above eq.~\eqref{relDR}, a particular modified dispersion relation is valid in a particular observer's reference frame (or rather: a set of concordant frames). In this context, it should be mentioned that in the condensed matter models, it is natural to assume that the preferred reference frame is the laboratory frame, since this is the frame in which the privileged observer (the experimenter) measures things. In the laboratory frame, eq.~\eqref{modDR-omega-k} becomes
\begin{equation}\label{bog}
(\omega-vk)^2= c^2k^2 + f^{(4)}c^2k^4/k_L^2~,
\end{equation}
which immediately leads to the Bogoliubov dispersion relation for a BEC~\eqref{dispersion_xi} by setting $k_L=2/\xi$, with $\xi$ the healing length of the condensate, and where $v$ represents the flow velocity of the backgrounf fluid and $c$ the co-moving speed of sound. 
%
%
In a relativistic context, the question of preferred reference frames is more subtle. For example, in the case of a black hole configuration, assuming that Lorentz invariance is broken, there are at least two natural candidates for preferred reference frames: the rest frame of an observer free-falling into the black hole, and the rest frame of the black hole itself, which in the fluid analogy  correspond essentially to an observer co-moving with the fluid and to the laboratory frame, respectively. We will see in the next chapter that this subtlety can play an important role. Note also that most of the work in the EFT framework uses the rest frame determined by the cosmic microwave background radiation to constrain the possible Lorentz-violating parameters.

A second remark is that the previous equation~\eqref{bog} and the observation that $k_L=2/\xi$ illustrate an important lesson from condensed matter models, namely that naive dimensional estimates in the domain of quantum gravity should be handled with care. Indeed, as we mentioned above, in quantum gravity phenomenology, it is usually assumed that Lorentz violation, if it occurs, should occur approximately at the Planck scale. However, condensed matter models teach us that the Lorentz violation scale $E_L$ and the Planck scale $E_{Pl}$ are not necessarily related. Indeed, in a BEC, the Lorentz violation scale $E_L$ is determined by the healing length $\xi$, which depends ultimately on the $s$-wave scattering length. The analogue of the Planck scale $E_{Pl}$, on the other hand, can be thought of as the scale at which the granularity of the background becomes apparent. In this sense, then, the analogue Planck scale is determined by the interatomic distance $a_i$, which depends essentially on the density: $a_i \sim n^{-1/3}$. It can then be seen intuitively from the weakness of the interatomic interaction in dilute gases that $E_L\ll E_{Pl}$ for all BECs, but the important point is that $E_L$ and $E_{Pl}$ are in principle unrelated. This point might well extend to quantum gravity: different types of quantum gravity phenomenology could be characterised by different, mutually independent energy scales, which are not necessarily accessible to an observer who is limited to the effective low-energy physics. Our previous mention of observational constraints on Planck-scale Lorentz violating operators in the EFT framework just before eq.~\eqref{modDR-f4} could then be rephrased by saying that observations strongly support $E_L\gg E_{Pl}$.

A third remark is that the sign of $f^{(4)}$ in eq.~\eqref{modDR-f4}--\eqref{bog} determines the propagation at high energies: $f^{(4)}<0$ stands for subluminal propagation, i.e., high-energy modes propagate slower than low-energy ones, whereas the BEC model that we are interested in gives $f^{(4)}>0$ and hence superluminal propagation: the high-energy modes propagate faster than the low-energy ones.

A fourth and final remark is the following. Apart from the intrinsic interest of knowing whether Lorentz invariance is a fundamental symmetry of our universe or not, it is of course also to be hoped that phenomenological approaches to quantum gravity will allow us to discriminate between various theoretical frameworks for quantum gravity. Unfortunately, in these theoretical frameworks or `top-down' approaches for quantum gravity, it is currently not really clear whether Lorentz invariance is fundamental or effective,
and in the latter case, how its breaking scale is related to the
Planck scale. For example, while many string theory scenarios
axiomatically incorporate Lorentz invariance, it has been argued that
in certain situations, violations of Lorentz invariance may occur in a
way consistent with world-sheet conformal
invariance~\cite{Mavromatos:2007xe}, thus leading to acceptable string
theory backgrounds. In the context of loop quantum gravity,
in~\cite{Gambini:1998it} it has been argued that quantum effects
should modify the relativistic dispersion relations, although the
issue seems far from settled (see~\cite{Ashtekar:2007px} for some
general remarks). It is arguably only in scenarios of emergent gravity
based on condensed matter
analogies 
that the situation is
clear: Lorentz invariance is a low-energy effective symmetry, and so
it is certainly expected to break at some scale, although not necessarily
related to (and therefore possibly much higher than) the Planck
scale~\cite{Klinkhamer:2005cp}.

To sum up, the possibility of Lorentz symmetry violations at high energy due to quantum gravity effects is an active branch of research, both theoretically and experimentally. There are several frameworks to define such violations. The one that we discussed here is based on effective field theory. There are numerous examples of real physical systems, in particular condensed matter systems such as BECs, where this approach is the natural one to describe the effective low-energy Lorentz symmetry and its high-energy breaking. This allows us to link studies of gravitational analogies in BECs, such as the ones presented so far in this thesis, directly to quantum gravity phenomenology with superluminal dispersion relations. It also means that we can take the complementary approach and study the effect of Lorentz violation such as it occurs in BECs (and such as it might occur in quantum gravity) directly based on modified dispersion relations. We will use this second approach in the following chapter.

\clearpage
\chapter[Hawking radiation in a collapse scenario]{Hawking radiation in a collapse scenario with superluminal dispersion relations}
\label{S:HR}
In this chapter, we analyse the Hawking radiation process due to collapsing
configurations in the presence of superluminal modifications of the
dispersion relation. With such superluminal dispersion relations, the
horizon effectively becomes a frequency-dependent concept. In
particular, at every moment of the collapse, there is a critical frequency
above which no horizon is experienced. We show that, as a consequence,
the late-time radiation originating from the collapse suffers strong modifications, both quantitative
and qualitative, compared to the standard Hawking picture. Concretely,
we show that the radiation spectrum becomes dependent on the critical frequency, on the surface gravities associated with different
frequencies, and on the 
measuring time. Even if the critical frequency
is extremely high, important modifications can still show up.

\section{Introduction}
Hawking radiation is the process whereby a gravitational collapse which leads to the formation of a black hole is predicted to produce late-time thermal radiation due to the quantum creation of particles with an approximately Planckian (black-body) spectrum~\cite{Hawking:1974rv,Hawking:1974sw}. The calculation consists essentially in writing down Bogoliubov transformations to estimate the particle creation between inequivalent vacua, as we will detail further on. In 1976, Unruh developed a technique for replacing the collapse by a stationary spacetime representing an eternal black hole, by imposing adequate boundary conditions on the past horizon (the so-called `Unruh state'). This leads to exactly the same result as in the collapse scenario while alleviating some of its technical difficulties~\cite{Unruh:1976db}. 

Several alternative ways have since been developed to derive Hawking radiation (see~\cite{Carlip:2008wv} for a recent selective overview), for example through the stress-energy tensor trace anomaly~\cite{Christensen:1977jc} or via a tunneling method~\cite{Volovik:1999fc,Parikh:1999mf}. Most of these derivations are based on stationary black holes, and it is not always clear how they are physically related to Hawking's original collapse scenario. Nevertheless, it is remarkable that they all seem to lead to essentially the same result. However, one may wonder whether there is no hidden assumption in all these derivations, for example the implicit requirement of Lorentz invariance up to arbitrary frequencies. Such an implicit assumption about the physics at arbitrarily high frequencies was certainly the case in Hawking's original derivation, as was pointed out almost immediately after its publication, for example in~\cite{Gibbons:1975pq}, and which has become known as the ``transplanckian problem'' with respect to Hawking radiation.\footnote{A related transplanckian problem exists in cosmology.} It was also the main motivation for Unruh's 1981 paper~\cite{Unruh:1980cg}, in which he pointed out that some of the high-frequency issues related to black holes can be studied by analogy with the propagation of sound waves in a fluid flow. In spite of the transplanckian problem, Hawking radiation was and still is considered a crucial landmark in quantum field theory in curved spacetimes, and its implications for the connection (predicted earlier by Bekenstein~\cite{Bekenstein:1973ur}) between black hole physics and thermodynamics are an essential touchstone for the development of quantum gravity.\footnote{It is perhaps useful to point out that Hawking radiation {\it does not} depend on the dynamical aspects of gravitation, i.e., it is a purely kinematic effect, or in other words: a prediction of quantum field theory in curved spacetimes in the presence of black hole horizons. The thermodynamical aspects of black hole physics, on the other hand, {\it do} crucially depend on the dynamical aspects of gravity, i.e., ultimately, on the Einstein equations. This is essentially because the first thermodynamic law of black holes, which establishes the relation between changes in its mass $M$ and changes in the surface $A$ of its horizon: \mbox{$\d M=(\kappa/8\pi)\,\d A$} (for a Schwarzschild black hole with surface gravity $\kappa$), depends on the relation between the mass and the geometry of the black hole, and thus on the Einstein equations. A similar problem applies to the zeroth law, which establishes the constancy of the surface gravity over the horizon. Thus analogue black holes are expected to allow simulation of Hawking radiation, but not of their thermodynamic aspects~\cite{Visser:1997ux,Visser:2001kq}.}

Here, we will stay quite close to Hawking's original derivation, and modify it carefully in a way that allows us to calculate the late-time black hole radiation for a collapsing scenario, but now in the case of superluminal dispersion relations at high frequencies. 

The motivation for such an analysis is threefold. First, as we indicated in the foregoing intermezzo~\ref{S:intermezzo}, from the point of view of quantum gravity phenomenology, increasing
attention has focused on the consideration that maybe Lorentz
invariance is not a fundamental law, but an effective low-energy
symmetry which is broken at high energies. From extrapolation of current experiments, we know that
there exist stringent bounds on the most commonly expected types of
Lorentz violations at the Planck scale. Nevertheless,
even violations at much higher energy scales might still significantly
affect black hole physics. On the other hand, it could also be that Hawking radiation is recovered 
precisely because of this relation between the (hypothetical) Lorentz violation scale and the Planck scale. In order to better understand this issue, a simple way of modelling a wide range of Lorentz violating effects
(and a quite natural one from the point of view of condensed matter analogies)
consists in modifying the dispersion relations at high
energy. This modification can be subluminal or
superluminal, depending on whether high-frequency modes move slower or
faster than low-energy ones.

This brings us to the (closely related) second motivation, namely the intrinsic question of the robustness of Hawking's prediction, regardless of whether the high-frequency theory describing our universe will indeed turn out to incorporate superluminal corrections. Even if our universe turns out to be perfectly Lorentz invariant, this question is still interesting as a matter of principle, namely with regard to how inevitable or fundamental Hawking radiation really is, rather than an `accidental' consequence of Lorentz invariance. The question of the robustness of Hawking's prediction with respect to modifications of the transplanckian physics has
been tackled principally within the same framework that we are using, namely by analysing effective field theories such that at
high energies a modification of the dispersion relation is 
incorporated (see however \cite{Agullo:2006um} for a different take on 
the problem).
Historically, however, attention has mainly been given to subluminal dispersion
relations. Important contributions such as
\cite{Unruh:1980cg,Jacobson:1991gr,Jacobson:1993hn,Unruh:1994je,Brout:1995wp,
Corley:1996ar,Corley:1996nw,Corley:1997pr}, and more recently
\cite{Unruh:2004zk,Schutzhold:2008tx} seem to have settled the robustness of
Hawking radiation with respect to subluminal modifications
of the frequency spectrum, at least in the case of stationary black-hole scenarios, although this conclusion still rests on certain
assumptions, usually related to the behaviour of the fields near the horizon. In any case, it is important to remember that this only solves part of the transplanckian problem with regard to Hawking radiation. Indeed, subluminal modifications gradually dampen the influence of ultra-high frequencies, and so they do not explore arbitrarily large frequencies. So even assuming that it has been demonstrated that Hawking's result can be recovered in the stationary case in the presence of subluminal modifications, the question remains whether it is also robust with respect to a (non-damped) modification of the transplanckian physics in a collapse scenario. Superluminal modifications differ conceptually from subluminal ones, in that they gradually magnify the influence of ultra-high energies, and thereby offer an interesting test-case for the transplanckian robustness of Hawking radiation. 

A third motivation for our study, as already mentioned several times throughout this thesis, comes from a possible connection with experiment. 
%
BECs are expected to be good candidates for a possible experimental
verification
of Hawking radiation~\cite{Garay:2000jj,Garay:1999sk,Barcelo:2001ca,Barcelo:2000tg,Schutzhold:2006,Wuster:2007nf,Balbinot:2007de,Carusotto:2008ep,Wuester:2008kh,Cornell:2009}. Furthermore, we have shown that dynamically stable black hole configurations are possible in a BEC, thereby removing a major potential complication at the moment of carrying out such an experiment. But since the physics of BECs automatically leads to a superluminal dispersion relation at high energies, the
question is again which kind of modifications are to be expected in the laboratory realisation of a BEC black hole with respect to the standard Hawking picture.

Of the above-mentioned works on the robustness of Hawking radiation, a
few have also tried to address superluminal modifications
\cite{Corley:1997pr,Unruh:2004zk,Schutzhold:2008tx}. Various problems make this
case quite different from the
subluminal one. These problems can be related to the fact that the
horizon becomes frequency-dependent when modifying the dispersion
relations. This is also true in the subluminal case, though
qualitatively in a very different way. With superluminal
modifications, the horizon lies ever closer to the singularity for
increasing frequencies, and asymptotically coincides with it. This causes the
`apparent' interior of the black hole (the interior of the zero-frequency horizon) to be exposed to the outside world. Since it
seems unreasonable
to impose a condition arbitrarily close to the singularity as long as
we do not have a solidly confirmed quantum theory of gravity, most of
the approaches used for subluminal dispersion are invalid, or at least
questionable, for superluminal dispersion. Moreover, since it seems
reasonable to expect that quantum effects will remove the general
relativistic singularity, an overall critical frequency might appear above which
no horizon would be experienced at all.

In the analysis that follows we will try to avoid making any
further assumptions about the physics near the singularity. For instance
we will analyse the characteristics of the radiation at retarded times at which
the singularity has not yet formed, and consider both final options, either with or without a singularity. Our approach will be based on a derivation of the radiation very much in the spirit of Hawking's original calculation through the
relation between the asymptotic past ray trajectories and the
asymptotic future ones in the case of a collapsing
configuration. The only assumptions about these asymptotic extremities
are the standard ones, namely a Minkowski geometry in the asymptotic
past, and flatness at spatial infinity also in the asymptotic
future. The language we will use is related to the fluid analogy for
black holes, which provides the intuitive picture of the spacetime
vacuum flowing into the black hole and getting shredded in its centre.

Our main results can be summarised as follows. Three crucial elements
distinguish the late-time radiation with superluminal dispersion
relations from standard relativistic ones. First, at any instant there
will be a critical frequency above which no horizon has yet been
experienced. This critical frequency will induce a finite limit in the
modes contributing to the radiation, which will therefore have a lower
intensity than in the standard case, even if the critical frequency is
well above the Planck scale. Second, due to the effective
frequency-dependence of the horizon, the surface gravity will also
become frequency-dependent and the radiation will depend on the
physics inside the black hole. Unless special conditions are imposed
on the profile to ensure that the surface gravity is nonetheless the
same for all frequencies below the critical one, the radiation
spectrum can also undergo a strong qualitative modification. Depending
on the relation between the critical frequency and the Lorentz violation
scale, the radiation from high frequencies is no longer negligible
compared to the low-frequency thermal part, but can even become
dominant. This effect becomes more important with increasing critical
frequency. Finally, a third effect is that the radiation originating from the collapse process will
extinguish as time advances.

The remainder of this chapter has the following structure. In section
\ref{SS:model}, we will describe and motivate the classical geometry of our model and the concrete kinds of
profiles that we are interested in. Note that these configurations are related, but not identical to the ones discussed in the chapters~\ref{S:BHconfigurations}--~\ref{S:QNMs}.
In section \ref{SS:SDR}, we will briefly review how
Hawking's standard result can be obtained in the formalism of our choice for the
case of standard relativistic dispersion relations. This calculation will be
adapted in section \ref{SS:MDR} to superluminal dispersion relations, and we will
analytically obtain a formula for the late-time radiation for this case. Then,
in section \ref{SS:graphics}, we will present graphics obtained from this
analytic formula by numerical integration. These graphics will illustrate the results mentioned in the previous paragraph, which we will discuss in more depth and compare with other results in the recent literature in section \ref{SS:conclusion}.

\section{Classical geometry}
\label{SS:model}

We will study radiation effects in simple (1+1)-dimensional collapsing configurations.
It is well known that in the Hawking process, most of the radiation is produced
in the $s$-wave sector. Hence, a spherically symmetric treatment, effectively
(1+1)-dimensional, suffices to capture the most relevant aspects of the process.
We will work in Painlev\'e-Gullstrand coordinates, where
\begin{eqnarray}%
\d s^2=-[c^2-v^2(t,x)]\d t^2 - 2v(t,x)\d t\, \d x +\d x^2~,
\label{acoustic-metric}
\end{eqnarray}%
and make the further simplification of taking a constant $c$ (we will
always write $c$ explicitly to differentiate it from the
frequency/wave-number dependent $c_k$ that will show up in the
presence of a superluminal dispersion relation). In this manner all
the information about the configuration is encoded in the single
function $v(t,x)$. The Painlev\'e-Gullstrand coordinates have the
advantage, compared to the Schwarzschild form, of being regular at the
horizon. Moreover, they suggest a natural interpretation in terms of
the language of acoustic models. In such a model, as before and throughout this thesis, $c$ represents the
speed of sound and $v$ is the velocity of the fluid (which corresponds
to the velocity of an observer free-falling into the black hole).
Also as before, we will consider the fluid to be left-moving, $v
\leq 0$, so that the outgoing particles of light/sound move towards
the right.

The most relevant aspects of the analysis presented hereafter
depend only on the qualitative features of the fluid profile. However, to 
justify some specific calculations later on it is helpful to use the following 
concrete profiles, see Figs.~\ref{Fig:v-constant-kappa} and
\ref{Fig:v-increasing-kappa}. Consider a velocity profile $\bar{v}(x)$
such that \mbox{$\bar{v}(x\to +\infty)=0$}, \mbox{$\bar{v}(x=0)=-c$} and further
decreasing monotonically as \mbox{$x\to -\infty$}. In the fluid image, the
fluid nearly stands still at large distances and accelerates inwards,
with a sonic point or horizon at $x=0$. The further decrease in
$\bar{v}$ can either be linear until a constant limiting value is almost
achieved, as in
Fig.~\ref{Fig:v-constant-kappa}, or of the form 
\begin{eqnarray}%
\bar v(x)=-c\sqrt\frac{2M/c^2}{x +2M/c^2}~,
\end{eqnarray}%
again up to a constant limiting value, as in
Fig.~\ref{Fig:v-increasing-kappa}. Let us recall that this second
velocity profile corresponds to the Schwarzschild line element with $M$ 
the central mass, as can be seen by reparameterising the time coordinate and using \mbox{$r=x+2M/c^2$} as radial coordinate~\cite{Visser:1997ux}.

\begin{figure}
\begin{center}
\includegraphics[width=0.6\columnwidth]{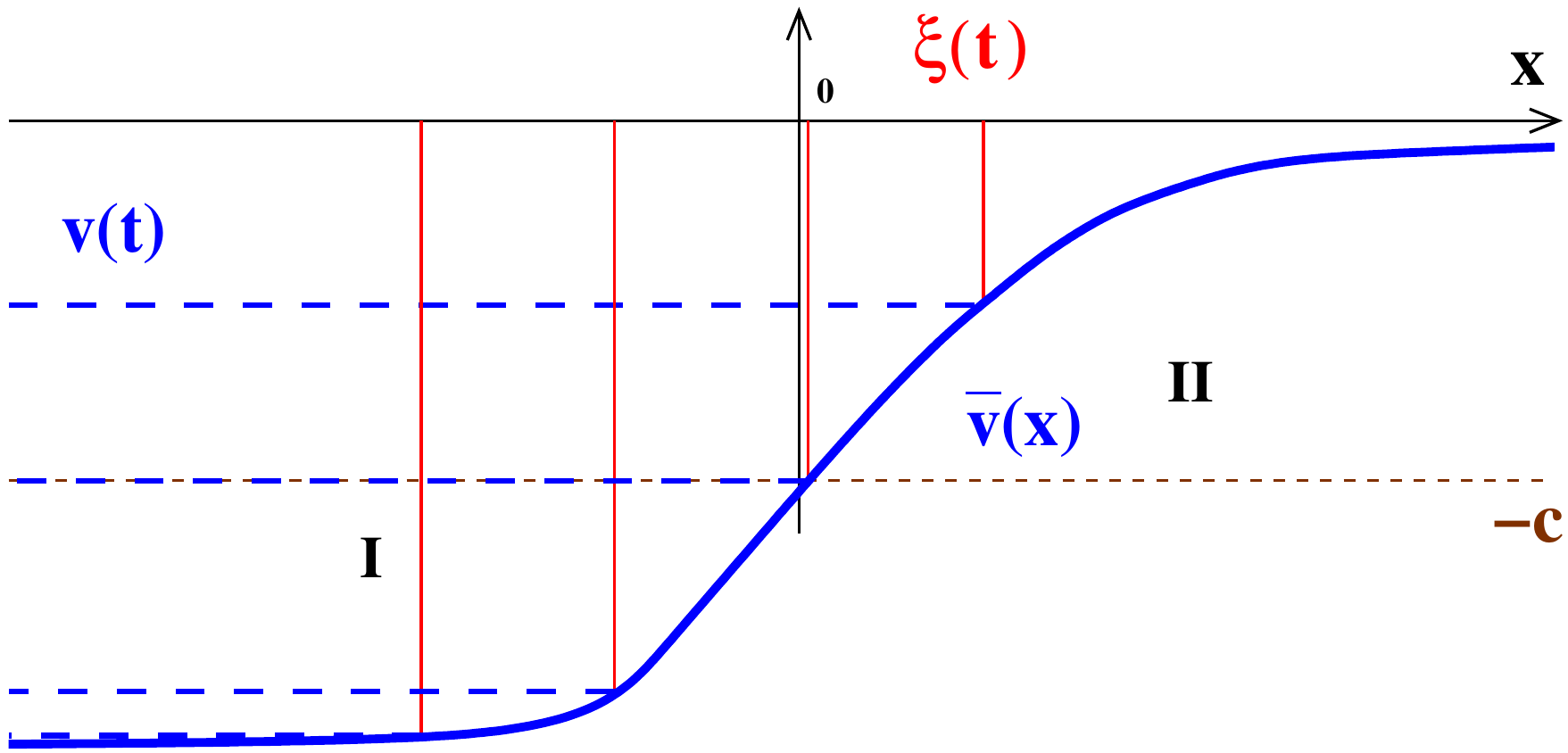}
\end{center}
\figcaptionl{Fig:v-constant-kappa}{Velocity profile of a black hole with a linear slope of the velocity $\bar{v}(x)$, and hence a constant surface gravity $\kappa_{\omega'}$, from the classical (zero-frequency) horizon at $x=0$ down to some predefined limiting value. The auxiliary function $\xi(t)$ separates each instantaneous velocity profile $v(t,x)$ into a dynamical \mbox{region I}: $v(t,x)=v(t)=\bar{v}(\xi(t))$ and a stationary region II: $v(t,x)=\bar{v}(x)$.}
\end{figure}

\begin{figure}
\begin{center}
\includegraphics[width=0.6\columnwidth]{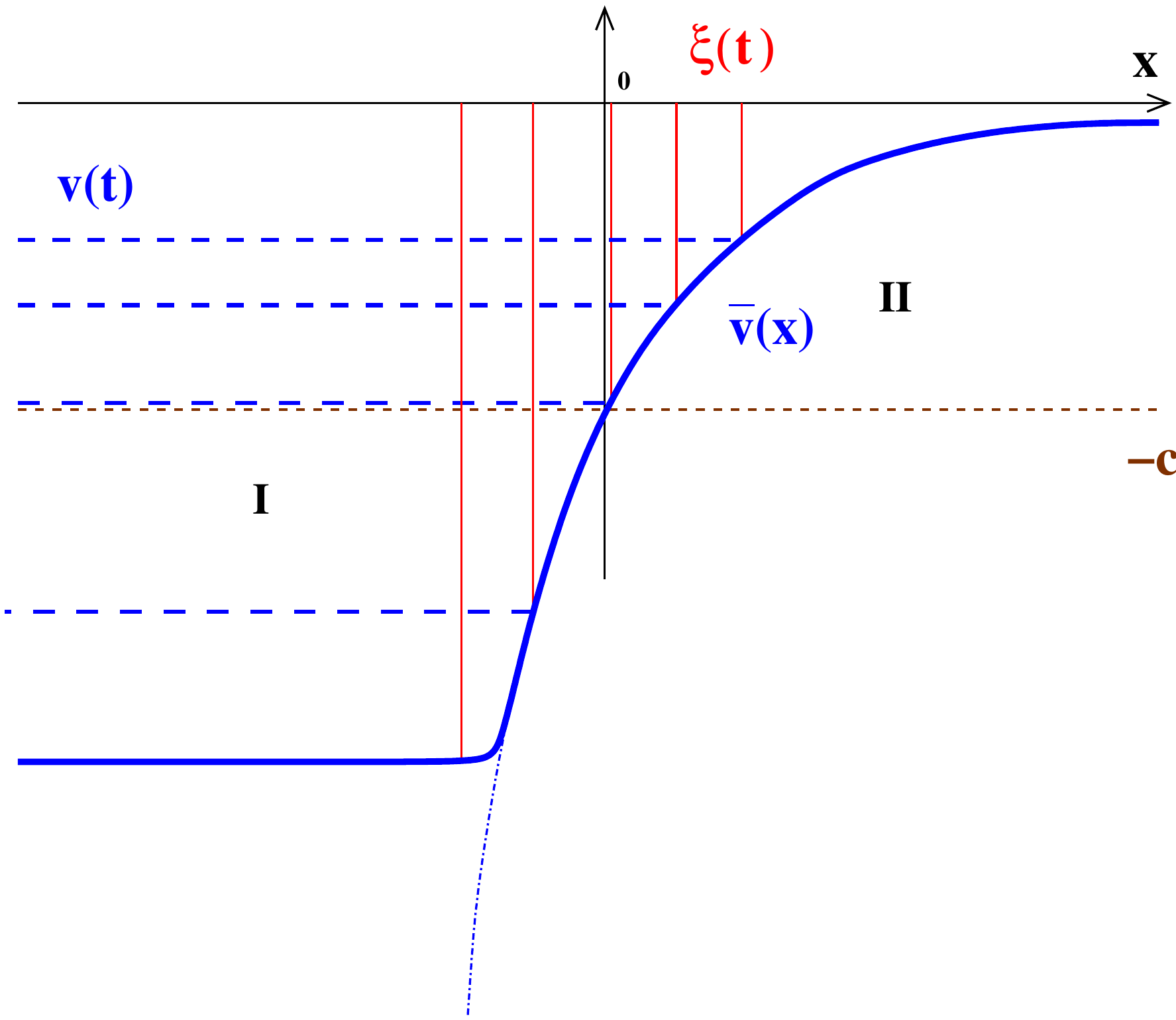}
\end{center}
\figcaptionl{Fig:v-increasing-kappa}{Velocity profile characteristic
of a Schwarzschild black hole. The slope of $\bar{v}(x)$ increases from the classical horizon at $x=0$ leftwards up to a constant limiting value, leading to a surface gravity $\kappa_{\omega'}$ which increases with the frequency.}
\end{figure}

Up to here, we have described a stationary profile. Now, to incorporate
the dynamics of the collapse, let us introduce a
monotonically decreasing function $\xi(t)$, \mbox{$\xi(t \to -\infty) \to
+\infty$}, and define $v(t,x)$ as
\begin{eqnarray}\label{profile}
v(t,x)~ = 
\left \{ 
\begin{array}{lll}
\bar{v}(\xi(t)) &~\text{if}~ &x \leq \xi(t)~,\\
\bar{v}(x) &~\text{if}~ &x \geq \xi(t)~.
        \end{array}
\right.
\end{eqnarray}
Imagining the collapse of a homogeneous star, the function $\xi(t)$
represents the distance from the star surface to its gravitational (Schwarzschild) radius. The dynamical configuration that
we obtain
consists of a series of snapshots, see Fig.~\ref{Fig:snapshots}. In each snapshot, $|v(t,x)|$
increases (i.e., the fluid accelerates) from $\bar{v}=0$ for $x\to
+\infty$, up to a point $x_0=\xi(t)$, and then remains constant as $x$
further decreases towards $-\infty$. For consecutive snapshots, the
point $x_0=\xi(t)$ moves leftwards, so $v(t,x)$ covers an ever
larger part of $\bar{v}(x)$.

\begin{figure}
\begin{center}
\includegraphics[width=0.9\columnwidth]{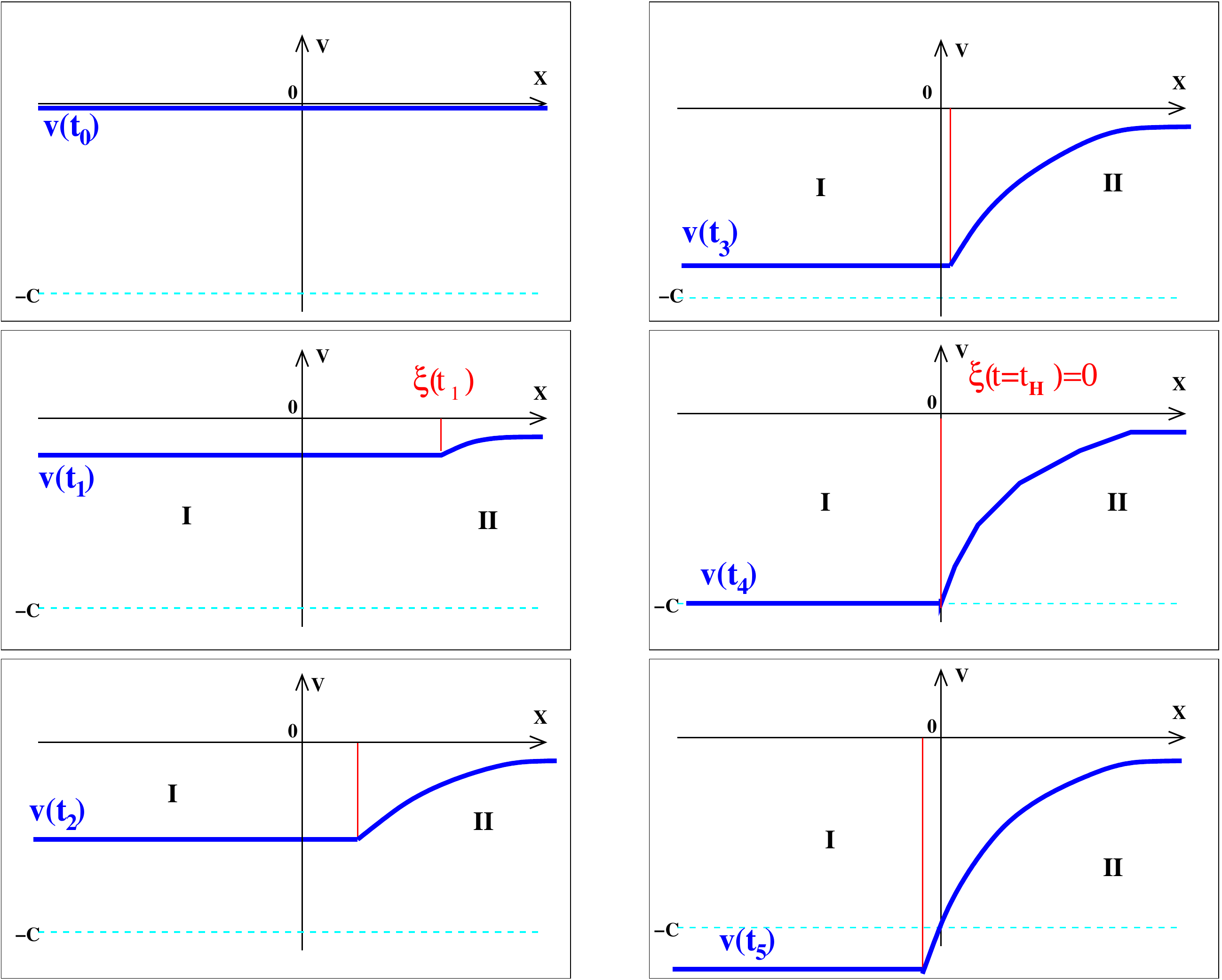}
\end{center}
\figcaptionl{Fig:snapshots}{[From top to bottom and left to right] Various snapshots of the dynamical evolution of a collapsing configuration with variable surface gravity, as in Fig.~\ref{Fig:v-increasing-kappa}. The classical horizon is formed at $t_4$.}
\end{figure}

From the point of view of an outgoing (right-moving) particle of
light/sound, the configuration is nicely split up into, first, a
dynamical or $t$-dependent region, and then a stationary $x$-dependent
region, as defined in eq.~\eqref{profile}.\footnote{In a physically
realistic model, the apparent kink in the profile where the transition
between both regions takes place will of course be smoothed out.}

The difference between the two types of profiles,
Figs.~\ref{Fig:v-constant-kappa} and \ref{Fig:v-increasing-kappa}, can
best be explained in terms of the surface gravity. When modifying the
dispersion relation, the horizon becomes a frequency-dependent
concept: each frequency experiences a different horizon, 
as we will see explicitly in section~\ref{SS:MDR}. In
particular, for superluminal modifications, the horizon forms later
(i.e., at higher values of $|\bar{v}|$, or more negative values of
$x$) for increasing frequencies. The surface gravity will then also
become frequency-dependent:
\begin{eqnarray}\label{def-kappa}
\kappa_{\omega'}\equiv c \bigg|{\d \bar v \over \d x} \bigg|_{x=x_{H,\omega'}}~,
\end{eqnarray}
where $x_{H,\omega'}$ is the frequency-dependent location of the
horizon.  This surface gravity will be seen in section~\ref{SS:MDR} and
\ref{SS:graphics} to play a crucial role. This explains our choice of 
the two types of profiles: In the first profile,
Fig.~\ref{Fig:v-constant-kappa}, we have considered a linear velocity
profile $|\bar{v}(x)|$ from the classical or zero-frequency horizon
($c=|v|$) down to some given limiting value, $|v|_{\rm max}$, from
where it stays constant. This maximum velocity defines a critical
frequency $\omega'_c$ as we will discuss in more detail in
section~\ref{SS:MDR}. Then the surface gravity $\kappa_{\omega'}$ will
be frequency-independent up to this critical frequency $\omega'_c$ after
which it will rapidly vanish. In the second profile, see
Fig.~\ref{Fig:v-increasing-kappa}, we have taken a velocity profile
typical for a Schwarzschild black hole, and therefore
$\kappa_{\omega'}$ increases with $\omega'$, again up to a predefined
limiting value corresponding to the horizon for a critical frequency
$\omega'_c$. 

In the context of the fluid analogy, it seems obvious
that some mechanism will avoid the formation of a singularity, while in gravity it is also usually expected that quantum effects will ultimately resolve the general relativistic singularity at the centre of a black hole. But in
any case, we will also take into account the limit for an infinite
critical frequency, which corresponds to the eventual formation of a singularity, and is characterised by a velocity profile with 
$|v|_{\rm max} \to \infty$.

\section{Standard Dispersion Relations}
\label{SS:SDR}

In this section we will briefly review a way of deriving Hawking's
formula for the radiation of a black hole with standard (relativistic)
dispersion relations of the form \mbox{$\omega^2=c^2k^2$}. 
For the sake of simplicity, we will only consider a massless scalar field.  
We will summarise the main steps: we start with a sketch of the general idea and then go from the Klein-Gordon inner product and the
definition of the Bogoliubov $\beta$ coefficients, over the relation
between past and future null coordinates, to the black-body radiation
in the wave-packet formulation. Our aim is to present the key points of the procedure in
such a way that they can easily be adapted to superluminally
modified dispersion relations---the subject of the next sections.

\subsection{General idea}
Hawking radiation can be understood as an example of mode mixing between positive and negative-energy modes in a dynamical curved spacetime~\cite{Hawking:1974sw,Birrell:1982ix,Fabbri:2005mw}.

In flat spacetime, due to the high amount of symmetry present, all inertial observers agree on the definition of the vacuum and of what constitutes a particle. But this is no longer the case in curved spacetime. In particular, the definition of positive-energy modes $\psi^{(+)}$ with respect to some choice of time coordinate $t$ is that they should be eigenfunctions of the operator $\partial_t$ such that 
\begin{eqnarray}
\partial_t \psi^{(+)}=-i\omega\psi^{(+)}~~ \text{for}~~\omega>0~.  
\end{eqnarray}
This definition assumes the existence of a timelike Killing vector field associated with $t$. In Minkowski spacetime, the Poincar\'{e} group guarantees that all inertial observers agree on the definition of positive-energy modes. However, in curved spacetime the Poincar\'{e} group is no longer a symmetry of the spacetime, and so the previous conclusion does not hold in general. We will therefore compare the vacua defined by two different observers.

Assume two complete orthonormal sets of modes $\{\psi_i'\}$ and $\{\psi_j\}$, and express the scalar field $\phi(x)$ in terms of each of them:
\begin{eqnarray}
\phi(x)&=\sum_i[a_i\psi_i'(x)+a^\dag_i\psi_i'^*(x)]=\sum_j[b_j\psi_j(x)+b^\dag_j\psi_j^*(x)]~.
\end{eqnarray}
Each decomposition defines a corresponding vacuum, $|0_I\rangle$ and $|0_{II}\rangle$ respectively, through
\begin{eqnarray}
&a_i|0_I\rangle&=0~~~~~~~~(\forall i)\\
&b_j|0_{II}\rangle&=0~~~~~~~~(\forall j)
\end{eqnarray}
Since both sets are complete and orthonormal, one can write:
\begin{eqnarray}\label{alpha-beta}
\psi_j=\sum_i(\alpha_{ji}\psi_i'+\beta_{ji}\psi_i'^*)~;~~~~~~b_j=\sum_i(\alpha^*_{ji}a_i+\beta^*_{ji}a_i^\dag)~,
\end{eqnarray}
where $\alpha_{ji}=(\psi_j,\psi_i')$ and $\beta_{ji}=-(\psi_j,\psi_i'^*)$ are the so-called Bogoliubov coefficients. 

If $~\beta_{ji}\neq 0~$ then there will be mode mixing: a mode expressed in terms of $\psi_j$ will necessarily contain contributions from modes in terms of $\psi_i'^*$. This implies
\begin{eqnarray}
b_j|0_{I}\rangle=\sum_i\beta^*_{ji}a_i^\dag|0_{I}\rangle \neq 0 ~.
\end{eqnarray}
The vacuum $|0_I\rangle$  associated with $\{\psi_i'\}$ is then not a vacuum state for an observer associated with $\{\psi_j\}$. One can calculate the amount of particles that such an observer will see in the `vacuum' state $|0_I\rangle$ as follows. From the number operator $N_j=b_j^\dag b_j$, and assuming that all one-particle states are properly normalised, one obtains $\langle 0_I|N_j|0_I\rangle = \sum_i|\beta_{ji}|^2 \neq 0$. The vacuum $|0_I\rangle$  associated with $\{\psi_i'\}$ therefore contains a number $\sum_i|\beta_{ji}|^2$ of $j$-particles in $\{\psi_j\}$.

When indexing the modes with a continuous index such as the frequency, as we will do in the remainder of this chapter, instead of the discrete indices $i$ and $j$, this sum becomes an integral. But similar technical complications left aside, the brief discussion given here captures the essence of particle creation in curved spacetimes. 

The particular case of Hawking radiation is obtained by associating $\{\psi_i'\}$ to an asymptotic past observer, before the onset of the collapse, and $\{\psi_j\}$ to an asymptotic future observer. For both observers, $\partial_t$ is (asymptotically) a timelike Killing vector. The observer in the past can then define the $\psi_i'$ as the positive-energy modes and $\psi_i'^*$ as the negative-energy ones with respect to $t$, and likewise for the future observer. Even though, for both observers, the spacetime is asymptotically stationary, the dynamical evolution of the spacetime inbetween means that they will not agree on the definition of the vacuum. In particular, there will be particle creation as observed by the second (asymptotic future) observer originating in the quantum state which the first (asymptotic past) observer would call the vacuum state. 

Note that the definition of the Bogoliubov coefficients assumes that there is a well-defined inner product, which we will discuss now.

\subsection{Inner product}
The d'Alembertian or wave equation for a massless scalar field in 1+1
dimensions for the metric~\eqref{acoustic-metric} with constant $c$
can be written as
\begin{eqnarray}%
&&(\partial_t+\partial_x v)(\partial_t+v\partial_x)\phi = c^2 \partial_x^2
\phi~.
\end{eqnarray}
This conformally invariant theory is equivalent to the
dimensionally-reduced 3+1 spherically-symmetric theory if one neglects
the backscattering due to the angular-momentum potential barrier
(responsible for the so-called grey-body factors).

In the space of solutions of this wave equation, we can define the Klein-Gordon
pseudo-scalar product
\begin{eqnarray}%
(\varphi_1,\varphi_2) \equiv 
-i \int_\Sigma \d\Sigma^\mu ~ 
\varphi_1 \stackrel{\leftrightarrow}{\partial_\mu}\varphi_2^*~,
\end{eqnarray}%
which is independent of the choice of the spatial slice $\Sigma$. For a
\mbox{$t=\text{constant}$} slice, and in particular for \mbox{$t\to +\infty$,} this
becomes
\begin{equation}\label{varphi1-varphi2}%
(\varphi_1,\varphi_2) =
-i \int \d x ~ 
\left[\varphi_1 (\partial_t+v\partial_x )\varphi_2^* 
-\varphi_2^* (\partial_t+v\partial_x )\varphi_1\right]~.
\end{equation}%
We can define future null coordinates $u(t,x)$ and $w(t,x)$, such that, when \mbox{$t,x\to +\infty$},
\begin{eqnarray}%
u(t,x) \to t - x/c~,&&~~
w(t,x) \to t + x/c~.
\end{eqnarray}%
%
%
Writing the inner product in terms of these null coordinates, gives, for the
limit $t \to +\infty$,
\begin{eqnarray}\label{scalar_product}%
(\varphi_1,\varphi_2) = &&\hspace{-4mm}
-{i c \over 2}\left\{
\int_{-\infty}^{+\infty} \d u  ~ 
\left[\varphi_1 \partial_u\varphi_2^* 
-\varphi_2^* \partial_u\varphi_1\right]_{w=+\infty}
\right.
\nonumber
\\
&&\hspace{-10mm}
\left.
+\int_{-\infty}^{+\infty} \d w ~ 
\left[\varphi_1 \partial_w\varphi_2^* 
-\varphi_2^* \partial_w\varphi_1\right]_{u=+\infty}
\right\}~.
\label{KG-product}
\end{eqnarray}%
Note that, in the derivation of this formula, we have only
made use of coordinate transformations, without relying on their
null character or on geometrical tools such as conformal diagrams or
the deformation of Cauchy surfaces. This is important because in
the case of modified dispersion relations, such geometrical concepts
become problematic and actually can only be maintained in the context
of a rainbow geometry---if at all.

Similarly we could have used past null coordinates $(U,W)$, which obey, when $t\to -\infty$,
\begin{eqnarray}%
U(t,x) \to t - x/c~,&&~~
W(t,x) \to t + x/c~,
\end{eqnarray}%
%
%
to calculate the inner product in the asymptotic past.

\subsection{Bogoliubov $\beta$ coefficient}

The right-moving positive-energy solutions associated with the 
asymptotic past and with the asymptotic future, normalised in the Dirac-delta sense, can be expressed as 
\begin{eqnarray}\label{modes-past+future}%
\psi'_{\omega'}={1 \over \sqrt{2\pi c ~ \omega'}} e^{-i\omega' U}~,&&
\psi_{\omega}={1 \over \sqrt{2\pi c ~ \omega}} e^{-i\omega u}~,
\end{eqnarray}
respectively, where we use primes to indicate asymptotic past values.

%
%
The mode mixing relevant for the Hawking process occurs in the 
right-moving sector. Therefore, we only need to calculate 
the first term in the scalar product~\eqref{KG-product}.
Then, plugging eq.~\eqref{modes-past+future} into the definition of $\beta$ given just below eq.~\eqref{alpha-beta} and 
integrating by parts we obtain the simple expression
\begin{eqnarray}\label{beta}%
\beta_{\omega \omega'} =
{1 \over 2 \pi} \sqrt{\omega \over \omega'}
\int \d u  ~  e^{-i \omega' U(u)} e^{-i \omega u}~,
\end{eqnarray}%
so that all the information about the produced radiation is contained in the 
relation $U=U(u) \equiv U(u,w\to +\infty)$.

\subsection{Relation $U(u)$}
\label{SS:relation_U(u)}

For a standard relativistic dispersion relation, it is well known (see, e.g.,~\cite{Birrell:1982ix}) that the relation between $U$ and $u$ for a configuration that forms a horizon 
(in our case at $x=0$) can be expressed at late times as \mbox{$U=U_H-Ae^{-\kappa u/c}~$} (note that we use a subscript $H$ for all quantities associated with the horizon), where $U_H$, $A$ and the surface gravity 
\begin{eqnarray}\label{standard-kappa}
\kappa \equiv c\bigg| \frac{\d \bar{v}}{\d x}\bigg|_{x=0}
\end{eqnarray}
are constants.

We can define a threshold time $u_I$ at which an asymptotic
observer will start to detect thermal radiation from the black hole.
This retarded time corresponds to the moment at which the function
$U(u)$ enters the exponential regime.
We can then rewrite the previous expression, valid for $u>u_I$, as
\begin{eqnarray}\label{exp-shift}%
U=U_H-A_0 e^{-\kappa (u-u_I)/c}~.
\end{eqnarray}
Plugging this relation into eq.~\eqref{beta} and integrating in $u$ gives
\begin{eqnarray}\label{U-u_step1}%
\beta_{\omega\omega'} &=
&{1 \over 2\pi }\sqrt{\omega \over \omega'} \;
{c \over \kappa}\exp[- i\omega' U_H] 
\exp[-i\frac{c\omega}{\kappa}\ln(\omega'A_0)]\nonumber\\
&&~~\times \exp(-i\omega u_I)\exp(-{\pi c\omega \over 2 \kappa})
\Gamma(ic\omega/\kappa )~.
\end{eqnarray}%

\subsection{Wave packet formulation}
\label{SS:wave-standard}

In order to obtain physically sensible results, it is a good
precautionary measure to replace the monochromatic rays used so far by wave packets (see e.g.~\cite{Hawking:1974sw} or the
discussion in~\cite{Fabbri:2005mw}). Positive-energy wave packets can be
defined as
\begin{eqnarray}
P_{\omega_j,u_l}(\omega) \equiv \bigg\{ 
\begin{tabular}{ll}
${e^{i\omega u_l} \over \sqrt{\Delta \omega}}$ &~~~
$- {1 \over 2}\Delta \omega  <\omega -\omega_j< {1 \over 2}\Delta
\omega $\\
~~~$0$ &~~~~{\rm otherwise}~,
\end{tabular}
\end{eqnarray}
where $u_l \equiv u_0+2\pi l/\Delta \omega$ with $u_0$ an overall reference and
$l$ an integer phase parameter. The central
frequencies of the wave 
packets are \mbox{$\omega_j \equiv j\Delta \omega$}, with $\Delta \omega$ their width.

Then, from the expression \mbox{$\beta_{\omega_{j},u_{l};\omega'} \equiv
\int \d\omega \;
\beta_{\omega\omega'}
P_{\omega_j,u_l}(\omega)~,$} 
and assuming that the wave packets are sufficiently narrow
($\Delta\omega\ll\omega_j$), we obtain
\begin{eqnarray}\label{U-u_step2}
|\beta_{\omega_{j},u_{l};\omega'}|^2 
\approx \frac{c\Delta \omega}{2\pi\omega'\kappa}\frac{\sin^2(z-z_l)}{(z-z_l)^2} 
{1 \over \exp({2\pi c\omega_j \over \kappa})-1}~,
\end{eqnarray}
where we have defined
\begin{eqnarray}
z=\frac{c\Delta\omega}{2\kappa}\ln\omega' A_0~;&~~~~~~~~
z_l=\frac{\Delta\omega}{2}(u_l-u_I)~.
\end{eqnarray}
Finally, integration in $\omega'$ gives the number of particles with frequency
$\omega_j$ detected at time $u_l$ by an asymptotic observer:
\begin{eqnarray}\label{N-standard}%
N_{\omega_j,u_l}& = &\int_0^{+\infty}\d\omega'~
|\beta_{\omega_j,u_l;\omega'}|^2~
\\
& \approx &{1 \over \exp(2\pi c\omega_j/\kappa) -1}~,
\end{eqnarray}%
which reproduces Hawking's formula (in the absence of backscattering) and 
corresponds to a Planckian or black-body spectrum with temperature $T_H=\kappa/(2\pi c)$.

Note that the essence of the transplanckian problem is contained in this last step, since the integration must indeed be carried out up to infinite frequency to recover the standard Hawking result, an issue which we will further discuss in section~\ref{SS:wave-mod}.

\section{Superluminally modified dispersion relations}
\label{SS:MDR}

In this section we will indicate how the late-time radiation
originating from the formation of a black hole through a collapse process can be
calculated in the case of superluminal dispersion relations.

We have discussed earlier in this thesis how superluminal dispersion relations can arise as a phenomenological description in quantum gravity scenarios with Lorentz violations at high energies, see chapter~\ref{S:intermezzo}, and how they are also naturally found in BECs (chapter~\ref{S:preliminaries}). From a theoretical point of view, one can also  introduce superluminally modified dispersion relations by adding a
quartic term to the wave equation:
\begin{eqnarray}\label{modified_laplace}%
(\partial_t+\partial_x v)(\partial_t+v\partial_x)\phi = 
c^2 \left(\partial_x^2 +{1 \over k_L^2}\partial_x^4 \right)\phi~,
\end{eqnarray}
where $k_L$ (the Lorentz symmetry breaking scale) is the scale at which non-relativistic deviations in the associated dispersion relation
\begin{eqnarray}%
(\omega - vk)^2= c^2k^2\left(1+\frac{k^2}{k_L^2}\right)~
\label{dispersion}
\end{eqnarray}%
become significant. The reasons for using this relation in particular were discussed in chapter ~\ref{S:intermezzo}. Let us just recall that in Bose--Einstein condensates, $k_L=2/\xi$, with
$\xi$ the healing length of the condensate. We also repeat that in quantum gravity, this Lorentz violation scale is usually identified with the Planck scale, although analogue gravity models indicate that the two are not necessarily related. In any case, qualitatively,
our results will not depend on the specific form of the deviations from the
relativistic dispersion relation but on their
superluminal character.

This dispersion relation leads to a modification in both the phase velocity $v_{ph}$ and the group velocity $v_g$. 
%
\begin{figure}
\begin{center}
\includegraphics[width=0.6\columnwidth]{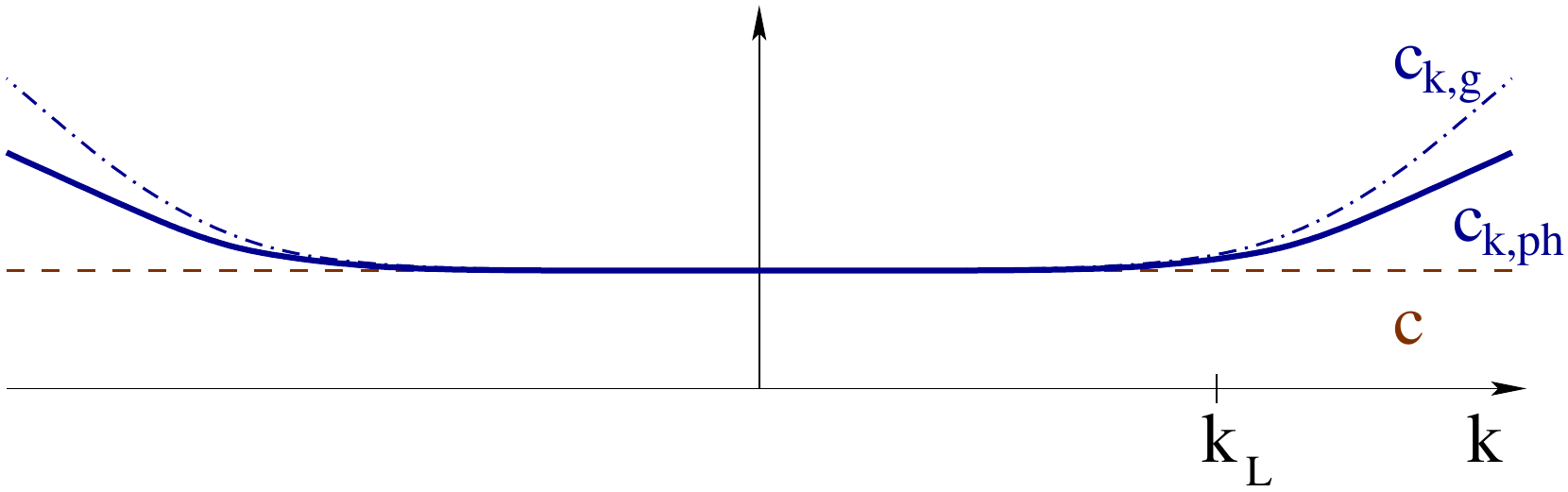}
\end{center}
\figcaptionl{Fig:c_k}{Behaviour of the effective 
phase $c_{k,ph}$ and group $c_{k,g}$ speeds of light/sound with respect to the wave number $k$. Due to the non-relativistic dispersion relation, the effective velocities become `superluminal' for $k>k_L$ (where $k_L$ is the Lorentz violation scale).
}
\end{figure}
%
For a right-moving wave, we have
\begin{eqnarray}
v_{ph} \equiv \frac{\omega}{k} =c_{k,ph}+v~, &~~~~~~~~~&
v_g\equiv \frac{\d \omega}{\d k} =c_{k,g}+v~,
\end{eqnarray}
where we have introduced the effective $k$-dependent phase and group speeds of light/sound
\begin{eqnarray}\label{group_speeds}
c_{k,ph}=c\sqrt{1+\frac{k^2}{k_L^2}}~, &~~~~~~~~~~&
c_{k,g}=c\frac{1+2\frac{k^2}{k_L^2}}{\sqrt{1+\frac{k^2}{k_L^2}}}~,
\end{eqnarray}
respectively (see Fig.~\ref{Fig:c_k}). Both $c_{k,ph}$ and $c_{k,g}$ become larger than $c$ (i.e., become `superluminal') as $k$ increases above $k_L$.

At first sight, it seems obvious that the
ray equation should be defined in terms of the group velocity
$v_g$. Nevertheless, the question of whether the velocity relevant for
Hawking radiation is the group or the phase velocity seems to be
tricky~\cite{Schutzhold:2008tx,Visser:2007du,Visser:2007nx}. For example,
\eqref{stationary} below suggests that the phase velocity might be
relevant. We limit ourselves to remark that $c_{k,g}$ and $c_{k,ph}$
show the same qualitative behaviour. Hence our results are independent
of this issue and we will simply write $c_k$ (or $c_{k(\omega')}$
when wishing to emphasise the frequency-dependence). There will then be a horizon, which becomes frequency-dependent, when $c_k+v=0$,
irrespectively of whether $c_{k,g}$ or $c_{k,ph}$ is used for $c_k$.
Moreover, since $c_k$ becomes arbitrarily high for increasing wave
number $k$, given a certain $|v|_{\rm max}$ at a particular instant of
time, there will be a critical $\omega'_c$ such that waves with an
initial frequency $\omega'>\omega'_c$ do not experience a horizon at
all. The only exception to this rule occurs when the velocity profile
ends in a singularity $\bar{v}\to -\infty$, which implies $\omega'_c
\to \infty$.

Our aim is to calculate the black hole radiation with superluminally
modified dispersion relations. We will now repeat the main steps of
section~\ref{SS:SDR}, and point out where and how these modifications
must be taken into account.

\subsection{Generalisation of inner product}
\label{SS:generalization_inner_product}

The essential point with regard to the pseudo-scalar
product~\eqref{varphi1-varphi2} is the following. Its explicit form
for $t=$constant is not changed by the presence of the $\partial_x^4$
term in the wave equation (see also the discussion in~\cite{Corley:1998rk}). Indeed, it is still a well-defined inner
product, and in particular it is still a conserved quantity, since
\begin{eqnarray}
\partial_t(\varphi_1,\varphi_2)=\int \d x~
\varphi_1\stackrel{\leftrightarrow}{\partial^4_x}\varphi_2^*~,
\end{eqnarray}
which can be seen, by repeated integration by parts, to vanish under the usual
assumption that the fields die off asymptotically. Note that the
modification of the dispersion relation singles out a preferred time
frame: the `laboratory' time $t$. Changing to another time $\tilde t$
will in general lead to a mixing between $t$ and $x$, and hence the
simple relations given here would no longer be valid.

Using the preferred time $t$, and making exactly the same change of
coordinates as in the case of standard dispersion relations, we can
again transform the inner product, evaluated at $t \to +\infty$, into
the expression~\eqref{scalar_product} in terms of $u$ and $w$. 
As in the standard case, only the first term is relevant for the Hawking process:
\begin{eqnarray}\label{scalar_U}%
-{i c \over 2}
\int_{-\infty}^{+\infty} \d u  ~ 
\left[\varphi_1 \partial_u\varphi_2^* 
-\varphi_2^* \partial_u\varphi_1\right]_{w=+\infty}~.
\end{eqnarray}

Note that we are now using $u$ and $w$ merely as a perfectly good set of
(auxiliary) coordinates, in order to cast the inner product into a
useful form. However, here they do not have the null character they had in the case of standard dispersion relations.

\subsection{Rainbow null coordinates}
\label{SS:Rainbow}

Let us define some sets of spacetime functions that will prove
to be useful in what follows. Given a fluid profile of the type 
described in section~\ref{SS:model}, we can integrate the ray equation 
\begin{eqnarray}%
{\d x \over \d t}=c_{k(\omega')}(t,x)+v(t,x)~,
\end{eqnarray}
starting from the past left infinity towards the right. The
ray has an initial frequency $\omega'$ and an associated initial
wave number $k'$, from which we can deduce the value of $c_{k'}$. In
the left region, where the velocity profile is dynamic but
position-independent ($v(x,t)=v(\xi(t))$, with $\xi(t)$ the auxiliary function introduced in section~\ref{SS:model}), $k=k'$ can be considered as fixed
while the frequency changes (this is what happens in a mode
solution of equation~\eqref{modified_laplace} in this region). Then,
we can define the function
\begin{eqnarray}%
{\cal U}_{\omega'}(t,x)= 
{1 \over \omega'}\int {\bar \omega}(t)\d t -{k' \over \omega'} x~,
\end{eqnarray}
where ${\bar \omega}(t)$ is the instantaneous frequency of the
particle at each time $t$, defined through the dispersion
relation and such that \mbox{${\bar \omega}(t \to -\infty)=\omega'$}.

When the ray reaches the kink it passes into a stationary region in
which the velocity profile only depends on the position ($v(t,x)=\bar v(x)$). At
the kink, the ray still has the initial wave number $k'$ and
a frequency $\omega$. In its propagation towards the right this
frequency now remains fixed while its wave number becomes a decreasing
function $\bar k (x)$ of the position, such that $\bar k (\xi(t_0))=k'$, with $t_0$ the moment at which the kink is crossed. The final frequency of the ray will then simply be $\omega$ and its final wave number $k=
\lim_{x \to +\infty} \bar k(x)$. In this region, then,
the function ${\cal U}_{\omega'}(t,x)$ can be expressed as 
\begin{eqnarray}\label{stationary}%
{\cal U}_{\omega'}(t,x) = t - \int {{\bar k}(x) \over \omega} \d x~.
\end{eqnarray}
Note here that $\omega$ and $\bar k(x)$ both depend on the initial $\omega'$.

The same can be done by integrating the ray equation starting from the
future. In this way we can define $u_\omega$ functions. (The same procedure can be used to define ${\cal W}_{\omega'}$ and $w_{\omega}$). It is worth noting that ${\cal U}_{\omega'}$ and $u_\omega$ are not null coordinates in the usual geometric sense, since they are frequency-dependent, but they nevertheless share many properties with null coordinates.

\subsection{Bogoliubov $\beta$ coefficient}

The calculation of the inner product~\eqref{scalar_U} involves
the limit $w \to +\infty$ (equivalently, $w_\omega \to +\infty$). 
%
%
%
%
We can therefore
change variables in the inner product from $u$ to \mbox{$u_{\omega}= u_{\omega}(u)\equiv u_\omega(u,w\to +\infty)$}. The
combination of the derivative and the integral means that the form of
\eqref{scalar_U} is preserved in the new integration variable $u_{\omega}$.
We then can write the Bogoliubov $\beta$ coefficients relevant for the Hawking process as
\begin{equation}
\beta_{\omega \omega'}=
-{i c \over 2}
\int_{-\infty}^{+\infty} \d u_{\omega}  
[\psi'_{\omega'} \partial_{u_{\omega}}\psi_{\omega} 
-\psi_{\omega} \partial_{u_{\omega}}\psi'_{\omega'}
]_{w_{\omega}=+\infty}~.
\label{KG-modified-product}
\end{equation}

Now, assuming that the profiles vary slowly (in scales much larger than the
Planck distance), the right-moving positive-energy modes associated with
past and future infinity can be approximated by the following simple
expressions:
\begin{subequations}
\begin{eqnarray}
\psi'_{\omega'} &\approx& {1 \over \sqrt{2\pi~c~\omega'}} e^{-i\omega' ~{\cal
U}_{\omega'}}~,\label{modes_MDR_P}\\
\psi_{\omega} &\approx& {1 \over \sqrt{2\pi~c~\omega}} e^{-i\omega ~u_{\omega}}~,\label{modes_MDR_F}
\end{eqnarray}
\end{subequations}
so that the Bogoliubov coefficients read
\begin{eqnarray}\label{beta_MDR}%
\beta_{\omega \omega'} \approx
{1 \over 2 \pi} \sqrt{\omega \over \omega'}
\int \d u_{\omega}  ~  
e^{-i \omega' \mathcal{U}_{\omega'}(u_{\omega})} e^{-i \omega u_{\omega}}~.
\end{eqnarray}%

In analogy with the standard case, all information about the radiation
is seen to be encoded in the relation $\mathcal{U}_{\omega'}(u_{\omega})$. In
this expression $\omega'$ is the initial frequency of a ray at the past
left infinity and $\omega=\omega(\omega')$ its final frequency when
reaching the future right infinity.

The approximation just discussed for the Bogoliubov coefficients amounts to
considering profiles that vary sufficiently slowly both with $x$ and $t$. This is
equivalent to considering large black holes. In general the quartic
term in the wave equation, or equivalently the quartic modification of
the dispersion relation, introduces a new source of backscattering, on
top of the usual angular-momentum potential barrier which we have
already neglected. In our approximation, this additional
backscattering (beyond the standard grey-body factors) has been
neglected. For large black holes this contribution will in any case be
very small as has been observed in numerical simulations~\cite{Schutzhold:2008tx}. In addition we are also neglecting any reflection caused
by the kink. However, let us remark that given an approximate scheme
for calculating $\mathcal{U}_{\omega'}(u_{\omega})$ for general
profiles $v(t,x)$, the same $\mathcal{U}_{\omega'}(u_{\omega})$
obtained from a profile with a kink would be obtainable from one (or
several) specific $v(t,x)$, this time perfectly smooth and thus
causing no further backscattering. Our results, which rely only on the
specific form of the relation $\mathcal{U}_{\omega'}(u_{\omega})$, are
therefore valid beyond the specific configurations with a kink
presented in this chapter.

\subsection{Relation $\mathcal{U}_{\omega'}(u_{\omega})$}
\label{SS:U-vs-u-super}
Our next task is to calculate the uniparametric family of functions $\mathcal{U}_{\omega'}$ and the relation between $\mathcal{U}_{\omega'}$ and $u_{\omega}$ for different configurations. As explained in section~\ref{SS:Rainbow}, the relation $\mathcal{U}_{\omega'}(u_{\omega})$ is obtained by integrating the
ray equation \mbox{$\d x/\d t=c_k+v$} using the profiles discussed in
section~\ref{SS:model}. These can be described by means of
stationary profiles $\bar{v}(x)$ and an auxiliary function $\xi(t)$,
see the definition of $v(t,x)$ in eq.~\eqref{profile}. We will use a straightforward extension of the procedure established in~\cite{Barcelo:2006np} (see also~\cite{Barcelo:2006uw} for a summary) for a relativistic dispersion relation. Care must be taken, however, with the quantities that depend on the frequency $\omega'$. In particular:
\begin{itemize}
\item We will denote by $x_{H,\omega'}$ and $t_{H,\omega'}$ the 
position and the time at which the horizon associated with a particular 
initial frequency $\omega'$ is formed.
\item The surface gravity $\kappa_{\omega'}$, defined in
\eqref{def-kappa}, allows to write, for all $\omega'<\omega'_c$ and for $x$
close to $x_{H,\omega'}$:
 \begin{eqnarray}\label{vbar}
 \bar{v}(x) \approx -c_k+\frac{1}{c}\kappa_{\omega'}(x-x_{H,\omega'})~,
 \end{eqnarray}
up to higher-order terms in $x-x_{H,\omega'}$.
\item In a similar way, $\xi(t)$ can be linearised near $t_{H,\omega'}$ by introducing an \mbox{$\omega'$-dependent} parameter $\lambda_{\omega'}$:
 \begin{eqnarray}\label{xi}
 \xi(t)  \approx x_{H,\omega'}-\lambda_{\omega'}(t-t_{H,\omega'})~,
 \end{eqnarray}
again up to higher-order terms.
\end{itemize}
Let us consider laboratory times $t>t_{H,0}\equiv t_{H,\omega'=0}$ such that $\xi(t)$
has already crossed the classical horizon at \mbox{$x=0$}: \mbox{$\xi(t>t_{H,0})<0$}.
As we explained earlier we can define a critical frequency $\omega_c'$ as the
minimum initial frequency such that $c_k+\bar
v(\xi(t))> 0$ for all $\omega'>\omega_c'$, or, in other words, the minimum
frequency which at that particular time has not experienced any horizon yet.

We first integrate the ray equation in the dynamical part of the profile, where $v=\bar{v}(\xi(t))$, for rays crossing the kink just before the formation of the horizon. The integral in this region runs between the asymptotic past and the point $(t_{0,\omega'},x_{0,\omega'})$ at which the kink separating the dynamical and the stationary regions is crossed for this particular frequency:
\begin{eqnarray}
\int_{-\infty}^{t_{0,\omega'}}\d t\left[\bar{v}\left(\xi(t)\right)+c_{k(\omega')}\right] =\int_{-\infty}^{x_{0,\omega'}}\d x~.
\end{eqnarray}
Writing \mbox{$c_{k'}=\lim_{t\to
-\infty}c_{k(\omega')}$} for the speed of propagation in the asymptotic past (and by extension in this whole first region, since $k=k'$ remains constant, see section~\ref{SS:Rainbow}), using the property \mbox{${\cal U}_{\omega'}(t,x) \to t-x/c_{k'}$ for $t,x \to -\infty$}, and noting that \mbox{$x_{0,\omega'}\equiv\xi(t_{0,\omega'})$}, this leads to
\begin{eqnarray}
{\cal U}_{\omega'}
= t_{0,\omega'} -\frac{\xi(t_{0,\omega'})}{c_{k'}}+ \frac{1}{c_{k'}}\int_{-\infty}^{t_{0,\omega'}}\bar{v}(\xi(t)) \d t~,
\label{U_vs_xi}
\end{eqnarray}
which can be written as
\begin{eqnarray}\label{U_vs_U_H}
{\cal U}_{\omega'}
\approx {\cal U}_{H,\omega'} +
\frac{\lambda_{\omega'}}{c_{k'}}(t_{0,\omega'}-t_{H,\omega'})~,
\end{eqnarray}
for small values of $|t_{0,\omega'}-t_{H,\omega'}|$, where
\begin{equation}
{\cal U}_{H,\omega'} \equiv t_{H,\omega'} -\frac{\xi(t_{H,\omega'})}{c_{k'}}+
\frac{1}{c_{k'}}\int_{-\infty}^{t_{H,\omega'}}\bar{v}(\xi(t)) \d t~
\end{equation}
is the ray constituting the horizon associated with the frequency $\omega'$.

In the stationary part of the profile, where $v=\bar{v}(x)$, we can proceed in a similar way and write
\begin{equation}\label{u_omega-step1}
u_{\omega}= \lim_{x_f; t_f\to +\infty}\left (t_{0,\omega'} - \frac{x_f}{c_{k,f}}+ \int_{\xi(t_{0,\omega'})}^{x_f}\frac{\d x}{\bar{v}(x)+c_{k(\omega')}}\right )~,
\end{equation}
where $c_{k,f}=\lim_{t\to +\infty}c_{k(\omega')}$ is the speed of propagation in the asymptotic future. Using the identity
\begin{equation}
\frac{\xi(t_{0,\omega'})}{c_{k,f}}-\frac{x_f}{c_{k,f}}=\int_{x_f}^{\xi(t_{0,\omega'})}\frac{\d x}{c_{k,f}}~,
\end{equation}
eq.~\eqref{u_omega-step1} can be cast in the following form:
\begin{equation}
u_{\omega}=t_{0,\omega'}-\frac{\xi(t_{0,\omega'})}{c_{k,f}} + \int^{+\infty}_{\xi(t_{0,\omega'})}\d x\left[ \frac{1}{\bar{v}(x)+c_{k(\omega')}}-\frac{1}{c_{k,f}}  \right]~.
\end{equation}
We now focus on the integral on the right-hand side and split the integration interval into three parts:
\begin{eqnarray}
\int_{\xi(t_{0,\omega'})}^{+\infty}=\int_{\xi(t_{0,\omega'})}^{x_I}+\int_{x_I}^{x_{II}}+\int_{x_{II}}^{+\infty}~,
\end{eqnarray}
with $x_I$ close to $\xi(t_{0,\omega'})$ and $x_{II}$ sufficiently large so that, effectively, $\bar{v}(x)\to 0$ and $c_{k(\omega')}\to c_{k,f}$ for $x=x_{II}$. The value of the third integral now approximately vanishes. The value of the first integral evaluated at its upper integration boundary $x_I$, together with the complete second integral, can be absorbed into a largely irrelevant bulk integration constant $C_1$. What remains is the behaviour of the first integral around its lower integration boundary $\xi(t_{0,\omega'})$. We are interested in its behaviour for small values of $|t_{0,\omega'}-t_{H,\omega'}|$. 
Using~\eqref{vbar} to approximate the integrand, we obtain
\begin{equation}
u_{\omega} \approx t_{0,\omega'}-\frac{\xi(t_{0,\omega'})}{c_{k,f}}+C_1
-\frac{1}{c_{k,f}}\int^{\xi(t_{0,\omega'})}\frac{c_{k,f}}{\frac{1}{c}\kappa_{\omega'}(x-x_{H,\omega'})}\d x~,
\end{equation}
from which we conclude that 
\begin{eqnarray}\label{u_vs_xi}
\xi(t_{0,\omega'})-x_{H,\omega'} = C_{\omega'}e^{-\kappa_{\omega'}u_{\omega}/c}~.
\end{eqnarray}

Both regions are connected by combining eqs.~\eqref{U_vs_U_H} and~\eqref{u_vs_xi}.
Making use of eq.~\eqref{xi}, this gives
\begin{eqnarray}\label{U-u_MDR}
{\cal U}_{\omega'}={\cal U}_{H,\omega'} - A_{\omega'} e^{-\kappa_{\omega'}
u_{\omega}/c}~.
\end{eqnarray}
This relation is valid for frequencies for which a horizon is
experienced, i.e.\ for \mbox{$\omega'<\omega'_c$} and times $u_{\omega}>u_{I,\omega'}$, where $u_{I,\omega'}$ is the
threshold time defined in section~\ref{SS:relation_U(u)} which,
unsurprisingly, has become frequency-dependent. Again, as in the case
of standard dispersion relations---section \ref{SS:relation_U(u)}---we
can write
\begin{eqnarray}\label{A_0-I_omega}
{\cal U}_{\omega'} = {\cal U}_{H,\omega'} 
- A_0 e^{-\kappa_{\omega'} (u_{\omega}-\bar u_{I,\omega'})/c}~,
\end{eqnarray}
where $\bar u_{I,\omega'}$ is essentially $u_{I,\omega'}$ 
(with possible higher-order corrections). Assuming that the collapse
takes place rapidly, it is a good approximation to replace
$\bar u_{I,\omega'}$ by $\bar u_{I,\omega'_c}$. Actually, as we will see
shortly, this is a conservative estimate, in the sense that it
slightly underestimates the superluminal correction to the radiation
spectrum.

\subsection{Wave packet formulation}
\label{SS:wave-mod}
The relation ${\cal U}_{\omega'}(u_{\omega})$ we are considering, interpolates
between a linear behaviour at early times and an exponential behaviour
at late times.  It has the same form as the relation $U(u)$ for
standard dispersion relations, and so we can continue following the
steps of the standard case. In particular, the equivalent of eq.~\eqref{U-u_step1} is obtained by integrating out $u_\omega$. Note that we use the subscript $\omega$ to emphasise that the $u_\omega$ are not the null coordinates of the standard case, but this should not be interpreted as an explicit function of $\omega$ and so does not complicate the integration steps in $u_\omega$ and $\omega$. However, when integrating $|\beta_{\omega_{j},u_{l};\omega'}|^2$ in $\omega'$ to obtain $N_{\omega_{j},u_{l}}$, see section~\ref{SS:wave-standard}, we must carefully consider the
frequency-dependence of the relevant terms, i.e., of ${\cal
U}_{H,\omega'}$, $\bar{u}_{I,\omega'}$ and $\kappa_{\omega'}$. The term
carrying ${\cal U}_{H,\omega'}$ is moduloed away in eq.~\eqref{U-u_step2}, and we have replaced $\bar{u}_{I,\omega'}$ by $\bar{u}_{I,\omega'_c}$, so the only relevant frequency-dependent factor that we are left with is the surface gravity $\kappa_{\omega'}$.

Moreover, because of the critical frequency $\omega'_c$ in the horizon formation
process, a finite upper boundary will also be induced in the integral. Indeed, frequencies $\omega'>\omega'_c$ do not contribute
to the radiation at all, since they do not experience a horizon. This
is a delicate but crucial point. It was already observed long ago
by Jacobson~\cite{Jacobson:1991gr} that trying to solve the
transplanckian problem naively by imposing a cut-off frequency would
seemingly extinguish Hawking radiation on a relatively short time scale. In our
case, however, this cut-off is not imposed {\it ad hoc}, but appears
explicitly because of the superluminal character of the system at high
frequencies. Moreover, the critical frequency, and hence the upper boundary
induced in the integral, depend directly on the physics inside the
horizon. Indeed, given a certain velocity profile, and in particular
its behaviour near the centre of the black hole, the critical
frequency can be calculated by setting $c_k=|v|$ in eq.~\eqref{group_speeds} and extracting the corresponding critical frequency from the dispersion relation~\eqref{dispersion}. We will see this effect graphically in
section~\ref{SS:graphics}.

In analogy with eqs.
\eqref{U-u_step2} and~\eqref{N-standard}, we now obtain the number of particles detected for each frequency $\omega_j$ as
\begin{eqnarray}\label{N-z}
N_{\omega_{j},u_{l}}&=&\int_0^{\omega_c'} \d\omega'~
|\beta_{\omega_{j},u_{l};\omega'}|^2
\\
&\approx& {c\Delta\omega \over 2\pi}
\int_0^{\omega_c'} {\d\omega' \over \omega'}
{1 \over \kappa_{\omega'}}
{\sin^2\left[{\kappa_0 \over \kappa_{\omega'}}(z -z_{l,\omega'})\right] 
\over \left[{\kappa_0 \over \kappa_{\omega'}}(z -z_{l,\omega'})\right]^2}\;{1 \over \exp({2\pi c\omega_j
\over \kappa_{\omega'}})-1}\nonumber,
\end{eqnarray}
where now \mbox{$z= {c\Delta \omega \over 2 \kappa_0}\ln \omega' A_0$}, with
\mbox{$\kappa_0\equiv\kappa_{\omega'=0}$} (which corresponds to the standard $\kappa$ of
\eqref{standard-kappa}), and \mbox{$z_{l,\omega'}= {\kappa_{\omega'} \over \kappa_0} {\Delta \omega \over 2}(u_l-\bar u_{I,\omega'_c})$}.
Changing the integration variable from $\omega'$ to $z$, we finally obtain the central expression in our analysis:
\begin{equation}\label{N-y}
N_{\omega_{j},u_{l}}=
\frac{1}{\pi}\int_{-\infty}^{z_c} \d z 
{\kappa_0 \over \kappa_{\omega'}}
{\sin^2\left[{\kappa_0 \over \kappa_{\omega'}}(z -z_{l,\omega'})\right] 
\over \left[{\kappa_0 \over \kappa_{\omega'}}(z -z_{l,\omega'})\right]^2}\;
{1 \over \exp({2\pi c\omega_j \over \kappa_{\omega'}})-1}~,
\end{equation}
where the upper integration boundary $z_c= {c\Delta \omega \over 2 \kappa_0}\ln \omega'_c A_0$. Note that, as we indicated earlier, the use
of $\bar u_{I,\omega'_c}$ in the definition of $z_{l,\omega'}$ is a conservative
stance. Indeed, strictly speaking, we should write
$z_{l,\omega'}\propto(u_l-\bar u_{I,\omega'})$. For a fixed $z_c$, a smaller
value of $z_{l,\omega'}$, and hence a larger value of $\bar u_{I,\omega'}$,
means that a larger part of the central peak of the integrand will be
integrated over. Since we are replacing $\bar u_{I,\omega'}$ by
the upper bound $\bar u_{I,\omega'_c}$, we are overestimating the resulting radiation, i.e., underestimating the
modification with respect to the standard Hawking radiation.

The expression~\eqref{N-y} brings out the two crucial factors mentioned earlier, and a third, corollary one.
\begin{itemize}
\item First, it shows the dependence of the total radiation on the critical frequency
$\omega'_c$ (through the integration boundary $z_c$ induced by it), as discussed just before obtaining formula~\eqref{N-z}.
\item Second, it shows the importance of the frequency-dependent $\kappa_{\omega'}$
(to be compared with the fixed $\kappa$ of the standard case). Given a concrete
profile $v(t,x)$, the frequency-dependence of $\kappa_{\omega'}$ can be derived
explicitly, as we will discuss next.
\item Finally, as a corollary of the first point, it shows that, as $u_l$ (and hence $z_{l,\omega'}$) increases, a smaller part of the central peak of the integrand will be integrated over, so the radiation will die off as $u_l$ advances.
\end{itemize}
It is also a simple exercise to consider what this result implies for the case of a constant surface gravity $\kappa_{\omega'}=\kappa_0$. From~\eqref{N-y}, one obtains immediately that the exponential Planckian factor leaves the integral, whereas the remaining terms lead to a prefactor smaller than one.

\subsection{Surface gravity}
A careful analysis has shown that, so far, most formulas for the
standard case could be adapted to superluminal dispersion relations by replacing the relevant magnitudes with their
frequency-dependent counterparts. For example,
$\beta_{\omega\omega'}$ was obtained by replacing $U$ and $u$ by
${\cal U}_{\omega'}$ and $u_{\omega}$, and in particular $U_H$, $u_I$
and $\kappa$ by ${\cal U}_{H,\omega'}$, $\bar{u}_{I,\omega'}$ and
$\kappa_{\omega'}$, respectively.
Given a concrete profile, we can explicitly deduce the relation
between $\kappa_{\omega'}$ and $\omega'$, with the Lorentz violation scale $k_L$
as a parameter, as follows. The horizon for a particular initial
frequency $\omega'$ is formed when 
\begin{eqnarray}
1+{k_H^2 \over k_L^2}={|v(x_{H,\omega'})|^2 \over c^2}~,
\end{eqnarray}
i.e., when $v_{ph}=c_k+v=0$, where we have used the phase velocity for
concreteness. But again, qualitatively, our results would be similar if
we take the group velocity $v_g$ instead of $v_{ph}$.

Taking into account that $k_H=k'$, the dispersion relation in the asymptotic past can be written as
\begin{eqnarray}\label{omega-v_H}
\omega'^2= 
|v(x_{H,\omega'})|^2 k_L^2 \left({|v^2(x_{H,\omega'})|^2 \over c^2}-1\right)~.
\end{eqnarray}
Given a concrete profile, $x_{H,\omega'}$ can then be obtained and $\kappa_{\omega'}$
calculated explicitly. 

For the profiles of the first type, see Fig.~\ref{Fig:v-constant-kappa}, the result is trivial. Indeed, $|\bar{v}(x)|$ increases linearly between the horizon corresponding to $\omega'=0$ and the one for $\omega'_c$, so we obtain a constant $\kappa_{\omega'}=\kappa_0$ for all $\omega'<\omega'_c$.

For a Schwarzschild profile as in Fig.~\ref{Fig:v-increasing-kappa}, on the other hand,
\begin{eqnarray}
\bar v(x)= 
-c\sqrt{c^2/2 \kappa_0 \over x+c^2/2 \kappa_0}~,
\end{eqnarray}
so we obtain
\begin{eqnarray}
\kappa_{\omega'}&\equiv &c \bigg|{\d \bar v \over \d x} \bigg|_{x=x_{H,\omega'}}
= \kappa_0 \left({|\bar v_H| \over c}\right)^{3/2}\nonumber\\
&=&\kappa_0 {1 \over 2\sqrt{2}}\left(
1+\sqrt{1+4\frac{\omega'^2}{c^2 k_L^2}}
\right)^{3/2}~.
\end{eqnarray}
Note that $\kappa_{\omega'}$ remains nearly constant until frequencies of the order of magnitude of the Lorentz violation scale are reached, and then starts to increase rapidly, see Fig.~\ref{Fig:kappa-omega}. As we will see graphically in the next section, this can have an important qualitative influence on the radiation spectrum.

\begin{figure}
\begin{center}
\includegraphics[width=0.45\textwidth]{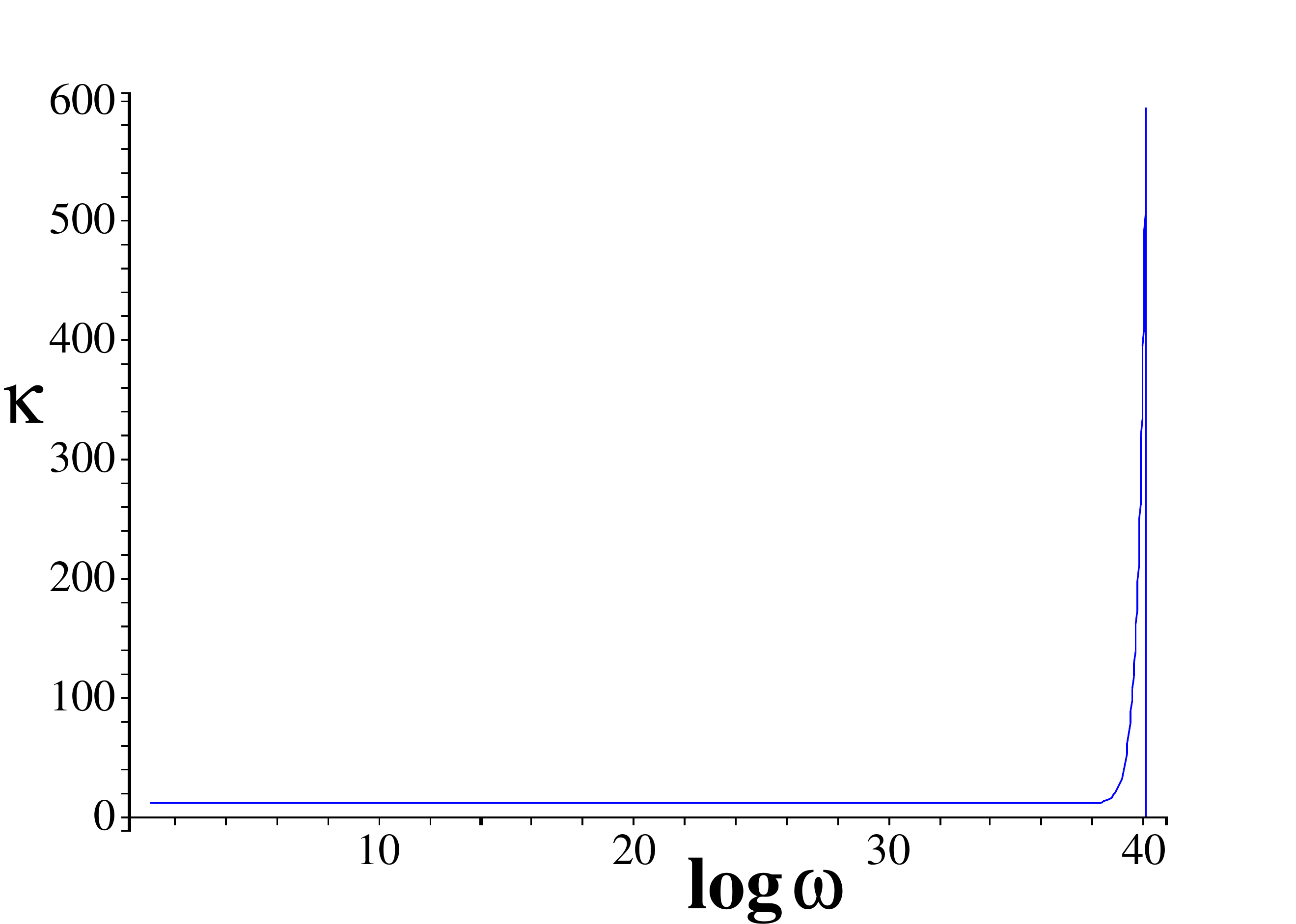}
\end{center}
\figcaptionl{Fig:kappa-omega}{Surface gravity $\kappa_{\omega'}$ with respect to the (logarithm of the) frequency for a Schwarzschild-type black hole as in Fig.~\ref{Fig:v-increasing-kappa}, with the Lorentz violation scale $k_L=10^{39}~{\rm t}^{-1}$ and the critical frequency $\omega'_c=13\times 10^{39}~{\rm t}^{-1}$
(where ${\rm t}$ represents an arbitrary time unit.)}
\end{figure}

We now have all the tools necessary to compute and plot
\eqref{N-y} for different parameters of the profiles described in section~\ref{SS:model}.

\section{Graphical results and discussion}
\label{SS:graphics}
We have integrated eq.~\eqref{N-y} numerically using the Gauss-Chebyshev quadrature method. The results are plotted in
Figs.~\ref{Fig:cut-off}--\ref{Fig:u_l} and perfectly illustrate the three important factors that we deduced theoretically in section~\ref{SS:wave-mod}. Note that in all the figures we have plotted $E\equiv\omega^3\times N$ against $\omega$ to make visual comparison with the usual thermal spectra in 3+1 dimensions easier.

Fig.~\ref{Fig:cut-off} shows the influence of the critical frequency
$\omega'_c$ for a profile with constant $\kappa_{\omega'}$, as in
Fig.~\ref{Fig:v-constant-kappa}, and $u_l =\bar
u_{I,\omega_c'}$ (i.e., immediately after the horizons have formed for all $\omega'<\omega'_c$). The constant $\kappa_{\omega'}$ guarantees that the
form of the Planckian spectrum is approximately preserved. In particular, the peak frequency is not shifted. However, the intensity of the radiation decreases with decreasing critical frequency. Actually, only for
extremely high critical frequencies is the original Hawking spectrum
recovered. For a critical frequency still well above the Lorentz violation scale, the
decrease can be significant. For example, for $\omega'_c=10^{61} c
k_L$, the peak intensity decreases by nearly 30\%, while for
$\omega'_c=c k_L$, the decrease is approximately 40\%. At the other end, note
that extremely low critical frequencies still leave a significant amount of
radiation. For example, for $\omega'_c=10^{-139} c k_L$, still 20\% of
the original peak radiation is obtained. So Hawking radiation receives significant contributions from an extremely wide range of frequencies.

\begin{figure}[ht]
\begin{center}
\includegraphics[width=0.60\columnwidth]{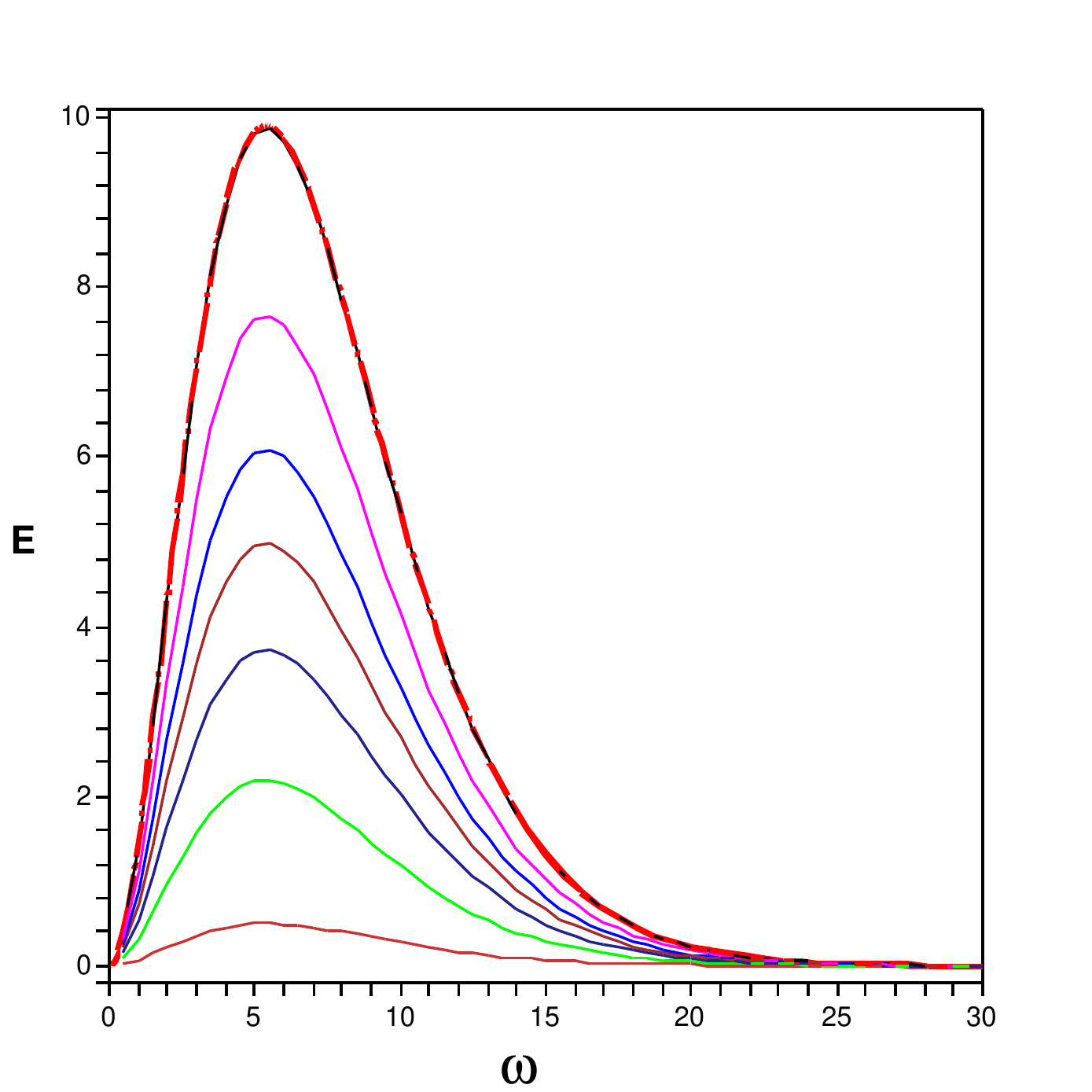}
\end{center}
\figcaptionl{Fig:cut-off}{Influence of the critical frequency on the radiation spectrum for a black hole with velocity profile such that the surface gravity is constant, as in Fig.~\ref{Fig:v-constant-kappa}, and different values of the critical frequency. We have chosen $c=1$, $u_l=u_I=0$, $U_H=A=1\,{\rm t}$, $\kappa_0=12~{\rm t}^{-1}$ and the Lorentz violation scale $k_L=10^{39}~{\rm t}^{-1}$ (where ${\rm t}$ represents an arbitrary time unit), which amounts to considering a solar-mass black hole. From top to bottom we have plotted 
$\omega'_c=10^{3000}, 10^{100}, 10^{39}, 1, 10^{-39}, 10^{-100}, 10^{-300}$. Note that the standard Hawking spectrum coincides perfectly with the upper curve, which effectively corresponds to the absence of a critical frequency.}
\end{figure}

Fig.~\ref{Fig:increasing-kappa} shows the radiation spectrum for a
Schwarzschild-like profile and hence increasing $\kappa_{\omega'}$, as
in Fig.~\ref{Fig:v-increasing-kappa}, in particular for critical
frequencies $\omega'_c$ close to the Lorentz violation scale.  On top of the
general decrease of the standard Planckian part of the spectrum
(approximately 40\%, as in the previous case) due to the finite integration boundary
induced by the critical frequency, the fact that the surface gravity
is now frequency-dependent leads to an important qualitative change of
the spectrum. The high-frequency tail of the spectrum is totally
transformed. Actually, if the critical frequency is sufficiently
higher than the Lorentz violation scale, the dominant source of radiation lies in
the high-frequency region. Note that this effect is truly a
consequence of the modification of the physics for frequencies above
the Lorentz violation scale. This can be appreciated by noticing that, at
$\omega'_c=0.1c k_L$, the whole tail-modifying effect has disappeared
and the usual Planckian form of the thermal radiation spectrum is approximately recovered, although with a lower intensity, as the quantitative decrease described above still occurs.

\begin{figure}[ht]
\begin{center}
\includegraphics[width=0.60\columnwidth]{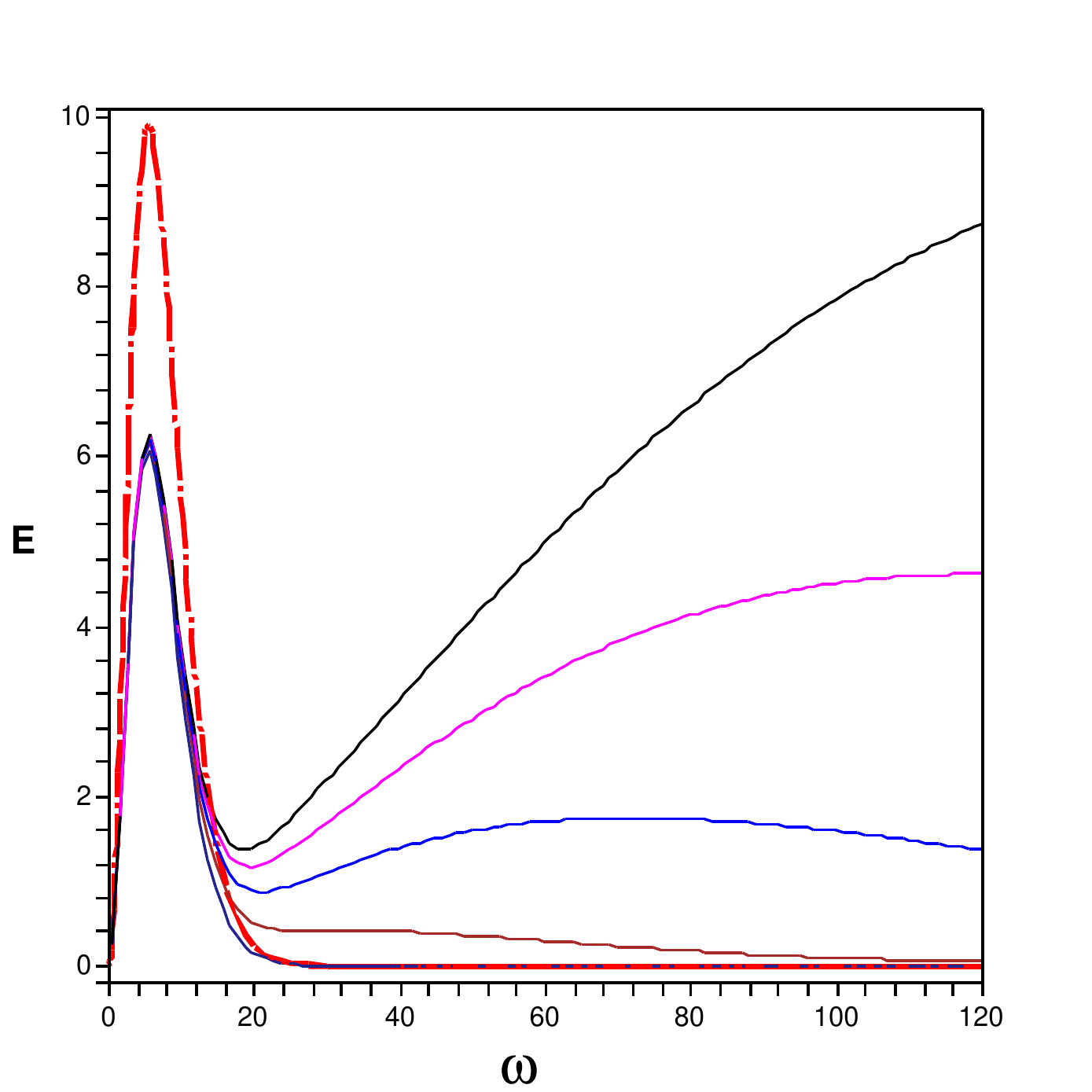}
\end{center}
\figcaptionl{Fig:increasing-kappa}{Influence of a frequency-dependent surface gravity $\kappa_{\omega'}$ on the radiation spectrum for a black hole with Schwarzschild-type velocity profile (surface gravity increases with frequency), as in Fig.~\ref{Fig:v-increasing-kappa}, and different values of the critical frequency around the Lorentz violation scale $k_L$: \mbox{$\omega'_c/c k_L = 13, 10, 7, 4, 0.1$} (from top to bottom). The standard Hawking spectrum is depicted in dashed--dotted line for comparison. Numerical values of $\kappa_0$, $k_L$ etc as in Fig.~\ref{Fig:cut-off}.
}
\end{figure}

Finally, Fig.~\ref{Fig:u_l} shows the influence of the measuring time
$u_l$ (measured with respect to $\bar{u}_{I,\omega'_c}$) for a profile
of the second type (increasing $\kappa_{\omega'}$). It is clearly seen
that the radiation originating from the collapse process dies off with time, and actually dies off rather
fast. For a solar-mass black hole and an $\omega'_c$ of the order 
of the Planck scale, the radiation would last less than a second.
Note that this effect is a corollary of the existence of a
critical frequency $\omega'_c$, since in its absence, the integration
boundaries in eq.~\eqref{N-y} would be infinite, and so the integral would
be insensitive to a change $u\to u+\Delta u$. 

This decrease with time is not wholly unexpected~\cite{Jacobson:1991gr,Gibbons:1975pq}. 
Indeed, the redshift in the frequency suffered by a wavepacket moving away from the black hole is exponential, in analogy with the relation between the past and future null coordinates (see eqs.~\eqref{exp-shift} and~\eqref{U-u_MDR}). In terms of original frequencies $\omega'$ and final frequencies $\omega$, this redshift can be written as
\begin{equation}
\omega\propto \omega' e^{-c^3u/4M}~,
\end{equation}
in units such that $G=1$, and where $u$ is the time of arrival at the asymptotic future infinity. Therefore, when tracing modes with a fixed final frequency $\omega$ back in time, the corresponding original frequency $\omega'$ is exponentially blueshifted with increasing $u$. In the case of superluminal dispersion, this implies that, as time advances, one inevitably reaches a moment when original frequencies above the critical frequency would be needed in order to sustain the radiation. But since these ultra-high-frequencies do not see a horizon, they do not suffer such an exponential shift, hence they simply do not contribute to the Planckian radiation.

The combined effect of Fig.~\ref{Fig:increasing-kappa} and
Fig.~\ref{Fig:u_l} leads to the following qualitative picture for the
further collapse towards a singularity once the initial or classical
horizon has formed. As the interior gradually uncovers a larger
portion of the Schwarzschild geometry, two competing processes will
take place. On the one hand, the spectrum acquires ever larger
contributions associated with higher and higher temperatures. On the
other hand, the overall magnitude of the spectrum is damped with
time. The question of which process dominates could depend on the fine
details of the dynamics of the collapse, and might be further
complicated by backreaction effects, which we have not considered in
our analysis.

\begin{figure}[t]
\begin{center}
\includegraphics[width=0.60\columnwidth]{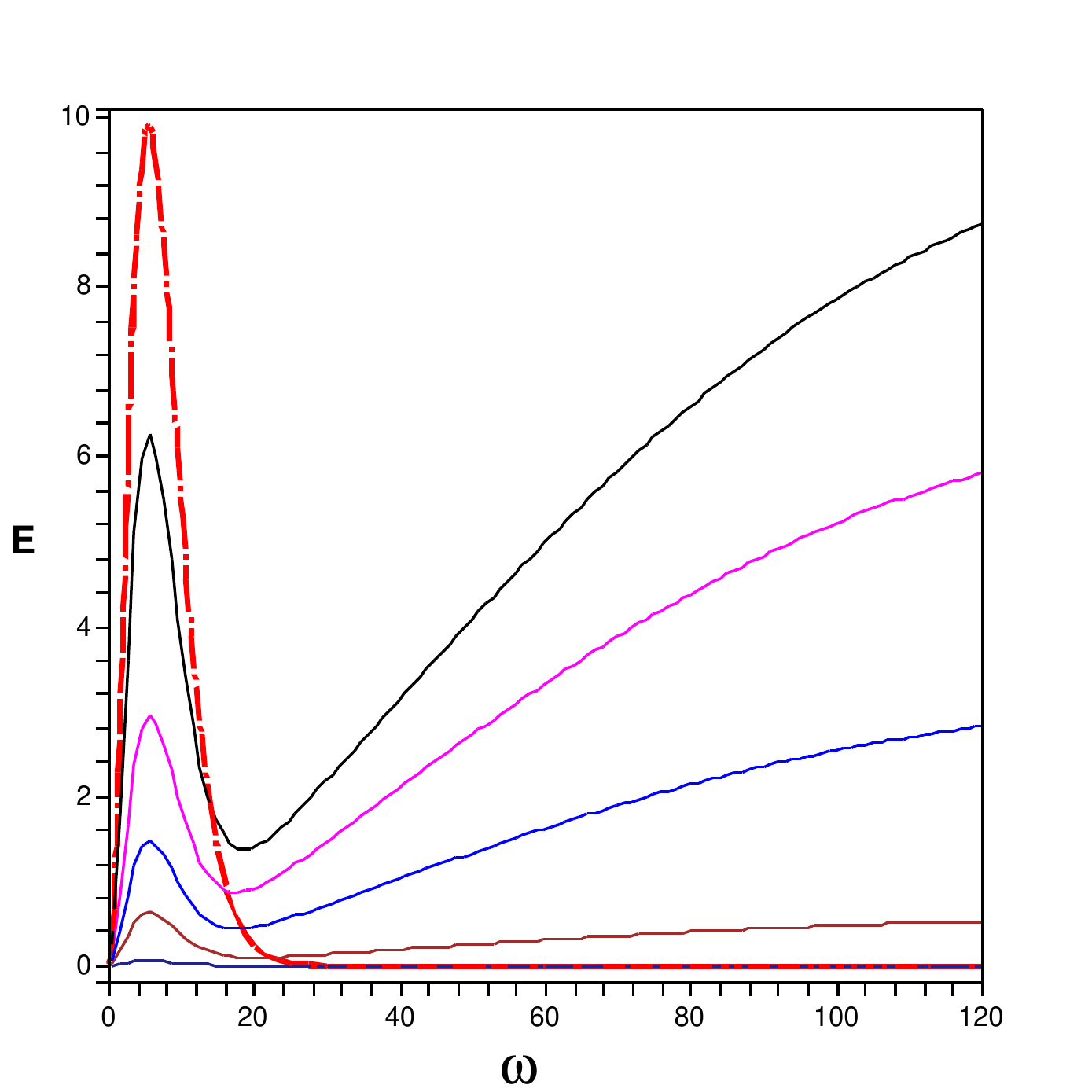}
\end{center}
\figcaptionl{Fig:u_l}{Influence of the measuring time $u_l$ on the radiation spectrum, for a Schwarzschild-type velocity profile, see Fig.~\ref{Fig:v-increasing-kappa} and $\omega'_c/ck_L=13$, compare with Fig.~\ref{Fig:increasing-kappa}.
Different values of $u_l$ (from top to bottom): $u_l=0,22,35,50,500~{\rm t}$
with ${\rm t}$ denoting an arbitrary time unit. The standard Hawking spectrum is depicted in dashed--dotted line for comparison. Numerical values of $\kappa_0$, $k_L$ etc as in Fig.~\ref{Fig:cut-off}.
}
\end{figure}

It is anyhow clear, particularly for the case of a Schwarzschild collapse, that the relation between the critical frequency and the Lorentz violation scale can play a crucial role. Since the critical frequency is associated with the level at which the collapse process saturates, it is reasonable to expect that, for the case of a {\it gravitational} collapse, this critical frequency will ultimately be related to the Planck scale $E_{Pl}$. Indications this expected relationship can be found both in dimensional estimates for quantum gravity based on quantum mechanics, special relativity and general relativity~\cite{Garay:1994en}, and in toy models, for example within the context of loop quantum gravity~\cite{Bohmer:2007wi}. Since current observations seem to indicate that Lorentz violations at the Planck scale are strongly suppressed (if present at all), and hence $E_L\gg E_{Pl}$, it seems that at least the {\it Planckian form} of the black hole radiation is rather robust, even if there turns out to exist superluminal dispersion at high frequencies. The full intensity of a truly Planckian spectrum, however, is only recovered if the critical frequency $\omega_c'\to +\infty$, in other words, if there is a singularity at the end of the collapse, and if moreover the surface gravity is constant for all frequencies (which, we again insist, in a superluminal scenario is not the case for a Schwarzschild collapse). 

However, this robustness of the Planckian form of the spectrum seems to be more `accidental' than intrinsic (unless a fundamental explanation can show why $E_L\gg E_{Pl}$ should necessarily be the case). To illustrate this, remember that for BECs, we argued in chapter~\ref{S:intermezzo} that the analogue of the Planck scale $E_{Pl}$ is related to the interatomic distance, which provides an intrinsic short-distance cutoff, whereas the Lorentz violation scale $E_L$ is determined by the healing length $\xi$, such that $E_L\ll E_{Pl}$. It is easy to see, then, that the mere creation of a black hole horizon in a BEC already implies that the critical frequency must lie above the Lorentz violation scale.\footnote{A straightforward calculation taking for example $v=2c$ and using the phase velocity~\eqref{group_speeds} and the dispersion relation~\eqref{dispersion} gives $\omega'_c=4\sqrt{3}ck_L$. From Fig.~\ref{Fig:increasing-kappa}, we see that this might already be sufficient to detect the ultraviolet contributions.} 
Hence we expect ultraviolet contributions to spontaneously show up in any Schwarzschild-type collapse experiment in a BEC in which the collapse sufficiently crosses the zero-frequency horizon. In practical terms, in the model that we are considering, this involves an increasingly accelerating background flow velocity, see Fig.~\ref{Fig:v-increasing-kappa}, up to at least a few times the speed of sound. On the other hand, if the background flow velocity has a linear slope, as in Fig.~\ref{Fig:v-constant-kappa}, then such ultraviolet contributions should be absent.

In any case, care should be taken when proclaiming the robustness of Hawking radiation in collapsing configurations. Indeed, our results show that, while it might be possible to recover the standard Hawking spectrum in some particular limit, it is certainly not the general rule in the case of superluminal dispersion.

\section{Summary and discussion}
\label{SS:conclusion}
We have discussed the Hawking radiation for a collapsing configuration with
superluminal dispersion relations. Our approach followed the lines of Hawking's original calculation through the relation between the ray trajectories in the asymptotic past and those in the
asymptotic future. We did therefore not need to impose any condition or assumption at the horizon. Rather, our main assumptions were the usual ones with regard to a gravitational collapse, namely a Minkowski geometry in the asymptotic past, and asymptotic flatness also in the future.

An additional implicit assumption in our calculations is the following. We have assumed that the 
number of degrees of freedom of the quantum theory associated with 
the collapsing configurations is equal to that of the field theory in a 
stationary Minkowskian spacetime, which is the starting point of the dynamical evolution of our profiles. In this context, note that for a constant subsonic flow, 
given a real frequency $\omega$, the modified scalar field equation has 
two independent normalisable solutions (with the standard caveats associated with the normalisation of plane waves), whereas the other two solutions become divergent in some region and hence must be discarded. However, in a constant 
supersonic flow, depending on the frequency range, one can find either two or 
four independent normalisable solutions (see Fig.~\ref{F:superDR}), with a transition point in which there are three solutions (of which one is degenerate). Upon quantisation,
these theories have a different number of degrees of freedom. Since a stationary black hole 
configuration consists of a subsonic and a supersonic region, the number of normalisable solutions for a given frequency can then either be two or three, depending on the frequency under consideration. But there always exist frequencies with three associated normalisable solutions. So the field theory over a stationary black hole configuration has more degrees of freedom than that for a Minkowskian (i.e., purely subsonic) spacetime. Our assumption can then be rephrased as follows. We assume that the introduction of dynamics does not change the number of degrees of freedom of the initial theory, and so we consider only how the two initial Minkowskian degrees of freedom are distorted by the dynamics of collapse, and how this influences the standard mechanism for Hawking radiation in such a collapse process.

\begin{figure}[t]\centering
\includegraphics[width=0.70\textwidth,clip]{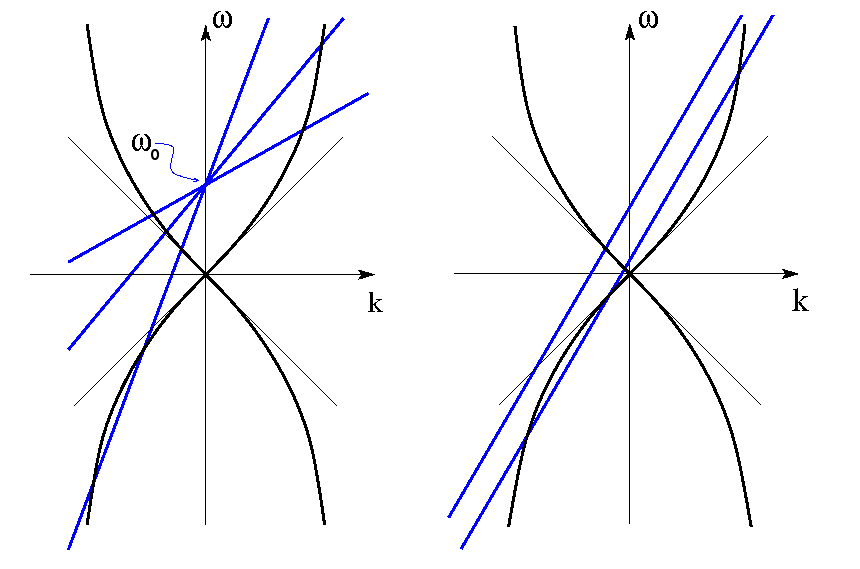}
\figcaptionl{F:superDR}{Superluminal dispersion relation for various values of $v$ (left) and of $\omega$ (right), scaled at $c=1$. The intersection points between the line $\omega-vk$ (in blue) and the curves $\pm ck\sqrt{1+k^2/k_L^2}$ (in black) mark the real (normalisable) mode solutions for a given $\omega$. For a given frequency $\omega_0$ (left), there are always 2 normalisable solutions in a subsonic regime \mbox{($|v|<c$)}, but there can be up to 4 in a supersonic regime \mbox{($|v|>c$)}. For a given supersonic flow (right), depending on the value of $\omega$, there can be 2 or 4 normalisable solutions.}
\end{figure}

Our results can be summarised as follows.

Modifications of the dispersion relation
cause the horizon, and various associated quantities such as the surface
gravity $\kappa$, to become frequency-dependent. In particular, a critical
frequency $\omega'_c$ naturally appears such that frequencies higher than $\omega'_c$
do not experience a horizon. More generally, it also means that the standard
geometric concepts traditionally used to study black holes must be handled with
care. Nevertheless, through a detailed analysis, we have seen that the equations
related to the late-time radiation can be adapted quite straightforwardly from
standard (relativistic) to superluminal dispersion relations. 

We analytically derived an approximate equation for the particle production at
late times~\eqref{N-y} with superluminal modifications of the dispersion relations. This equation clearly shows that important
modifications in the late-time radiation should be expected, first, due to the
existence of the critical frequency $\omega'_c$ and the finite upper boundary it induces in the integration, and second, due to the
frequency-dependence of $\kappa_{\omega'}$. We integrated eq.~\eqref{N-y}
numerically, and plotted the resulting spectrum, thereby confirming these expectations. 

We have seen that the standard Hawking spectrum is recovered only in a very
particular case: in the limit when the critical frequency goes to infinity (i.e.,
when the profile for the velocity $|v|$ goes to a singularity) and moreover the
surface gravity $\kappa_{\omega'}$ is constant (linear velocity profiles of the type of
Fig.~\ref{Fig:v-constant-kappa}). For lower critical frequencies, as long as $\kappa$
remains constant, the Planckian form of the Hawking spectrum is roughly maintained and the peak remains at the same frequency, but
the peak intensity decreases rapidly with decreasing $\omega'_c$. A non-negligible radiation persists, however, even for extremely low critical frequencies. This clearly establishes the first point, namely the importance of the critical frequency $\omega'_c$, which illustrates the statement made in the introduction that with superluminal dispersion relations, the interior of the black hole is probed and significantly affects the Hawking process.

For a Schwarzschild collapse, the velocity profile generally leads to a frequency-dependent surface gravity $\kappa_{\omega'}$, increasing with the frequency $\omega'$. This means that the standard Hawking result
cannot be recovered for a Schwarzschild black hole with superluminal dispersion
relations, even when the velocity profile has a singularity. Actually, when the
velocity profile has a limiting value, the same quantitative decrease of the
Hawking part of the spectrum as before shows up. Moreover, if the critical frequency
$\omega'_c$ is above the Lorentz violation scale, a drastic qualitative change of the
radiation spectrum takes place, and for sufficiently high $\omega'_c$ (roughly a
few times the Lorentz violation scale) the high-frequency part of the spectrum even becomes
dominant. This shows the importance of the surface gravity, which again illustrates the role played by the interior of the black hole. We have indicated that the critical frequency is likely to be related to the Planck scale, or its analogue (associated with the interatomic distance) in the case of BECs. Observational limits on $E_L$ from cosmology (namely, $E_L\gg E_{Pl}$, see chapter~\ref{S:intermezzo}) then suggest that astrophysical black holes would not suffer such strong ultraviolet (UV) contributions. However, this does not mean that the Hawking radiation is intrinsically robust to them. Since for BECs, the relation between $E_L$ and $E_{Pl}$ is precisely the opposite ($E_{Pl}\gg E_L$), the phononic radiation from a collapsing experiment in a BEC could in principle show such important UV contributions.

Finally, we have also seen that, as a corollary of the existence of a critical frequency and of the finite upper boundary induced by it in the thermal response function, the radiation spectrum of the collapse dies off as time advances. This effect can be understood by the fact that frequencies are exponentially redshifted when moving away from the black hole. Therefore, when tracing back the outgoing radiation at a time $u$ to the original frequencies that created it, these original frequencies increase exponentially with $u$. Inevitably, as soon as this tracing-back process reaches the critical frequency, the radiation is effectively switched off. This {\it cut-off} effect has long been understood to be one of the crucial aspects of the transplanckian problem related to Hawking radiation. Here, however, the cut-off is not imposed by hand to solve the transplanckian problem, but appears naturally because of the superluminal dispersion.

A few observations might be useful to connect our work with
existing results on Hawking radiation and its sensitivity to modified
dispersion relations. The most general observation is that the
`robustness' of Hawking radiation, which is often considered to be a
well-established result, actually depends crucially on a series of
assumptions, usually imposed at or near the horizon. Such assumptions might be reasonable in the case of subluminal dispersion relations. But for the case of superluminal
dispersion relations in a collapse scenario, as we have shown explicitly, the assumptions
needed to reproduce the standard Hawking result depend on the physics
inside the (zero-frequency) horizon, and moreover in a way which is
not compatible with the Schwarzschild geometry. 

In particular, in~\cite{Corley:1997pr}, it was shown for a stationary scenario that Hawking radiation is robust with respect to superluminal modifications of the dispersion relations, provided that
positive free fall frequency modes were in their ground state just before crossing the horizon. However, it was also admitted that it is not clear whether this is the physically correct quantum state condition. In~\cite{Unruh:2004zk}, three explicit assumptions were given for the previous condition to hold: freely falling frame, ground
state and adiabatic evolution. In slightly more detail: 
\begin{enumerate}
 \item If the Lorentz symmetry is violated in such a way that there is a preferred reference frame, then this should be the freely falling frame (and not the rest frame of the black hole, for example).
 \item High-frequency excitations must start off in their ground state (with respect to the freely falling frame) before leaving the region infinitesimally close to the horizon on their way out.
 \item The dynamics of all internal degrees of freedom should evolve much faster than any external time dependence, such that no high-frequency modes are excited by these external variations.
\end{enumerate}
In the same paper~\cite{Unruh:2004zk}, it is similarly pointed out that these
assumptions might fail for a superluminal dispersion relation, since (as we also mentioned in the introduction) for high-frequency modes this implies that one needs to make an assumption about the physics at the singularity. 

Rather than imposing any
conditions on the state near the horizon, and hence ultimately at the singularity, we have fixed the {\it
initial} geometry to be Minkowskian and the quantum field to be in the natural associated vacuum state, and evolved this into a black
hole configuration. Since our results are seemingly in contradiction with the
ones obtained in~\cite{Corley:1997pr} and~\cite{Unruh:2004zk}, it is worth examining in which sense the
conditions stated by those authors are violated in our approach. If one considers a collapse scenario, for example of a BEC, in a laboratory setting,
then it is quite natural to assume that the `freely falling frame' condition
will be violated. Indeed, the superluminal modification
is in this case associated with the existence of a privileged external reference
frame: the black hole rest frame or lab frame (as we noted in sec.~\ref{SS:generalization_inner_product}), and not the freely falling frame, as assumed
in~\cite{Unruh:2004zk}. Note that this violation of the free-fall frame condition automatically implies a
violation of the ground state condition in the sense in which this condition is
formulated in~\cite{Unruh:2004zk}, namely as the ground state ``with respect to the freely
falling frame''. 

Our results then suggest the following picture. The
low-energy modes experience the classical geometry, are therefore dragged along
in the free-falling frame and so at the horizon they occupy the vacuum state associated with this free-falling frame, namely the Unruh vacuum. Hence they  contribute to the black hole radiation in the
traditional Hawking way. However, the ultra-high-frequency modes (above the critical
frequency) do not see the horizon. Hence they do not couple to the classical
geometry of the collapse, but rather remain connected to the external or
laboratory frame, and therefore pass through the black hole (nearly) undisturbed. The ground state of these high-frequency modes is then not the vacuum associated with the freely falling frame, but the Boulware vacuum associated with the original Minkowski geometry, or in other words, with the stationary reference frame of the lab. So these frequencies above the critical frequency do not contribute to the thermal output spectrum. The overall radiation is a
convolution of the contributions from all the different initial frequencies,
where the surface gravity for each frequency can be interpreted as leading to an effective
weighting factor. Depending on the internal physics of the black hole, this
leads to either a spectrum of the traditional Hawking form but with a reduced
intensity (in the case of a constant surface gravity), or a modified spectrum
where the high-frequency contributions dominate (as for a Schwarzschild profile,
provided that the critical frequency lies sufficiently above the Lorentz violation scale). 

In any case, what our analysis shows 
with respect to the situation treated in~\cite{Corley:1997pr} and~\cite{Unruh:2004zk} is the following. The robustness of Hawking radiation discussed by those authors is the robustness with respect to a superluminal modification under the precise conditions occurring in the standard case, in which all modes occupy the ground state associated with the free-falling frame just before crossing the horizon. This assumption implies a stationary configuration in which all modes, regardless of their frequency, occupy the Unruh vacuum. Our analysis precisely shows that the superluminality can lead to a spontaneous breaking of these conditions. The natural evolution of a collapsing configuration is such that the frequencies above the critical frequency $\omega'_c$ are not in the Unruh vacuum associated with the freely falling frame, but remain in the Boulware vacuum associated with the initial asymptotic Minkowski condition. It should then come as no surprise that this can indeed have an important impact on the resulting radiation spectrum.

An additional remark, already mentioned at the beginning of this section, is that we have only considered the evolution of the two Minkowskian degrees of freedom, whereas \cite{Corley:1997pr,Unruh:2004zk} deal with stationary configurations, and hence with a higher number of degrees of freedom. It would certainly be interesting to see if it is possible to connect both situations in a consistent way. This would require the description of a possible transition from two to three degrees of freedom, which is not within the direct scope of the formalism that we have used, since this is essentially based on a ray-tracing method, see section~\ref{SS:U-vs-u-super}. Our approach would obviously not be valid in the case of a stationary black hole configuration, since there the rays would simply break down at the horizon. Here, however, we have traced the rays that escape just before the formation of the horizon, and therefore reach the asymptotic infinity without ever having encountered a turning point that would signal such a breakdown. We therefore emphasise that our calculation considers exactly the same mechanism that produces Hawking radiation in the standard (Lorentz-invariant) collapse case. 

Our approach is therefore {\it conservative} in the sense that we have studied how the standard mechanism is deformed in the case of superluminal dispersion. The results that we have obtained are {\it analytical}. Indeed, although we have illustrated our main points graphically through numerical integration, the qualitative content of our results was obtained by inspection of formula~\eqref{N-y} (or~\eqref{N-z}). The fact that we have limited ourselves to study the Minkowskian degrees of freedom, or in other words: that we have used a ray-tracing method, means that our method is also {\it approximative}, since we have not taken into account the possible transition to a higher number of degrees of freedom.

Our result should therefore be understood in a dual sense. 

First, we have established that the standard mechanism of Hawking radiation in a collapse scenario is generally not robust under superluminal dispersion. Moreover, the equivalence between collapsing and stationary black holes (at least, in the straightforward extension from the standard case) is no longer valid in general, since the Unruh state is not necessarily the natural end-state of the gravitational collapse process.

Second, it would be extremely interesting to see to what extent it is possible to go beyond the approximation that we have used, and see to what extent all degrees of freedom can be treated in a consistent way. In this sense, our work will hopefully stimulate further investigation of these issues.

A final observation concerns the recent
article~\cite{Carusotto:2008ep}, in which the authors numerically
simulate the creation of an acoustic horizon in a BEC, and analyse
the production of correlated pairs of phonons through the so-called
truncated Wigner method. Our findings for the case of a constant
surface gravity seem to be in qualitative disagreement with the
discussion presented in~\cite{Carusotto:2008ep}, since the authors of
this paper assert having observed a stationary Hawking flux. The source of this apparent discrepancy might reside in the fact that the correlation function which they study is normalised, and
therefore probes the form of the spectrum (which we also find stationary with time),
but does not provide details about the net amount of particle production. Alternatively, it could be that the numerical method used in~\cite{Carusotto:2008ep} automatically includes the creation of an additional, `non-Minkowskian' degree of freedom. In that case, it would be extremely interesting to see whether we can understand the exact physical mechanism that compensates for the decrease in the radiation coming only from the Minkowskian ones, such as in our analysis. A last remark is the following. If an experiment is realised in which the analogue Hawking radiation in a BEC can indeed be detected, for example along the lines suggested in~\cite{Carusotto:2008ep}, this would certainly be a major advance (as well as a tremendous source of publicity). Additionally, it might help us to gain further insight into the (so far theoretical) discussion about the importance of the additional modes, possible ultraviolet contributions and so forth.


\clearpage
\chapter[Emergent gravity]{Emergent gravity from condensed matter models}
\label{S:emergent-gravity}
\section{Introduction}
The various case studies that we presented so far were either derived from or at least inspired by the gravitational analogy in condensed matter systems, and in BECs in particular. To interpret these case studies as quantum gravity phenomenology, as we did, assumes that Lorentz invariance will turn out to be broken by quantum gravity in a way which can, in a first approximation, be modelled by a superluminal modification of the relativistic dispersion relations at high energies. In this final chapter, we will partly address the question of how seriously the condensed matter analogy can or should be taken, and indicate some conceptual issues related to the construction of a model for emergent gravity based on the condensed matter analogy.

A naive but nevertheless crucial observation in this respect is the following~\cite{Volovik:2008dd}. The fundamental `quantum gravity' theory is generally assumed to have the Planck level as its characteristic scale. Expressed as a temperature, this Planck level lies at approximately $10^{32}$ K. On the other hand, almost all of the observable universe has temperatures that barely exceed the cosmic background radiation temperature of a few Kelvins. Even the interior of a star such as the sun is more than 20 orders of magnitude colder than the Planck temperature, while the highest energies that are planned to be produced at the Large Hadron Collider are still roughly 15 orders of magnitude lower than the Planck scale. So the degrees of freedom of quantum gravity, independently of their fundamental structure, are probably effectively frozen out in most of our universe, just like in a condensed matter system in a low-temperature laboratory. The physics that we observe might then well be due to collective excitations that result from the---comparatively tiny---thermal (or other) perturbations of this vacuum. If this observation is indeed relevant for quantum gravity, then one crucial consequence would be the following. In a condensed matter system, there is no direct quantisation path from the perturbations or  `effective' degrees of freedom (the phonons and quasi-particles) to the `fundamental' degrees of freedom: the condensed matter atoms.  Likewise, a basic tenet of emergent gravity is that there might not be any direct path such as a quantisation procedure from the effective degrees of freedom of classical gravity to the fundamental ones of quantum gravity.

Moreover, it is important to realise that many of the low-temperature effects that occur in condensed matter systems are not specific to one or another type of atoms or molecules, but are generic to a wide range of similar systems, e.g., bosonic systems in the weakly interacting regime, or fermionic systems within the same low-temperature topology class in momentum space~\cite{Volovik:2006gt} (of which the Fermi-point topology seems to be the most relevant for our universe~\cite{Volovik:2003fe}, and is highly robust with respect to fluctuations in the microscopic structure), in combination with some general features such as the relation between the characteristic scales (the interatomic distance and the scattering length, for example). 
From this point of view, knowledge of the relations between the characteristic scales such as the Planck scale and the Lorentz symmetry violation scale might actually be more relevant to calculate the high-energy corrections to classical gravity than an exact description of the fundamental, `quantum-gravitational' degrees of freedom. In this sense, classical gravity should perhaps be interpreted much as thermodynamics~\cite{Padmanabhan:2007tm}: a set of rules for the collective behaviour at low temperatures of some degrees of freedom whose precise microscopic constitution can vary quite widely within a certain universality class. It is in this sense that we use the term `emergent gravity' here.

In the following, we will concentrate on one particular issue of this general idea of emergent gravity from condensed matter models, namely diffeomorphism invariance. Indeed, one problem that immediately comes to mind from a relativistic point of view with respect to the condensed matter models is that they are `background dependent' in a trivial way: the metric description that emerges for the motion of the phonons depends on the collective behaviour of the condensed atoms. In other words, the background is precisely determined by the condensed phase of the condensate. However, in the context of loop quantum gravity, for example, it has been argued that diffeomorphism invariance, understood as a requirement for the non-existence of any prior geometry, is an essential feature of general relativity, and should therefore also be an essential feature of quantum gravity~\cite{Rovelli:2004tv,Smolin:2005mq}. It seems that the background dependence of condensed matter models obstructs the possibility of extending them to a diffeomorphism invariant model for gravity. We will see how this issue can at least partly be resolved, and discuss its relation to the problem of recovering the Einstein equations in models of emergent gravity.

First, though, although quite remote from the nucleus of this thesis, it is probably worth briefly mentioning an important reason why condensed matter models have attracted interest in the domain of quantum gravity, namely the accelerated expansion of the universe, which seems to require the existence of some form of repulsive `dark energy'.

\section{Quantum gravity and dark energy}
\label{SS:dark_energy}

Quantum gravity is in the first place the endeavour to reconcile quantum mechanics and the general theory of relativity. There are many arguments why such a unification is needed, see e.g.~\cite{Carlip:2001wq}. The bulk of these go along the line of consistency arguments, for example that intrinsically classical fields are incompatible with quantum mechanics and could lead to violations of Heisenberg's uncertainty principle. However, it is far from clear whether any such argument could have real empirical consequences, even in principle~\cite{Carlip:2008zf}. Another common argument is that general relativity predicts singularities, not only in mathematically exotic cases but also in situations of physical relevance such as black holes and initial `big bang' singularities in cosmological models. It is then generally assumed that a quantum theory of gravity will relieve this problem and avoid the occurrence of singularities. Nevertheless, again, this argument is far from water-tight: first of all, it is not obvious that a quantum theory of gravity is really {\it needed} to cure the singularities of general relativity. For instance, purely algebraic extensions of general relativity could solve the singularity problem without any need for a quantisation procedure~\cite{lam2007}. Furthermore, it could well be that gravity is intrinsically a non-quantum phenomenon~\cite{Boughn:2008jx}, which simply does not apply at the energy scales at which the singularities are predicted to occur. The approach discussed in this thesis is close in spirit to other suggestions that gravity should not be quantised, for example thermodynamic interpretations of the Einstein equations~\cite{Jacobson:1995ab,Padmanabhan:2009ry} or Sakharov's induced gravity~\cite{Sakharov:1967pk} (about which we will have more to say in section~\ref{SS:sakharov}).

Recently, it has become clear that, apart from the quest for a unification of general relativity and quantum mechanics, there is at least one empirical issue of importance related to (quantum) gravity: the accelerated expansion of the universe. This seems to require the existence of some sort of repulsive energy, generically baptised `dark energy'~\cite{Padmanabhan:2002ji}. It has been argued that the problem of dark energy indicates the need for a reassessment of the traditional paradigm of quantum gravity, for example in the following way~\cite{Padmanabhan:2007xy}. First, although there is a myriad of alternative approaches to dark energy, the cosmological constant is in many ways the most sensible candidate, if only because none of the other approaches explains why the cosmological constant should be (approximately) zero, i.e., why the zero-point energy of the quantum matter fields is not huge, as a naive calculation from quantum field theory seems to indicate~\cite{Weinberg:1988cp}. Second, the equations of motion for matter fields are invariant under a shift of the matter Lagrangian ${\cal L}_m$ by a constant. But the gravitational sector is obviously not invariant under such a shift. In~\cite{Padmanabhan:2007xy}, it was concluded that we need to rethink gravity in a way which can be described in terms of an effective action which is explicitly invariant under a shift of the energy-momentum tensor $T_{\mu\nu} \rightarrow T_{\mu\nu} + \Lambda g_{\mu\nu}$. The condensed matter perspective offers a related point of view, which shows that the original intuition that dark energy or the cosmological constant is simply the zero-point energy of all the quantum fields, i.e., the energy of the quantum vacuum, might prove to be right after all. The essence of the argument is the following~\cite{Volovik:2004gi,Volovik:2005vr,Volovik:2006bh}. 

The value of the quantum vacuum energy in a
condensed matter system in equilibrium is regulated by
macroscopic thermodynamic principles. The vacuum energy density $\epsilon_{\rm vac}=E_{\rm vac}/V$ of a quantum many-body system is obtained from the expectation value of the corresponding Hamiltonian which describes the effective low-temperature quantum field theory in second quantisation: $\hat H_{QFT}=\hat H-\mu\hat N$. Here, $\hat H$ is a Hamiltonian of the form~\eqref{hamiltonian}, 
$\hat N$ is the atom number operator and the chemical potential $\mu$ can be understood as a Lagrange multiplier which accounts for the conservation of the number of atoms.\footnote{If the relativistic quantum field theories of the standard model of particles can indeed be interpreted as effective theories in a similar way, then the equivalent to this condition would of course {\it not} be a conservation of particle number, since the particles of the standard model correspond to the {\it quasiparticles} of the condensed matter analogy, but would rather be related to the number of microscopic degrees of freedom of spacetime.} So $\epsilon_{\rm vac}=\langle\hat H-\mu\hat N\rangle_{\rm vac}/V$.
The equation of state relating the energy density and the pressure of the vacuum of a quantum many-body system can then be obtained through the Gibbs-Duhem equation $E-\mu N-TS=-pV$, which for a pure vacuum state ($T=0$) gives simply $\epsilon_{\rm vac}=-p_{\rm vac}$. This equation of state corresponds to a cosmological constant. 

It seems reasonable to require that the universe should be describable as an isolated system, i.e., without recurring to any external quantities such as an external pressure $p_{\rm ext}$. This suggests the analogy with a liquid-like system, since liquids can be in a self-sustained equilibrium without external pressure. If $p_{\rm ext}=0$, then for a liquid-like pure vacuum state in equilibrium, also $p_{\rm vac}=0$. The natural value for $\epsilon_{\rm vac}$ at $T=0$ in such a system in equilibrium is then $\epsilon_{\rm vac}=0$. In other words, according to this argument, a pure vacuum state in equilibrium in the absence of external pressure is non-gravitating. At $T\neq 0$, thermal fluctuations lead to quasi-particle excitations, and hence to a matter pressure $p_M$. In the absence of external pressure, this matter pressure is compensated by a non-zero vacuum pressure in order to restore the equilibrium: $p_{\rm vac}+p_M=0$. The vacuum energy therefore naturally evolves towards the value $\epsilon_{\rm vac}= p_M$ in equilibrium~\cite{Barcelo:2006cs}, and thereby becomes gravitating. The system automatically adjusts itself to obey the macroscopic thermodynamic rules, and there is no need to know the precise microscopic constitution of the system to calculate the macroscopic quantities. 


From such a condensed-matter point of view, the cosmological constant mystery becomes a lot less unsurmountable: from having to explain why the cosmological constant is more than a hundred orders of magnitude smaller than the value based on a naive zero-point energy calculation~\cite{Weinberg:1988cp}, it is reduced to having to explain why it is slightly bigger than the equilibrium value which would exactly cancel the total (baryonic + dark) matter contribution: $\Omega_\Lambda\approx 0.7$ versus $\Omega_M\approx 0.3$~\cite{Padmanabhan:2002ji} in terms of density parameters $\Omega_i=\rho_i/\rho_c$, with $\rho_c$ the critical density for a flat universe (which seems to correspond very closely to the actual density of our universe). So, the condensed matter approach offers a qualitative framework to understand dark energy, at least in terms of the coincidence problem (why $\Omega_\Lambda\sim\Omega_M$) and the cosmological constant problem (why the zero-point energy of the vacuum is not huge). Whether it also allows to extract a concrete quantitative prescription for the {\it evolution} of dark energy is an open problem, to which a possible first step has been described in~\cite{Klinkhamer:2007pe,Klinkhamer:2008ns}.

We now turn to the main subject of this chapter: diffeomorphism invariance in emergent-gravity models.

\section[Diffeomorphism invariance: kinematical vs. dynamical]{Diffeomorphism invariance: kinematical versus\\ dynamical aspects}
One of the important lessons from analogue gravity is to remind us of the fact that general relativity can be split up into kinematical and dynamical aspects. The kinematical aspects of general relativity are encoded in the curved Lorentzian spacetime, i.e., a manifold with a metric signature $(-,+,+,+)$. The dynamical aspect of general relativity is essentially the interplay between this curved geometry and the matter distribution through the Einstein equations (``Geometry tells matter how to move, and matter tells geometry how to curve''~\cite{Misner:1974qy}). The crucial lesson from analogue gravity in this respect is that the emergence of a curved Lorentzian spacetime occurs naturally in a wide variety of systems, in particular in several types of condensed matter systems, i.e., when a many-body system is sufficiently cooled. So the curved Lorentzian aspect of general relativity is in a certain sense `natural'. However, it seems that what really makes our universe very different from the effective spacetimes that are obtained in analogue models are the Einstein equations. No real laboratory system (condensed matter or other) reproduces the Einstein equations. One of the main objectives of the emergent-gravity programme (or `paradigm', to use the Kuhnian terminology~\cite{Kuhn:1962}), is precisely to elucidate the origin and the naturalness of the Einstein equations.

When discussing background independence and diffeomorphism invariance, a similar distinction between kinematical and dynamical aspects can be made. Technically speaking, diffeomorphism invariance is the requirement that the laws of physics should be formally invariant under transformations defined by a diffeomorphism, i.e., an invertible bijective and smooth function $f$ with smooth inverse $f^{-1}$ defined on two differentiable manifolds. Diffeomorphism invariance (or {\it general invariance}, which should not be confused with general {\it covariance})\footnote{General {\it covariance} simply means that the laws of physics must transform properly under arbitrary coordinate transformations, which is essentially equivalent to the (physically irrelevant) requirement that they can be written in tensorial form. General or diffeomorphism {\it invariance} means that the laws of physics should be {\it invariant}, i.e.: not transform at all, under arbitrary coordinate transformations. See, e.g.,~\cite{Westman:2007wy} for a discussion. Of course, terminology should not become sacrosanct. However, at least historically, the confusion between both principles has been more than just a linguistic issue~\cite{Norton:1993}.} is understood by most contemporary relativists as an essential feature of general relativity. This is why quantum gravity programmes such as loop quantum gravity and related spin foam approaches have developed a canonical quantisation scheme for general relativity which is diffeomorphism invariant by construction~\cite{Ashtekar:2004eh,Rovelli:2008zza}. Diffeomorphism invariance encompasses the requirement that the laws of physics as experienced by an internal observer should not depend in any way on external or predefined structures. In particular, the lesson from general relativity is understood to be that there should be no prior geometry, i.e., that the geometry is fully defined by the matter distribution.

How the terms {\it internal observer} and {\it external structure} in the previous description should be interpreted in the framework of the condensed matter model will become clear in the remainder of this chapter. But there is already an important point to be made here. If analogue gravity has reminded us that general relativity encompasses two different aspects, kinematical and dynamical ones, then perhaps the same distinction should be made when discussing diffeomorphism invariance. Indeed, it is in principle perfectly possible to have a system in which part of the physical laws---in particular, the equations of motion governed by the geometry (taken as instantaneous or stationary)---are diffeomorphism invariant, whereas another part---the dynamics that governs the evolution of this geometry---is not. We will therefore discuss diffeomorphism invariance in condensed matter systems at these two levels: the kinematical level and the dynamical level.

\section{Diffeomorphism invariance at the kinematical level}
\label{SS:sakharov}
In general relativity, given a particular geometry, there are many metrics that correspond to this geometry. One could formalise this by saying that the geometry is defined by $[g_{\mu\nu}]$, the equivalence class of all diffeomorphism equivalent metrics. The question with respect to condensed matter models is then whether the effective spacetimes that are obtained are really described by geometries in the sense just defined, rather than by individual metrics that are distinguishable due to the presence of the background structure.

To reformulate the question in terms of background (in)dependence, remember that in condensed matter models, we are dealing with bi-metric systems. The atoms of the condensate can be thought of as obeying the Newtonian laws of physics, and in particular Galilean relativity, since $v\ll c_\text{light}$ for the velocity $v$ of the condensed matter fluid. The excitations inside the condensed matter systems, on the other hand, see a relativistic geometry. This relativistic geometry is in a certain sense embedded in the laboratory frame, which provides a fixed background, and hence a privileged coordinate system. Then, if background independence is apparently violated, how could such a setup lead to a diffeomorphism invariant theory? 

A naive answer to these questions is the following. An observer inside, for example, a submarine could perform an `acoustic' experiment of the Michelson-Morley type~\cite{Michelson:1887} in the water surrounding him to find out whether the submarine is at rest or moving with respect to the water. An acoustic signal should here be interpreted as any signal in the form of a relativistic massless field perturbation, moving at the speed of sound $c_s$ (rather than the speed of light $c_\text{light}$).\footnote{In this section, we will explicitly differentiate the speed of sound $c_s$ and the speed of light $c_\text{light}$ in order to avoid any possible confusion.} Then, in the simplest form of the Michelson-Morley experiment,\footnote{Modern versions such as~\cite{Muller:2003zzc} are of course technically much more refined than the original Michelson-Morley experiment, but the argument given here is unaffected by the concrete implementation.} the observer would take an interferometer with two perpendicular arms of length $L$ and send acoustic signals along the arms. Mirrors at the end of the interferometer arms would reflect the signals back to a common point, where the observer looks at the interference pattern. If the interference fringes move when the interferometer is rotated, then the velocity of the submarine is not equal along the directions of both arms, and hence the observer concludes that the submarine moves with respect to the `absolute' reference frame provided by the water. So, the observer can determine his state of movement with respect to the water, and diffeomorphism invariance is broken by the privileged reference frame provided by the background fluid. 

However, the previous argument is not a satisfactory answer to the problem that we are discussing in the context of condensed matter models. The reason lies essentially in the characteristics of the observer, and of the building blocks available for him to carry out his experiment. To refine the answer, we need to go back to the concept of a bi-metric theory. The exterior world is the world of the laboratory, and of its Newtonian physics, where the basic building blocks are the condensate atoms and their microscopic physics. The interior world consists of everything that can be described in terms of collective excitations (phonons, quasi-particles) emerging inside the condensed matter system. An internal observer is then somebody who is limited to the observation of the interior world, and hence to the manipulation of these collective excitations emerging inside the condensed matter system. If the observer in the submarine is really an internal observer in a condensed matter world, the building blocks available for him to build an interferometer would necessarily have to be made out of these same collective excitations emerging inside the condensed matter system. The fundamental signalling velocity of those building blocks is the speed of sound $c_s$ of the fluid system rather than the speed of light $c_\text{light}$ of the laboratory. 

Let us work out this idea and rebuild the Michelson-Morley experiment from scratch for an internal observer inside a condensed matter system (see~\cite{Barcelo:2007iu} for more details). We will take the simplest case of a homogeneous medium at rest in the laboratory frame. The associated internal or effective metric is then Minkowskian. The internal observer tries to build a `quasi-interferometer' from quasi-particles and wants to obtain arms that are as rigid as possible. 

When Einstein introduced special relativity~\cite{einstein1905}, he postulated rigid bars (and perfect clocks) without further ado, although he has later repeatedly declared his dissatisfaction with the aprioriness of this assumption~\cite{brown2005}. Let us see what kind of structures an internal observer such as the one we are contemplating would have to look for, when trying to construct rigid bars for his experiment. 

Phenomenologically speaking, we know that rigid structures get their rigidity from the fact that their constituents are arranged in a regular and stable way. In particular, the equilibrium distance $a_0$ between two constituents (atoms, for example) is determined by a minimum in the interaction energy $E(a)$ between both atoms:
\begin{eqnarray}
 \frac{dE}{da}=0 \; \Longrightarrow \; E_0,a_0~.
\end{eqnarray}
Rigidity assumes a sharp minimum at $a_0$ in the $E(a)$ function, such that a small deviation $\delta a$ from the equilibrium distance $a_0$ would lead to a large energetic disadvantage $\Delta E$. Generally speaking, this corresponds to the requirement that the interatomic potential must be, for example, of the Lennard-Jones type. Such potentials are typically realised in nature by the dipolar interaction between globally neutral objects composed of charged substructures. At large distances $a\gg a_0$, the forces between the objects will be approximately zero. When they are brought closer together, but still $a>a_0$, opposite charges will rearrange themselves and the two objects will attract each other, with a force which becomes stronger as $a$ decreases. However, if the objects are forced to a separation $a<a_0$, then the charges will redistribute and due to the repulsion between like charges, a strong repellent force between the objects will come into play. 

Note that we are using the term `charge' here in a generalised way. Any charge with the characteristics just described serves our purpose. In particular, in $^3$He-A, chiral fermionic quasiparticles arise at low temperature together with an effective electromagnetic field $A_\mu$~\cite{Volovik:2003fe}. This effective theory is described by a $U(1)$ charge symmetry identical to the one of quantum electrodynamics, but with massless fermions. This  effective, `quasi-electrical' charge is a perfect example of what we have in mind.

Since the observer wants to keep the bars at rest within the system, we furthermore assume that he has {\it massive} charged quasi-particles at his disposal. This mass could for example originate in multi-component condensed matter systems (see~\cite{Visser:2005ss} for the case of a two-component BEC), in quasiparticles with an intrinsic internal structure (e.g., a small pipe through which the background fluid is pumped~\cite{Milgrom:2006wf}), or from a symmetry-breaking mechanism like the Brout-Englert-Higgs mechanism~\cite{Englert:1964et,Higgs:1964pj} which is generally assumed to provide mass to the particles of the standard model.\footnote{It is perhaps useful to point out that the idea for the Brout-Englert-Higgs mechanism is an excellent example of a condensed matter analogy applied in high-energy physics, since it was inspired precisely by the effective mass acquisition process in superconductors~\cite{Anderson:1963pc,Nambu:1961tp}.}

So assume charged massive `elementary quasi-particles' are available to the internal observer,  which he can use as building blocks to compose globally neutral constituents. Also, assume that these quasi-particles interact through the use of photon-like signals moving at a velocity $c_s$, which can be described as relativistic collective excitations corresponding to an effective field $A_\mu$, as in the case of $^3$He-A mentioned before. Furthermore, we assume that if there are different types of elementary quasi-particles, they all interact with signals propagating at the {\it same} velocity $c_s$. This condition is related to the principle of equivalence in general relativity, and can easily be fulfilled if one assumes (as in the BEC model discussed previously throughout this thesis) that the velocity of these wave signals is determined by the properties of the background condensate, and not by the intrinsic properties of the signals themselves. 

One might argue that we of course know that all these conditions are satisfied, since what we here call the quasi-particles created by the hypothetical microscopic condensed-matter-like system are precisely what we, as internal observers of our universe, perceive as the `real' particles of the standard model, and hence their conglomerates interact through electromagnetic signals. The point is that {\it any} type of interaction with {\it any} type of charge and {\it any} type of mass-acquisition process would do, as long as there is a constant signalling speed (or at least an invariant limiting speed) and a Lennard-Jones type of potential. Hence the construction that the internal observer in the condensed-matter system adopts to build rigid bars will inevitably lead to equations of exactly the same form as those that we know from classical electrodynamics, as we will now illustrate.

Let us first consider a single massive quasi-particle with a charge $q$, acting as a source for the relativistic field $A_\mu$. This field will then necessarily obey the following equations:
\begin{eqnarray}
& \Box A_\mu - \partial_\mu (\partial^\nu A_\nu)= j_\mu~;
\\
& j_\mu= \left\{
-q\delta^3[\x-\x(t)], q (\v /c_s) \delta^3[\x-\x(t)]~.
\right\}
\end{eqnarray}
For a source at rest, and taking the Lorenz gauge ($\partial^\mu A_\mu=0$), the solution is
\begin{eqnarray}
A_0 = -\frac{q}{[(x-x_0)^2+(y-y_0)^2+(z-z_0)^2]^{1/2}}~; ~~~~ A_i=0~.
\end{eqnarray}
Now assume that the source moves at a velocity $v$, for example in the $x$ direction. Then, because of the finite speed of signal propagation $c_s$, the solution becomes
\begin{eqnarray}
 &&A_0(x) = 
-\frac{q \gamma_s}{[(\gamma_s(x-v t) - \gamma_s x_0)^2+(y-y_0)^2+(z-z_0)^2]^{1/2}}~; \nonumber
\\
&&A_x(x)= \frac{q \gamma_s (v/c_s)}{[(\gamma_s(x-v t) - \gamma_s x_0)^2
+(y-y_0)^2+(z-z_0)^2]^{1/2}}~;
\\
&&A_y(x)=A_z(x)=0~.\nonumber
\end{eqnarray}
From this solution, we see that the fields decay faster in the $x$ direction than in the orthogonal $y$ and $z$ directions. The ratio between both decays is given by an acoustic or sonic Lorentz factor
\begin{eqnarray}
\gamma_s=(1-v^2/c_s^2)^{-1/2}~.
\end{eqnarray}
Since the internal observer is building an interferometer arm, he will need to add more quasi-particles. If a single quasi-particle acquires a $\gamma_s$ factor in its direction of movement, then two particles in movement, aligned along the direction of movement, will experience a modified interaction energy potential given by $E'(a)=E(\gamma_s a)$, where the prime denotes quantities in the co-moving reference frame. As before, the equilibrium distance is given by the minimum of this potential:
\begin{eqnarray}
 0=\frac{d E'(a)}{d a} \Longrightarrow E'_0,a'_0~.
\end{eqnarray}
We obtain
\begin{eqnarray}
0=\frac{d E'(a)}{d a}= \gamma_s \frac{d E(a')}{d a'}~; ~~~~~~~~~ 
a'= \gamma_s a~,
\end{eqnarray}
which shows that the minimum now occurs when $a'=a_0$, i.e., when $a=\gamma_s^{-1} a_0$. So the `real' distance (in the sense of the distance measured in the absolute reference frame provided by the laboratory) between the two quasi-atoms has decreased by an acoustic Lorentz factor $\gamma_s$ due to their velocity with respect to the medium. 

This factor $\gamma_s$ is precisely the length contraction (the contraction of an interferometer arm oriented in the direction of motion) needed for the interference pattern to remain unaffected by a uniform motion. The conclusion is then the following.  A Michelson-Morley type of experiment using a quasi-interferometer
does {\it not} allow internal observers to distinguish between rest and (uniform) movement with respect to the condensed matter medium (remarks to this effect were also made in~\cite{Volovik:2003fe,Liberati:2001sd}). 

We have explicitly worked out the case of a homogeneous condensate, in which the internal geometry is Minkowskian. But the essence of the argument is the same for more complicated situations: the internal observer has no way of connecting to the microscopic external world, at least not using only internal geometrical tools. Therefore, from an internal point of view, there is no way of connecting to the background structure and discriminating between the various metrics that describe the same given geometry. In other words: diffeomorphism invariance is effectively realised at the kinematical level.

Note that we have assumed that the wave perturbations propagate at the constant speed of sound $c_s$. As we have seen in the case of BECs, high-frequency perturbations might acquire frequency-dependent speeds of propagation, which incorporate sub- or supersonic corrections with respect to the (low-frequency) speed of sound $c_s$. Therefore, diffeomorphism invariance at the kinematical level is respected to the same order at which Lorentz invariance is, but can be croken at high frequencies. In this sense, diffeomorphism invariance might be `emergent', at least at the kinematical level, in the sense of being valid at the effective or internal level, but not necessarily at the fundamental level, i.e., in the external world. To put it differently: if local Lorentz invariance is a low-energy effective symmetry, violated at some high energy, then the same is true for diffeomorphism invariance at the kinematical level.

\section{Diffeomorphism invariance at the dynamical level}
In the previous section we discussed diffeomorphism invariance at the kinematical level. This requirement implies that the internal phenomena or observers cannot see the external degrees of freedom, but only the internal geometry. We argued that this requirement is fulfilled in condensed matter models for emergent gravity. Similarly, diffeomorphism invariance at the dynamical level can be understood as the requirement that the dynamical equations for the geometry only depend on the internal degrees of freedom, and not on any external or pre-existing degrees of freedom. 

The simplest dynamical equations that are diffeomorphism invariant and describe gravity in a way which agrees with observation are the Einstein equations~\cite{Will:2005va}. The dynamical equations obeyed by the geometries of condensed matter systems are not diffeomorphism invariant, they are hydrodynamical. For example, in the case of BECs, the Gross--Pitaevskii equation depends on external or microscopic degrees of freedom such as the interatomic potential and the scattering length. 

More in general, since it is a key requirement of quantum gravity to reproduce the Einstein equations in some adequate limit, it is essential within the condensed matter approach for emergent gravity to understand how the Einstein equations could be recovered, and if this is not possible, to at least understand why. This is largely an open problem, although some promising steps have been taken, see e.g.~\cite{Volovik:2006ft}. Here, we will limit ourselves to some basic remarks with respect to the possibility of obtaining the Einstein equations based on an induction scheme {\it \`{a} la} Sakharov. We emphasise that, from the point of view advocated here, the problem of dynamical diffeomorphism invariance and the problem of recovering the Einstein equations are essentially identical (up to higher-order curvature terms).

The basic idea behind Sakharov's induced gravity proposal~\cite{Sakharov:1967pk} can be understood as follows~\cite{Visser:2002ew,Birrell:1982ix}. It is well known that the {\it classical} Einstein equations $G_{\mu\nu}=\kappa T_{\mu\nu}$ (with $\kappa=8\pi G/c^4$) can be obtained via an extremal procedure from an action 
\begin{equation}
S=S_{\Lambda}+S_\text{EH}+S_m~,
\end{equation}
where $\Lambda$ is the cosmological constant, $S_\text{EH}$ the Einstein-Hilbert action and $S_m$ the action for the matter. In particular, in units such that $\hbar=c=1$:
\begin{equation}
S=\int d^4x\sqrt{-g}\left[{-\Lambda-\frac{R}{16\pi G}+{\cal L}_m}\right],
\end{equation}
with $g$ the determinant of the metric, $R$ the Ricci scalar and ${\cal L}_m$ the matter Lagrangian.

By analogy, one can seek a quantity $W$, called the {\it effective action}, such that minimising it with respect to the inverse metric $g^{\mu\nu}$ reproduces the {\it semiclassical} Einstein equations
\begin{equation}
G_{\mu\nu}=\kappa\langle T_{\mu\nu}\rangle~,
\end{equation}
in which the gravitational field can be treated classically even though the matter fields may be quantised.

Assume that a Lorentzian manifold with metric $g_{\mu\nu}$ is given, with quantum fields propagating on it, but without any prior dynamics. Then, such an effective action can indeed be constructed at one-loop order, based, e.g., on path-integral quantisation techniques~\cite{Birrell:1982ix}. For a scalar field with mass $m$, for example, this one-loop effective action $W_1$ can be written as~\cite{Visser:2002ew}:
\begin{equation}\label{w1}
W_1=-\frac{1}{2}\ln\det\left(\square+m^2+\chi R\right)=-\frac{1}{2} \text{Tr}\ln\left(\square+m^2+\chi R\right),
\end{equation}
where $\chi$ represents the coupling between the scalar and the gravitational field.

One can isolate the potential high-frequency divergences in $W_1$  by using a regularisation technique, for example the introduction of an explicit ultraviolet (UV) cut-off $\alpha_\text{uv}$, and expanding this action in orders of the curvature. Comparing this to the generalised Lagrangian for Einstein gravity (including all allowed invariant curvature-squared correction terms)
\begin{equation}\label{generalised-Einstein}
S=\int d^4x\sqrt{-g}\left[{-\Lambda-\frac{R}{16\pi G}+K_1C_{\mu\nu\rho\sigma}C^{\mu\nu\rho\sigma} + K_2 R^2 +
{\cal L}_m}\right]~,
\end{equation}
where $C_{\mu\nu\rho\sigma}$ is the Weyl curvature tensor, one sees that regularising the divergences of the one-loop effective action $W_1$ {\it automatically} induces a term which can be interpreted as the cosmological constant, one which provides the Newton constant, and curvature-squared terms in both $C_{\mu\nu\rho\sigma}C^{\mu\nu\rho\sigma}$ and $R^2$. In particular, such a comparison leads to
%
%
%
\begin{eqnarray}
\Lambda&=&\Lambda_0-\sum_i(-1)^F f_\Lambda(\alpha_\text{uv},m_i,g_{\mu\nu})+...\label{lambda-sakharov}~,\\
\frac{1}{G}&=&\frac{1}{G_0}-\sum_i(-1)^F f_{1/G}(\alpha_\text{uv},m_i,g_{\mu\nu})+...~,
\end{eqnarray}
for the cosmological and gravitational `constants' $\Lambda$ and $G$, plus similar equations for the second-order curvature terms~\cite{Visser:2002ew}. In these equations, the subindex $0$ refers to an arbitrary reference metric $g_0$ (for example, Minkowski), the summation is over all particle types $i$, with the factor $(-1)^F$ indicating that a relative minus sign should be added for fermions, and the ellipses represent non-UV-divergent terms. 

Several strategies are possible on what to do with this result. Sakharov's own proposal~\cite{Sakharov:1967pk} implicitly assumes dominance of the one-loop terms such that $\Lambda_0$ and $1/G_0$ can be neglected in the above equations. Furthermore, he explicitly suggests using the experimentally known value of $G$ to estimate the ultraviolet cut-off $\alpha_\text{uv}$, and emphasises that the magnitude of the gravitational interaction depends on the masses of the particles and their equations of motion (through the original metric). An alternative strategy~\cite{Visser:2002ew} would be to aim for one-loop finitess of $\Lambda$ and $G$ (and, if possible, also of the second-order curvature terms), and use this requirement to extract conditions on the particle families $i$. However, this turns out to produce unrealistic constraints, certainly not satisfied by the standard model and its most common extensions. This is mainly due to the fact that the obtention of a finite $\Lambda$ requires not only a compensation between the number of fermionic and bosonic particle families, but also an incredible fine-tuning between their respective masses.

In any case, for our present purposes, the key point is that the Einstein equations (plus possible higher-order curvature correction terms) can be obtained automatically from quantum field theory via an integration over the matter degrees of freedom. This idea seems very appealing in the present context, and agrees well with the basic idea of gravity as a macroscopic emergent phenomenon.\footnote{Sakharov himself talks of a ``metrical elasticity'' of space~\cite{Sakharov:1967pk}.} Moreover, given the argument discussed in section~\ref{SS:dark_energy} that the cosmological constant is regulated by thermodynamical principles, the fine-tuning problem just mentioned with respect to the one-loop finiteness of $\Lambda$ would also automatically be solved.

However, let us repeat the assumptions stated above. We assumed a Lorentzian manifold with metric $g_{\mu\nu}$, irrespective of how this is obtained, but {\it without any prior dynamics}. An Einstein-like dynamics is then spontaneously induced. In condensed matter models, the condition of no-prior-dynamics is violated due to the presence of hydrodynamics. An apparent way out would be to define an `abstract' condensed matter system without any prior dynamics. However, in the approach discussed in this thesis, the effective Lorentzian spacetime is precisely obtained from the hydrodynamic equations (from the Gross--Pitaevskii equation in the case of BECs, or more generally from the Euler and continuity equations, see chapter~\ref{S:preliminaries}). Therefore, a direct extension of the current setup does not seem to resolve the problem. One would need to redefine a microscopic system from scratch, in such a way that the effective spacetime is obtained without needing to appeal to hydrodynamics, thereby allowing the Einstein equations to be obtained through Sakharov's mechanism. 

For example, in~\cite{Barcelo:2001tb,Barcelo:2001ah}, it was shown that an effective Lorentzian spacetime can be obtained quite generally from the linearisation of a first-order Lagrangian ${\cal L}(\phi,\partial_\mu\phi)$ in terms of a single scalar field $\phi$, and even from a
wide class of second-order hyperbolic partial differential equations, without recurring to any Euler or continuity equation. A formal condition was also derived for the induced dynamics of the system to be `Einstein-like'. However, this formal condition seems to require some fine tuning of which it is not clear how it could be implemented physically, while there also arise serious technical complications when considering several fields~\cite{Barcelo:2001cp}. Alternatively, the following reasoning might provide interesting clues.

In the context of fermionic systems, it has been argued that an induction procedure in the spirit of Sakharov's proposal can nevertheless be carried out, even in hydrodynamic systems~\cite{Volovik:2003fe,Volovik:2006ft,Volovik:2007vs}. Moreover, it is argued that the resulting dynamics depends on the relation between the characteristic scales in the system. More precisely, in some fermionic systems such as
$^3$He-A, apart from the Lorentz violation scale $E_L$, a second characteristic energy scale $\eplanck$ appears, which determines 
the scale below which  part of the fermions can form a bosonic
condensate and the quasiparticles see an effective gravitational
field. This energy scale $\eplanck$ can be interpreted as (the analogue of) the Planck scale ($\eplanck\equiv E_{Pl}$)~\cite{Klinkhamer:2005cp,Barcelo:2009-qng}. As in the case of BECs (see the brief discussion in chapter~\ref{S:intermezzo}), these two scales $E_L$
and $\eplanck$ are in principle unrelated. According to the argument in ~\cite{Volovik:2003fe,Volovik:2006ft,Volovik:2007vs}, the dynamical laws for this gravitational field can be obtained by an induced-gravity procedure in the spirit of
Sakharov's proposal. 

In general, this will
generate the Einstein equations plus non-diffeomorphism-invariant higher-order terms in
powers of  $\eplanck/E_L$. Since for all known real laboratory
condensed matter systems, $\eplanck\gg E_L$, these higher-order
terms are actually dominant and the dynamical equations satisfied by the
laboratory geometries are not diffeomorphism invariant, but
hydrodynamical. If however $\eplanck\ll E_L$,
then the induction procedure would produce the Einstein equations plus
higher-order
perturbative corrections in $\eplanck/E_L$. 

Note that this argument is in line with the experimental constraints mentioned in chapter~\ref{S:intermezzo}, namely that extrapolations from recent cosmic ray and other high-energy
experiments show that there exist stringent bounds on the most
commonly expected types of Lorentz violation at the Planck
level~\cite{Jacobson:2005bg,Maccione:2009ju}, and hence, if Lorentz violations occur at all, they probably occur at energies much higher than the Planck scale: $E_L\gg E_{Pl}$. This observation adds strength to the argument that we just sketched, and reinforces the idea of treating gravity in an emergent framework. However, it is probably fair to admit that we need a more detailed understanding of this line of argumentation before any strong conclusions can be drawn from it with respect to the naturalness of the emergence of the Einstein equations.

As a final note, the previous reasoning was defended in the context of systems where the fundamental degrees of freedom are fermionic, leading to fermionic quasi-matter and collective bosonic fields. However, an induction procedure such as Sakharov's can in principle also be carried out in a bosonic system. Whether the Einstein equations can indeed be derived in a purely bosonic system is largely an academic question, since the degrees of freedom of the effective matter would then in principle also be purely bosonic, contrarily to the case of our universe. However, in terms of the naturalness or universality of the Einstein equations, the question might have some relevance. We do currently not have a clear view on this question. On the one hand, the fact that our universe contains effective fermionic matter, and that the Einstein equations---at least at the classical level---essentially depend on this effective matter content for the interplay with the geometry, seems to indicate that a system with only bosons at the fundamental level (and certainly a relatively simple system such as a BEC, let alone in the weakly interacting regimes that we have discussed in this thesis) will not generate sufficient complexity to reproduce the Einstein equations.\footnote{A similar conclusion is reached by the authors of~\cite{Girelli:2008gc}, who study a toy model based on a BEC but with massive quasi-particles, and obtain that the analogue gravitational dynamics is encoded in the form of a modified Poisson equation. In~\cite{Girelli:2008qp}, the same authors propose an alternative way to obtain effective gravitational dynamics in an abstract toy model related in spirit to~\cite{Barcelo:2001tb,Barcelo:2001ah,Barcelo:2001cp} mentioned above. Interestingly, this model is based on a fundamentally Euclidean (and hence `timeless') theory, but nevertheless an effective Lorentzian signature is obtained for the perturbations. Although again it seems that there is not a sufficient amount of complexity being generated to obtain the Einstein equations, the effective dynamics is nevertheless explicitly diffeomorphism invariant. In particular the gravitational dynamics is obtained in the form of a scalar theory of gravity, namely Nordstr\"{o}m gravity.} On the other hand, however, in the light of our above argument on the relation between the Einstein equations and diffeomorphism invariance, it might be sufficient to define a sufficiently complex diffeomorphism invariant system, whether fermionic or bosonic, to obtain the Einstein equations. 

In any case, although it is not clear if Bose--Einstein condensates are a relevant model for the dynamical aspects of gravity, they certainly have a lot to teach us about its kinematical aspects.

\section{Summary}
In this chapter, we have discussed some aspects of a model for emergent gravity based on the condensed matter analogy. In such a model, classical gravity is thought of as an effective property emerging at low energies from the collective behaviour of a large number of microscopic constituents or `spacetime atoms'. The quantum-gravitational degrees of freedom would then not be obtainable directly (for example, through a quantisation procedure) from the classical degrees of freedom of gravity.

A first motivation for such an approach is the observation that most of our universe is extremely cold compared to the Planck scale, hence the quantum-gravitational degrees of freedom are probably effectively frozen out in the physics that we observe. A second motivation comes from the accelerating expansion of our universe, which seems to require some form of `dark energy'. We discussed the argument that, from a condensed matter point of view, the quantum vacuum energy is really the natural candidate for such a dark energy, and provides a framework to understand both the coincidence problem (why $\Omega_\Lambda\sim\Omega_M$) and the cosmological constant problem (why the zero-point energy of the vacuum is not huge).

The main part of this chapter was dedicated to a discussion of diffeomorphism invariance in emergent gravity. We argued that this issue should be split up into a kinematical and a dynamical aspect. With respect to the kinematical aspect, we showed that diffeomorphism invariance can emerge as a low-energy effective property at the same level as Lorentz invariance. We discussed how the dynamical aspect of diffeomorphism invariance is related to the problem of recovering the Einstein equations in a model based on the condensed matter analogy. We made some observations with respect to the possibility of achieving this through an induction procedure such as Sakharov's proposal. The essential problem, as we pointed out, is that Sakharov's procedure assumes the absence of any prior dynamics, whereas in the condensed matter analogy, the hydrodynamical equations are essential to obtain the emergence of an effective spacetime. 

We mentioned statements in the literature, in the context of fermionic systems such as $^3$He-A, which argue that a procedure similar to Sakharov's approach could be used to calculate the effective gravitational dynamics, even in a hydrodynamic system, and would lead to the Einstein equations plus higher-order correction terms, provided that the Lorentz violation energy scale lies far above the (analogue) Planck scale. There exists no known laboratory condensed matter system which fulfils this requirement. However, it is not clear whether this is due to some fundamental reason, or whether it is precisely an illustration of the fact that quantum gravity can be treated as a condensed-matter-like system, albeit---precisely for this relation between the characteristic scales---an exceptional one in comparison with laboratory condensed matter systems. 

Finally, we noted that such an induction procedure could in principle also be applicable to bosonic systems, although it is not clear at present whether this would be sufficient to recover the Einstein equations.

\clearpage
\chapter{General summary and outlook}
\label{S:summary}
We have discussed various aspects of the gravitational analogy in Bose--Einstein condensates (BECs), and of how such an analogy can serve as a source of inspiration in quantum gravity, both in terms of quantum gravity phenomenology and in an attempt to understand some more conceptual issues related to quantum gravity.

We started with a review that led us from mean-field theory in BECs to the emergence of an effective acoustic metric for linear perturbations of the background condensate in the hydrodynamic limit. We studied the validity of this hydrodynamic limit, and emphasised that it is violated at high frequencies and in the presence of a horizon. In such cases, the effective geometric or relativistic description breaks down, giving way to the underlying microscopic theory. This behaviour is similar to what is expected in gravity, with the important difference that, in the case of BECs, the underlying microscopic theory is well understood. Moreover low-temperature experiments with, for example, dilute alkali gases are in accordance with the theoretical treatment of BECs to a high degree of accuracy. We also stressed that the full dynamical aspect of general relativity (the Einstein equations) are not included in the analogy, at least not in its direct application in BECs, a problem to which we came back in the previous chapter.

A first, general and well-known conclusion is that some phenomena related to black hole physics, such as Hawking radiation, can in principle be simulated in BECs, both theoretically and experimentally.  Several of the issues that we discussed in this thesis directly affect Hawking radiation and its possible experimental realisation. Generally speaking, though, the original contribution of this thesis lies beyond the basic observation of the possible simulation of black-hole related phenomena. In particular, we focused on how the gravitational analogy in BECs can be extended to study some aspects of quantum gravity, both in the sense of quantum gravity phenomenology and in terms of constructing a toy model for emergent gravity.

The crucial element in the theory of BECs that makes it relevant for the study of quantum gravity phenomenology is the following. Beyond the hydrodynamic approximation, the first corrections to the relativistic description in a BEC are encoded in the form of higher-order, superluminal modifications of the dispersion relation. These superluminal corrections are associated with the existence of a privileged reference frame, and thereby lead to a breaking of the effective Lorentz symmetry. They therefore form an excellent testbed for quantum gravity scenarios which consider a similar possibility of high-energy violations of local Lorentz invariance, and in particular of those scenarios, for example based on effective field theory, where the first modifications of the dispersion relation that are allowed by observations are of exactly the same kind as those that occur in a BEC.

Two possible strategies could then be developed. The first one consists in elaborating the gravitational analogy in a BEC, proposing a concrete configuration, and studying some of its aspects inspired by relativistic phenomena of interest. 

We applied this first strategy in the following way. We presented a simple (1+1)-dimensional black hole configuration, and studied its dynamical stability based on the full Gross--Pitaevskii equation (i.e., incorporating the superluminal corrections to the dispersion relation). We found that such configurations are devoid of dynamical instabilities under rather generic boundary conditions inspired directly by those used in the study of general relativistic black holes. This opened the way to study their stable dynamical modes or quasinormal modes, and moreover confirmed that, possible experimental complications apart, it should indeed be possible to reproduce the phononic analogue of Hawking radiation in a BEC.

With respect to the quasinormal modes of black hole configurations, we found that in the particular (1+1)-dimensional setup that we studied, no quasinormal modes could exist in the hydrodynamical limit. We remarked that in general relativity, similarly, no quasinormal modes can exist in 1+1 dimensions, because of the conformally flat character of (1+1)-dimensional spacetimes in combination with the conformal invariance of the d'Alembertian wave equation. In our BEC model, a continuous spectrum of short-lived quasinormal modes was found to exist beyond the hydrodynamical limit, i.e., when using the full dispersion relation. This result is surprising in the sense that in general relativity, even in 3+1 dimensions, the quasinormal modes form a discrete spectrum. We pointed out that the origin of these quasinormal modes lies in the superluminal character of the dispersion relation, which leads to a permeability of the (zero-frequency) horizon. Ultimately, this means that these quasinormal modes are characteristic of a relaxation at the microscopic level, which expresses itself in the form of lifetimes of the order of the Lorentz violation scale (determined by the healing length, and hence ultimately by the scattering length, in the case of BECs). We therefore speculated that a similar continuous spectrum of short-lived `planckian' quasinormal modes could show up in the spectrum of gravitational black holes, if quantum gravity also turns out to incorporate superluminal high-energy corrections.

A second strategy consists in directly modifying the physics that describes phenomena related to black holes in a way which incorporates superluminal corrections at high energies. We applied this strategy to the Hawking radiation of a dynamically collapsing configuration. 

We stressed the fact that the superluminal dispersion causes the effective speed of light to become frequency-dependent, and hence also the horizon to become a frequency-dependent concept. In particular, the horizon for higher frequencies lies inside the zero-frequency or classical horizon, and moves towards the singularity as the frequency increases. As a corollary, several  quantities associated with the horizon, such as the surface gravity, can also become frequency-dependent. Moreover, the superluminal dispersion means that, at every moment of the collapse there is a critical frequency such that higher frequencies simply do not experience a horizon. If the collapse saturates at some level---in other words: unless a singularity is formed at the end of the collapse---then there also exists a global critical frequency. We stressed the fact that the violation of Lorentz invariance, and in particular the superluminal character of the dispersion relation, implies that the usual equivalence between a collapsing configuration and a stationary black-hole configuration (with adequate initial conditions on the horizon) is no longer valid, at least not in a way straightforwardly adapted from the Lorentz-invariant case. An easy way to see this is that, because of the frequency-dependence of the horizon just mentioned, such a construction would require imposing conditions arbitrarily close to the singularity. 

Three important consequences distinguished the outcome of our model from the standard (Lorentz-invariant) Hawking result. First, we saw that the existence of the critical frequency makes the overall radiation weaker than the standard Hawking spectrum. Second, the frequency-dependence of the surface gravity implies that the radiation spectrum can also undergo a modification in its form. In particular, if the Lorentz violation scale lies sufficiently above the critical frequency, important ultraviolet contributions can show up which are associated with frequencies whose effective horizon lies well inside the zero-frequency horizon (in other words: close to the singularity). Finally, the radiation coming from the collapse process is not stationary, but dies out with time. 

It is probably worth stressing that, in contrast to the extensive body of literature on the transplanckian problem in stationary black holes, our work provides the first calculation of the sensitivity of Hawking radiation to the influence of superluminal dispersion relations in a collapsing configuration. Moreover, the results that we discussed were obtained analytically, i.e., we have used numerical integration to illustrate them, but they can qualitatively be derived from the analytical formula that we obtained for the late-time radiation. This has necessarily involved some (physically motivated) approximations, in particular we have only traced the evolution of the original (Minkowskian) degrees of freedom. It would certainly be interesting to see whether a consistent treatment can be given which takes into account the possibility that the number of degrees of freedom abruptly increases during the collapse process, and how this would affect our results. One result that seems firmly established, though, is that the equivalence between collapsing configurations and stationary ones is no longer valid in the case of superluminal dispersion, at least not in a form straightforwardly extended from the standard case. In any case, apart from the intrinsic interest of our results, this should also be considered as an invitation to further study the robustness of Hawking radiation in collapsing configurations.

In the final part of this thesis, we illustrated how the basic ideas of the condensed-matter analogy can be applied to study some more conceptual problems related to quantum gravity. We gave several motivations for such an `emergent gravity' approach, in which classical gravity is treated as a low-energy effective description emerging from the collective behaviour of a large quantity of microscopic degrees of freedom (`spacetime atoms'). In particular, we mentioned two reasons to take such an approach seriously: the ultra-cold temperatures in most of the universe (compared to the Planck scale) and the accelerating expansion of the universe, i.e., dark energy.

We then discussed the question of diffeomorphism invariance in such a model for emergent gravity based on the condensed matter analogy. We argued that, at the kinematical level, diffeomorphism invariance is realised as a low-energy, effective symmetry, in the same way as Lorentz invariance. The heart of our argument consisted in pointing out that an internal observer inside a condensed matter system, confined to the manipulation of effectively relativistic acoustic signals and quasi-particles, would have no way to detect his absolute state of movement with respect to the laboratory frame. In other words, at the kinematical level, the internal laws of physics do not depend on any prior or external geometry, and so diffeomorphism invariance is indeed realised at this level.

At the dynamical level, the problem is somewhat more complicated. We argued that the question of dynamical diffeomorphism invariance is essentially identical to the problem of recovering the Einstein equations, and briefly discussed a possible strategy, namely an induction procedure {\it \`a la} Sakharov. We pointed out that a crucial element in this respect seems to be the relation between (the analogue of) the Planck energy scale $E_{Pl}$ and the Lorentz violation scale $E_L$. The fact that condensed matter systems do not reproduce the Einstein equations might simply be due to the fact that all known real laboratory condensed matter systems have $E_{Pl}\gg E_L$. It remains to be seen whether a fundamental reason can be given for this relation.

The global conclusions of our thesis are then the following. 

First, we have confirmed that there is no theoretical obstacle to the simulation of Hawking radiation in a BEC. Indeed, dynamically stable black-hole configurations are in principle possible. 

Second, we have seen that the superluminal corrections to the dispersion relation can cause strong qualitative modifications to relativistic phenomena. In the case of quasinormal modes, these modifications were quantitatively small. However, in the case of Hawking radiation in a collapsing configuration, we found that the changes could be important both qualitatively and quantitatively. It is perhaps worth pointing out that we avoided at all times to make assumptions about the general relativistic singularity in a gravitational black hole. Therefore, our results can be interpreted as a consequence of the influence of microscopic corrections to the horizon only, not to the singularity. In this sense, the conventional wisdom which states that quantum gravitational effects near the horizon can in general (i.e., except for extremely small black holes) only lead to small modifications with respect to classical gravity---namely, modifications of the order of the Planck length divided by the radius of the black hole---should be taken with some caution. Indeed, the modifications that we have found are essentially applicable to black holes of any size, provided that the dispersion relations acquire superluminal modifications at high energies. There is no need to overthrow the common wisdom to interpret this : the key issue is that superluminal modifications imply that there will always be some high frequencies which experience the black hole as extremely small, thereby introducing important corrections, while other, even higher frequencies simply see no black hole at all. 

Finally, a third global conclusion is that it seems very worthwhile exploring how far the idea that classical gravity might emerge from the collective low-frequency behaviour of some microscopic degrees of freedom, in a sense related to thermodynamics, can be taken, and perhaps even {\it should} be taken if we really want to understand some aspects related to quantum gravity.

An outline for future work is tentatively the following. 

First, several experimental condensed matter groups are showing an increasingly concrete interest in a possible experimental detection of Hawking radiation. Such an experiment would, at least in a first instance, be based on a stationary configuration, or on an externally imposed evolution such that the final state is enforced to resemble a stationary black hole. We can of course only be extremely enthusiastic about the possibility of such an experimental confirmation. At the same time, however, we strongly feel that more effort is needed to understand the precise relation between such stationary or externally enforced configurations and the dynamical formation of a black hole through a collapse process, before extracting any general conclusions from such an experiment about collapsing black holes, or before proclaiming the universal robustness of Hawking radiation. It might for example be interesting to go beyond the particular model that we have studied and examine the possible influence of superluminal dispersion on collapsing models that are closer to what is expected for astrophysical black holes and/or that could be simulated in a laboratory. As we mentioned before, the work presented in chapter~\ref{S:HR}, apart from the intrinsic interest of the results obtained, should therefore be considered as an onset and an invitation to further study these aspects.

Second, while important technical progress has been made in the traditional theoretical or `top-down' approaches to quantum gravity, none of these currently make any clear and definite predictions with respect to, for example, semiclassical corrections to general relativity. The approach to quantum gravity phenomenology that we have applied in this thesis therefore seems particularly relevant in bridging the gap between classical gravity and a possible `full' theory of quantum gravity, since it is based on a concrete and real physical example of how effective low-energy symmetries can be modified at high energies. It would, for example, be interesting to see if any quantitative predictions can be made with respect to the modification of the quasinormal mode spectrum due to possible superluminal corrections in an astrophysically realistic situation involving black holes.

Third and finally, and in our personal opinion perhaps most relevant, we believe that several of the problems encountered in quantum gravity are not (or at least: not only) of a technical kind, but are conceptual in origin. For example, none of the main approaches to quantum gravity offers a satisfactory explanation for the accelerating expansion of the universe. It therefore seems to us that there remains a lot to be learned from the condensed matter analogy, and in particular from how it can guide us in the construction of a model for emergent gravity. In this respect, the most urgent point is perhaps to acquire a more detailed understanding of the possible emergence of the Einstein equations from the collective behaviour of microscopic degrees of freedom.

As a final note, we have not explicitly defined what `the BEC paradigm for emergent gravity' precisely consists in. Perhaps curiously, although the term `paradigm' in the context of scientific practice was coined by Thomas Kuhn in {\it The Structure of Scientific Revolutions}~\cite{Kuhn:1962}, he nowhere gives an explicit definition of it. Even more curiously, in the second (1970) edition of the same book, Kuhn added a postscript in which he sets out to clarify ``the key difficulties (...) about the concept of a paradigm'', but again fails to provide an explicit definition. The closest he arrives at a definition is the following: ``The term `paradigm' is used in two different senses. On the one hand, it stands for the entire constellation of beliefs, values, techniques, and so on shared by the members of a given community. On the other, it denotes one sort of element in that constellation, the concrete puzzle-solutions which, employed as models or examples, can replace explicit rules as a basis for the solution of the remaining puzzles of normal sciences'' (\cite{Kuhn:1962}, 1970 edition, p.175). 

It should then come as no surprise that, just like the concept of a paradigm itself, the BEC paradigm for emergent gravity cannot be defined in a single, simple and precise sentence, but is a general framework of which we have set out to illustrate some of the most important aspects. However, the three points that we already mentioned in the introduction (possibility of experimental simulation, complementing theoretical top-down approaches for quantum gravity with bottom-up approaches rooted in `real' physics, and finally: take the idea of emergent gravity seriously) and repeated in our outline for future work  perhaps give an appropriate summary for the BEC paradigm, and by extension for the condensed matter paradigm for emergent gravity, while at the same time indicating that it is very much a paradigm under construction. Then again, as the popular saying indicates, {\it the road to success is always under construction,} and perhaps this is also valid for the road to quantum gravity. 

As Karl Popper remarked, ``Our knowledge can only be finite, while our ignorance must necessarily be infinite''~\cite{Popper:1963}.


\newpage
\cleardoublepage\addcontentsline{toc}{chapter}{References}



\end{document}